\documentclass[twocolumn,amsmath,amssymb,a4paper,prb,superscriptaddress,floatfix]{revtex4-1}
\usepackage[dvipdfmx]{graphicx}
\usepackage{natbib}
\usepackage{multirow}
\usepackage{amsmath}
\usepackage{bm}
\usepackage{mathrsfs}
\usepackage{url}
\usepackage{color}
\usepackage{ulem}
\usepackage{mathtools}
\usepackage{verbatim}
\usepackage{cleveref}
\usepackage{float}
\usepackage[export]{adjustbox}
\usepackage{algorithm}
\usepackage{algpseudocode}
\usepackage{braket}
\usepackage{empheq}
\usepackage{cancel}
\usepackage{simpler-wick}

\DeclareMathAlphabet{\mathpzc}{OT1}{pzc}{m}{it}

\begin{document}

\title{Ab initio calculations of the thermoelectric figure of merit, \\ within the relaxation time approximation}

\author{L. Chaput}
\email{laurent.chaput@univ-lorraine.fr}
\affiliation{Université de Lorraine, LEMTA, CNRS, UMR 7563}
\affiliation{Institut Universitaire de France}

\author{H. Miranda}
\affiliation{VASP Software GmbH}

\author{A. Togo}
\affiliation{Center for Basic Research on Materials (CBRM), National Institute for Materials Science (NIMS)}

\author{M. Engel}
\author{M. Schlipf}
\author{M. Marsman}
\affiliation{VASP Software GmbH}

\author{G. Kresse}
\affiliation{VASP Software GmbH}
\affiliation{University of Vienna, Faculty of Physics and Center for Computational Materials Physics}

\begin{abstract}
In this paper, we propose a computational framework, based on the VASP and phono3py computer codes, to obtain the thermoelectric figure of merit from the electron-phonon and phonon-phonon interactions using finite displacements in supercells.
Several numerical techniques are developed for efficiency. 

The method is applied to several thermoelectric materials. We found different behaviors for the lifetimes of the electrons in PbTe, PbSe, and in compounds of the half-Heusler and magnesium silicide family. This is traced back to the different frequencies of the phonons involved in the scattering around the Fermi level.  They have a lower frequency in  PbTe and PbSe.

The magnitude of the thermoelectric figures of merit we computed compare well with experiments, but the agreement is far from perfect. The role of the defects, not explicitly considered in our calculations, but abundant in thermoelectric materials, is discussed as a possible explanation. It is also shown that the choice of the exchange-correlation functional can strongly impact the results.
\end{abstract}

    \maketitle


\newpage

\section{Introduction}

The study of electronic and thermal transport properties, such as the electronic conductivity ($\sigma$), the thermopower ($S$), the Peltier coefficient ($\Pi$) and the thermal conductivity ($\kappa$) dates back to the beginning of the 19th century, often related with names such as Ohm, Seebeck, Peltier and Fourier.
About a century later, the electrical and thermal currents obtained from the coexistence of an applied electric field and a temperature gradient was investigated in a thermoelectric generator, whose goal was to produce an electrical current from a temperature difference. It was soon realized that the thermoelectric efficiency of a material in such a generator is maximal when the figure of merit $ZT=\frac{S^2 \sigma}{\kappa}T $ is maximal, even if the theory became fully transparent only with the derivation of Onsager's reciprocity relations and the simple justification of the Kelvin relation\cite{Taylor_Heinonen_2002,nye,groot}, $\Pi = T S^t$.

Several ab initio calculations of the thermoelectric figure of merit $ZT$ have been performed already\cite{SONG_2017,wu_2023,MA_2024,cappai_2025}, sometimes even including GW correction\cite{BRAHMA_2025} or the phonon drag effect\cite{Li_2022}. However, such calculations remain challenging because they require the computation of both the phonon-phonon and electron-phonon interactions.

During the last 10 years the computation of the phonon-phonon interactions has become possible\cite{togo2015,ShengBTE_2014}, and the computation of the lattice thermal conductivity may now be automatized using workflows. However, often, several methods still need to be combined to compute the electronic transport properties. One may first compute the electronic structure, perform a Wannier transform\cite{wannier90}, obtain the electron-phonon interaction\cite{RevModPhys.89.015003,EPW}, and then solve the transport equation\cite{BoltzTraP2,Protik_2022}. There are alternatives\cite{Li_2015,Brunin}, however, which encapsulate all those steps.

To be able to combine several methods has its own merit; it is sometimes the only possible approach to achieve a calculation. However, combining several methods also increases the complexity of the calculations. At each step the calculation may fail for a different reason. Moreover, different convergence parameters need to be considered for the different methods. This makes the calculations difficult to automatize.



In the present paper, we propose an ab initio calculation of the thermoelectric figure of merit based on the finite-difference method for both the calculation of the phonon-phonon and electron-phonon interactions. The computational methods that allow the determination of phonon-phonon and electron-phonon interactions have been discussed in detail in the literature\cite{togo2015,chaput2019,engel2022}. In the present work they are integrated in a common workflow to obtain the $ZT$.

The paper is organized as follows. In the first part of the paper (Sec.~\ref{transport_theory}), we briefly review transport theory to establish the notation for future reference. In Sec.~\ref{workflow}, the workflow of our calculations is explained, and it is applied to several thermoelectric materials in Sec.~\ref{materials}.

\section{Thermoelectric transport theory\label{transport_theory}}
\subsection{Thermoelectric transport for electrons}
In a semiclassical approximation, the electrical current ($\mathbf{J}^{e}$) and heat current ($\mathbf{J}^{e}_Q$) carried by the electrons can be written as
\begin{align}
\mathbf{J}^{e}&= \frac{1}{NV_0}\sum_{\mathbf{k}n} (-e)\mathbf{v}_{\mathbf{k}n}f_{\mathbf{k}n} \label{Je}\\    
\mathbf{J}^{e}_Q&= \frac{1}{NV_0}\sum_{\mathbf{k}n} (\epsilon_{\mathbf{k}n}-\mu)\mathbf{v}_{\mathbf{k}n}f_{\mathbf{k}n} \label{JQ}
\end{align}
where $V_0$ is the volume of the primitive cell in the crystal, and $N$ the number of such primitive cells. $-e$ is the charge of the electrons, and $\mu$ is the chemical potential.
$\epsilon_{\mathbf{k}n}$ and $\mathbf{v}_{\mathbf{k}n}$ are the energy and group velocity of the Bloch states $|\varphi_{\mathbf{k}n} \rangle$ with wave vector $\mathbf{k}$ and band index $n$. $f_{\mathbf{k}n}$ is the occupation of the state $|\varphi_{\mathbf{k}n} \rangle$ when an electrical field and a temperature gradient are applied to the crystal.

In our approach, $|\varphi_{\mathbf{k}n} \rangle$ are taken to be the Kohn-Sham states obtained from density functional theory. The chemical potential $\mu$ is computed from the number of electrons $n_0$ in the primitive cell, 
\begin{align}
    n_0 = \int_{-\infty}^{+\infty} d\epsilon \, \rho(\epsilon) f^0_{\mathbf{k}n} (\epsilon,T,\mu), \label{fermi}
\end{align}
where $f^0(\epsilon,T,\mu) $ is the Fermi-Dirac occupation function, 
\begin{align}
f^0(\epsilon,T,\mu) = \frac{1}{\exp{\frac{\epsilon-\mu}{k_B T}}+1},
\end{align}
with $k_B$ the Boltzmann constant and $T$ the temperature. $\rho(\epsilon)$ is the density of states, 
\begin{align}
 \rho(\epsilon)  = \frac{1}{N}\sum_{\mathbf{k}n} \delta(\epsilon-\epsilon_{\mathbf{k}n}).
\end{align}

In the following sections, when doped compounds will be considered, Eq. \ref{fermi} will be used to compute the chemical potential in the rigid band approximation. It means that this equation will be solved for $\mu$ taking for $n_0$  the number of electrons in the doped compound, but for $ \rho(\epsilon)  $ the density of states of the perfect crystal.

To compute the currents in Eqs.~\ref{Je} and \ref{JQ}, the velocity and the occupation function must be computed. The computation of the velocity is explained in appendix \ref{velocity} when the Kohn Sham equations are solved using the projector-augmented-wave (PAW) method\cite{PAW-Blochl-1994,VASP1993,VASP1994,VASP1996a,VASP1996b,VASP1999}. The occupation function $f_{\mathbf{k}n}(\mathbf{r})$ is obtained from a Boltzmann equation, 
\begin{align}
    \frac{\partial f_{\mathbf{k}n} }{\partial t} +\mathbf{v}_{\mathbf{k}n} \cdot \frac{\partial f_{\mathbf{k}n} }{\partial \mathbf{r}}+\frac{-e}{\hbar}\mathbf{E}\cdot \frac{\partial f_{\mathbf{k}n} }{\partial \mathbf{k}}&= \frac{\partial f_{\mathbf{k}n} }{\partial t} \Big|_{col}.
    \label{boltz}
\end{align}

\begin{widetext}

The right-hand side of the above equation is a collision integral and gives the change, per unit time, of the number of electrons in state $|\varphi_{\mathbf{k}n} \rangle$ due to collisions. In the present study,  we consider collisions caused by electron-phonon interactions. Therefore if we denote by $g(\mathbf{k}n,\mathbf{k}'n',\mathbf{q}j)$ the electron-phonon coupling strength of the  electron-phonon Hamiltonian
\begin{align}
H_{ep} &= \sum_{\mathbf{k}n}\sum_{\mathbf{k}'n'}\sum_{\mathbf{q}j} g(\mathbf{k}n,\mathbf{k}'n',\mathbf{q}j) (a_{\mathbf{q}j}+a^{\dagger}_{-\mathbf{q}j}) c^{\dagger}_{\mathbf{k}n}c_{\mathbf{k}'n'}, 
\end{align} 
where $a_{\mathbf{q}j}$ and $a^{\dagger}_{\mathbf{q}j}$ are the phonon annihilation and creation operators in mode $\mathbf{q}j$, and $c_{\mathbf{k}n}$ and $c^{\dagger}_{\mathbf{k}n}$ the electron annihilation and creation operators in Bloch state $\mathbf{k}n$, Fermi's golden rule gives
\begin{align}
\frac{\partial f_{\mathbf{k}n} }{\partial t} \Big|_{col}
   = \frac{2\pi}{\hbar}\sum_{\mathbf{k}'n'}\sum_{\mathbf{q}j}&|g(\mathbf{k}n,\mathbf{k}'n',\mathbf{q}j) |^2 
   \Big[\delta(\epsilon_{\mathbf{k}n}-\epsilon_{\mathbf{k}'n'}-\hbar \omega_{\mathbf{q}j})\Big( (1-f_{\mathbf{k}n})f_{\mathbf{k}'n'}n_{\mathbf{q}j} - f_{\mathbf{k}n}(1-f_{\mathbf{k}'n'})(1+n_{\mathbf{q}j})\Big)\nonumber\\
   +&\delta(\epsilon_{\mathbf{k}n}-\epsilon_{\mathbf{k}'n'}+\hbar \omega_{-\mathbf{q}j}) \Big((1-f_{\mathbf{k}n})f_{\mathbf{k}'n'}(1+n_{-\mathbf{q}j})-f_{\mathbf{k}n}(1-f_{\mathbf{k}'n'})n_{-\mathbf{q}j}\Big) \Big].  \label{col}
\end{align}
In the above equation the summations are performed over the first Brillouin zone, and the momentum conservation, $\mathbf{k}=\mathbf{k}'+\mathbf{q} \mod \mathbf{G}$, is included in the definition of $g(\mathbf{k}n,\mathbf{k}'n',\mathbf{q}j)$.
$n_{\mathbf{q}j}$ is the occupation function of phonons with mode $\mathbf{q}j$ and energy $\hbar \omega_{\mathbf{q}j}$, which at equilibrium equals the Bose-Einstein occupation function, 
\begin{align}
n^0(\omega,T) = \frac{1}{\exp{\frac{\hbar\omega}{k_B T}}-1}.
\end{align}

\noindent If the applied fields $\mathbf{E}$ and $\nabla T$ are small, the Boltzmann equation in Eq.~\ref{boltz} can be linearized. Writing  $f_{\mathbf{k}n}=f^0_{\mathbf{k}n}+\delta f_{\mathbf{k}n}$ and $n_{\mathbf{q}j}=n^0_{\mathbf{q}j}+\delta n_{\mathbf{q}j}$, with $f^0_{\mathbf{k}n}=f^0(\epsilon_{\mathbf{k}n},T,\mu) $ and $n^0_{\mathbf{q}j}=n^0(\omega_{\mathbf{q}j},T)$, and assuming $\delta f_{\mathbf{k}n}$ and $\delta n_{\mathbf{q}j}$
to be of order 1 in the applied fields, from chain rules over $\mathbf{r}$ and $\mathbf{k}$,
in steady state, we obtain\cite{marder}

\begin{align}
\Big( (\epsilon_{\mathbf{k}n}-\mu)\frac{\nabla T}{T}+e\bm{\mathcal{E}}\Big)\cdot  \mathbf{v}_{\mathbf{k}n}\Big(-\frac{\partial f^0_{\mathbf{k}n} }{\partial \epsilon_{\mathbf{k}n}}\Big)&= \frac{\partial f_{\mathbf{k}n} }{\partial t} \Big|_{lin}, \label{lin_boltz}
\end{align}
with $\bm{\mathcal{E}}=\mathbf{E}+\nabla \mu / e$ and
{\small
\begin{align}
\frac{\partial f_{\mathbf{k}n} }{\partial t} \Big|_{lin}
   =&- \frac{2\pi}{\hbar}\sum_{\mathbf{k}'n'}\sum_{\mathbf{q}j}|g(\mathbf{k}n,\mathbf{k}'n',\mathbf{q}j) |^2 
   \Big((1+n^0_{\mathbf{q}j}-f^0_{\mathbf{k}'n'})\delta(\epsilon_{\mathbf{k}n}-\epsilon_{\mathbf{k}'n'}-\hbar \omega_{\mathbf{q}j})
   +(f^0_{\mathbf{k}'n'}+n^0_{-\mathbf{q}j})\delta(\epsilon_{\mathbf{k}n}-\epsilon_{\mathbf{k}'n'}+\hbar \omega_{-\mathbf{q}j})\Big)  \delta f_{\mathbf{k}n} \nonumber\\
   &+\frac{2\pi}{\hbar}\sum_{\mathbf{k}'n'}\sum_{\mathbf{q}j}|g(\mathbf{k}n,\mathbf{k}'n',\mathbf{q}j) |^2 
   \Big((n^0_{\mathbf{q}j}+f^0_{\mathbf{k}n})\delta(\epsilon_{\mathbf{k}n}-\epsilon_{\mathbf{k}'n'}-\hbar \omega_{\mathbf{q}j})
   +(1+n^0_{-\mathbf{q}j}-f^0_{\mathbf{k}n})\delta(\epsilon_{\mathbf{k}n}-\epsilon_{\mathbf{k}'n'}+\hbar \omega_{-\mathbf{q}j})\Big)\delta f_{\mathbf{k}'n'}\nonumber\\
   &+\frac{2\pi}{\hbar}\sum_{\mathbf{k}'n'}\sum_{\mathbf{q}j}|g(\mathbf{k}n,\mathbf{k}'n',\mathbf{q}j) |^2 (f_{\mathbf{k}'n'}-f_{\mathbf{k}n})  \Big(\delta(\epsilon_{\mathbf{k}n}-\epsilon_{\mathbf{k}'n'}-\hbar \omega_{\mathbf{q}j})\delta n_{\mathbf{q}j}
   +\delta(\epsilon_{\mathbf{k}n}-\epsilon_{\mathbf{k}'n'}+\hbar \omega_{-\mathbf{q}j})\delta n_{-\mathbf{q}j}\Big). \label{lin}
\end{align}
}

To solve the Boltzmann equation, the phonon system is assumed to remain at equilibrium, $\delta n_{\mathbf{q}j}=0$, then the third line of the above equation vanishes. If one wishes to go beyond this approximation,  the phonon Boltzmann equation need to be solved self-consistently together with the one for electrons. This is, for example, needed to describe the phonon drag effect.

Even assuming the phonon system to be in equilibrium, the Boltzmann equation is still an integral equation, since the second line of the previous equation involves a sum over the Brillouin zone. It can be solved exactly, but in the present paper, we will use simpler approximations based on relaxation times.

The most common approximation is to simply neglect the second line of Eq.~\ref{lin}. It amounts to assume $\delta f_{\mathbf{k}'n'}=0$, or in other words, that the change in the number of electrons in any other state than the state under consideration, $\mathbf{k}n$, is not changed by the collisions. This is the usual relaxation-time approximation, now called self-energy relaxation-time approximation (SERTA) to emphasize that the collision integral can be written as
\begin{align}
\frac{\partial f_{\mathbf{k}n} }{\partial t} \Big|_\text{SERTA}
   =&-\frac{\delta f_{\mathbf{k}n}}{\tau^\text{SERTA}_{\mathbf{k}n}}
\end{align}
where $1/\tau^{\text{SERTA}}_{\mathbf{k}n}$ is twice the imaginary part of the Fan-Migdal self-energy, 
\begin{align}
\frac{1}{\tau^\text{SERTA}_{\mathbf{k}n}}
   =& \frac{2\pi}{\hbar}\sum_{\mathbf{k}'n'}\sum_{\mathbf{q}j}|g(\mathbf{k}n,\mathbf{k}'n',\mathbf{q}j) |^2  w_{\mathbf{k}n,\mathbf{k}'n'}\\ 
   &\times \Big((1+n^0_{\mathbf{q}j}-f^0_{\mathbf{k}'n'})\delta(\epsilon_{\mathbf{k}n}-\epsilon_{\mathbf{k}'n'}-\hbar \omega_{\mathbf{q}j})
   +(f^0_{\mathbf{k}'n'}+n^0_{-\mathbf{q}j})\delta(\epsilon_{\mathbf{k}n}-\epsilon_{\mathbf{k}'n'}+\hbar \omega_{-\mathbf{q}j})\Big).  \label{tau_SERTA_1}
\end{align}

In the above equation, the weighting factor $ w_{\mathbf{k}n,\mathbf{k}'n'}$ is $1$. In other words, the relaxation time is equal to the lifetime. However different approximations can be obtained considering different values for  $w_{\mathbf{k}n,\mathbf{k}'n'}$.
Two other usual approximations\cite{Li_2015,gunst_2016,Li_2020}, the momentum relaxation-time approximation (MRTA) and the energy relaxation-time approximation (ERTA) are derived in Appendix \ref{XRTA}.
In those cases, we have
  \begin{align}
&w_{\mathbf{k}n,\mathbf{k}'n'}=  1- 
\frac{\mathbf{v}_{\mathbf{k}n}\cdot \mathbf{v}_{\mathbf{k}'n'}}{|\mathbf{v}_{\mathbf{k}n}|| \mathbf{v}_{\mathbf{k}'n'}|} &\text{ for MRTA}, \label{tau_MRTA_1}\\
&w_{\mathbf{k}n,\mathbf{k}'n'}=  1- 
\frac{|\epsilon_{\mathbf{k}'n'}-\mu|}{|\epsilon_{\mathbf{k}n}-\mu|}
\frac{\mathbf{v}_{\mathbf{k}n}\cdot \mathbf{v}_{\mathbf{k}'n'}}{|\mathbf{v}_{\mathbf{k}n}|| \mathbf{v}_{\mathbf{k}'n'}|} &\text{ for ERTA}.  \label{tau_ERTA_1}
\end{align}

\noindent The MRTA approximation is derived assuming there is no temperature gradient. Its weighting factor makes large-angle scatterings important for the resistivity. The ERTA is obtained considering a temperature gradient alone. 
It is often assumed\cite{Ziman_1972,lavasani,Li_2020} that for heat conduction all scattering processes are equally effective. Therefore, this factor is sometimes neglected, and the SERTA is used instead.  This is the assumption which will be used in this paper. The differences between ERTA, SERTA and MRTA will be considered in more detail in a future work, as it requires a careful treatment for numerical stability\cite{lavasani,sohier}. For a coupled electric field and a thermal gradient applied on the system, those approximations become weaker. Finally, an even simpler approximation is the constant relaxation time approximation (CRTA), in which the relaxation time $\tau_{\mathbf{k}n}$ is assumed to be independent of the electronic state, i.e.\ $\tau_{\mathbf{k}n} = \tau$. A typical value is $\tau = 10^{-14}\,\text{s}$, which will be adopted whenever the CRTA is applied in this work.

Using one of the above mentioned relaxation-time approximations to solve Eq.~\ref{lin_boltz}, we obtain
\begin{align}
\delta f_{\mathbf{k}n}=-\tau_{\mathbf{k}n}\Big( (\epsilon_{\mathbf{k}n}-\mu)\frac{\nabla T}{T}+e\bm{\mathcal{E}}\Big)\cdot  \mathbf{v}_{\mathbf{k}n}\Big(-\frac{\partial f^0_{\mathbf{k}n} }{\partial \epsilon_{\mathbf{k}n}}\Big)
\end{align}
and therefore Eqs.~\ref{Je} and \ref{JQ} give

\begin{align}
\mathbf{J}^{e}&=  \mathcal{L}_{11} \bm{\mathcal{E}} + \mathcal{L}_{12} \frac{-\nabla T}{T}, \\
\mathbf{J}^{e}_Q&=  \mathcal{L}_{21}\bm{\mathcal{E}}+ \mathcal{L}_{22} \frac{-\nabla T}{T},
\end{align}

with 

\begin{align}
&\mathcal{L}_{11}= \frac{1}{NV_0}\sum_{\mathbf{k}n} (e^2)\tau_{\mathbf{k}n}\mathbf{v}_{\mathbf{k}n}\otimes  \mathbf{v}_{\mathbf{k}n}\Big(-\frac{\partial f^0_{\mathbf{k}n} }{\partial \epsilon_{\mathbf{k}n}}\Big), \label{L11} \\
&\mathcal{L}_{12}=\mathcal{L}_{21}= \frac{1}{NV_0}\sum_{\mathbf{k}n} (-e)(\epsilon_{\mathbf{k}n}-\mu)\tau_{\mathbf{k}n}\mathbf{v}_{\mathbf{k}n}\otimes  \mathbf{v}_{\mathbf{k}n}\Big(-\frac{\partial f^0_{\mathbf{k}n} }{\partial \epsilon_{\mathbf{k}n}}\Big), \label{L12}\\
&\mathcal{L}_{22}=\frac{1}{NV_0}\sum_{\mathbf{k}n} (\epsilon_{\mathbf{k}n}-\mu)^2\tau_{\mathbf{k}n} \mathbf{v}_{\mathbf{k}n}\otimes  \mathbf{v}_{\mathbf{k}n}\Big(-\frac{\partial f^0_{\mathbf{k}n} }{\partial \epsilon_{\mathbf{k}n}}\Big). \label{L22}
\end{align}

\noindent The Onsager coefficients $\mathcal{L}_{ij}$ can be written as
\begin{align}
\mathcal{L}_{ij} = \int d\epsilon \, \sigma(\epsilon) \Big(\frac{\epsilon-\mu}{-e} \Big)^{i+j-2} \Big(-\frac{\partial f^0 }{\partial \epsilon}\Big)     \label{eq:Lij}
\end{align}
with 
\begin{align}
\sigma(\epsilon) = \frac{e^2}{NV_0} \sum_{\mathbf{k}n}  \tau_{\mathbf{k}n}\mathbf{v}_{\mathbf{k}n}\otimes  \mathbf{v}_{\mathbf{k}n} \delta(\epsilon_{\mathbf{k}n}-\epsilon).  
\end{align}

\noindent Notice that, as for the density of states, a factor $2$ needs to be included for spin, if not included in the summation over bands. The transport function $\sigma(\epsilon)$ represents the conductivity of electrons with energy $\epsilon$, and therefore its unit is $(\Omega \cdot \text{m})^{-1}$. One could directly compute the Onsager coefficients from an accumulation over the Brillouin zone using Eqs. \ref{L11}, \ref{L12} and \ref{L22}. However, it is numerically advantageous to first compute the transport function  $\sigma(\epsilon)$ and then the Onsager coefficient $\mathcal{L}_{ij}$ through an integration over energy. Indeed,  $\sigma(\epsilon)$ can be computed using the tetrahedron method, which is a (linear) interpolation method and allows to use a coarser sampling of the Brillouin zone. To perform the integrals over energy, we use a Gauss-Legendre quadrature which is explained in Appendix \ref{GL}.
The electronic conductivity ($\sigma$), the thermopower ($S$) and the electronic part of the thermal conductivity ($\kappa_e$) are finally expressed as
\begin{align}
&\sigma = \mathcal{L}_{11},  \\
&S=\frac{1}{T} \mathcal{L}_{11}^{-1}\mathcal{L}_{12},\label{seebeck}\\ 
&\kappa_e= \frac{1}{T}(\mathcal{L}_{22}- \mathcal{L}_{21} \mathcal{L}_{11}^{-1}\mathcal{L}_{12}). 
\end{align}


\subsection{Thermoelectric transport for phonons}
In this section, we consider the lattice thermal conductivity $\kappa_L$. The details of the calculations are given in Ref.~\citenum{togo2015}. In particular, in this work, it is assumed that the phonons are only scattered by other phonons, and therefore the effect of the electron-phonon interactions on phonons is neglected.  The lifetime of the phonons is obtained from Fermi's golden rule on a Hamiltonian containing anharmonicity up to third order in the atomic displacements of the atoms around their equilibrium positions, 
\begin{align} &H_{pp}=\sum_{\mathbf{q}_1j_1}\sum_{\mathbf{q}_2j_2}\sum_{\mathbf{q}_3j_3}\Phi(\mathbf{q}_1j_1, \mathbf{q}_2j_2, \mathbf{q}_3j_3)(a_{\mathbf{q}_1j_1}+a^{\dagger}_{-\mathbf{q}_1j_1})(a_{\mathbf{q}_2j_2}+a^{\dagger}_{-\mathbf{q}_2j_2})(a_{\mathbf{q}_3j_3}+a^{\dagger}_{-\mathbf{q}_3j_3}). \label{Hpp}
\end{align}
The strength $\Phi$ of this interaction is given by the Fourier transform of the third derivative of the energy with respect to atomic positions, projected onto the eigenvectors of the phonons participating in the interaction.  As for the electron-phonon interaction, it includes the momentum conservation. The lifetime is then given by
\begin{align}
\frac{1}{\tau_{\mathbf{q}j}}&= \frac{
    36\pi}{\hbar^2}\sum_{\mathbf{q}_2j_2}\sum_{\mathbf{q}_3j_3}
    |\Phi(-\mathbf{q}j,\mathbf{q}_2j_2,\mathbf{q}_3j_3)|^2 \nonumber \\
    &\times
\Big[(n^0_{\mathbf{q}_2j_2}+n^0_{\mathbf{q}_3j_3}+1)\delta(\omega_{\mathbf{q}j}-\omega_{\mathbf{q}_2j_2}-\omega_{\mathbf{q}_3j_3})
    +2(n^0_{\mathbf{q}_2j_2}-n^0_{\mathbf{q}_3j_3})\delta(\omega_{\mathbf{q}j}-\omega_{\mathbf{q}_2j_2}+\omega_{\mathbf{q}_3j_3})\Big].
\end{align}

\noindent In the relaxation-time approximation, the lattice thermal conductivity is computed from
\begin{align}
\kappa_L = \frac{1}{NV_0}\sum_{\mathbf{q}j} C_{\mathbf{q}j}\tau_{\mathbf{q}j}\mathbf{v}_{\mathbf{q}j} \otimes \mathbf{v}_{\mathbf{q}j},  \label{LTC}
\end{align}
with the mode heat capacity $C_{\mathbf{q}j} = \frac{\partial  }{\partial T}\hbar \omega_{\mathbf{q}j}n^0_{\mathbf{q}j}$ and the group velocity $\mathbf{v}_{\mathbf{q}j}=\frac{\partial \omega_{\mathbf{q}j}}{\partial \mathbf{q}}$.

\end{widetext}

\section{Computational strategies\label{workflow}}

\subsection{Integration over energy}

\begin{figure}
    \centering
    \includegraphics[width=0.8\linewidth]{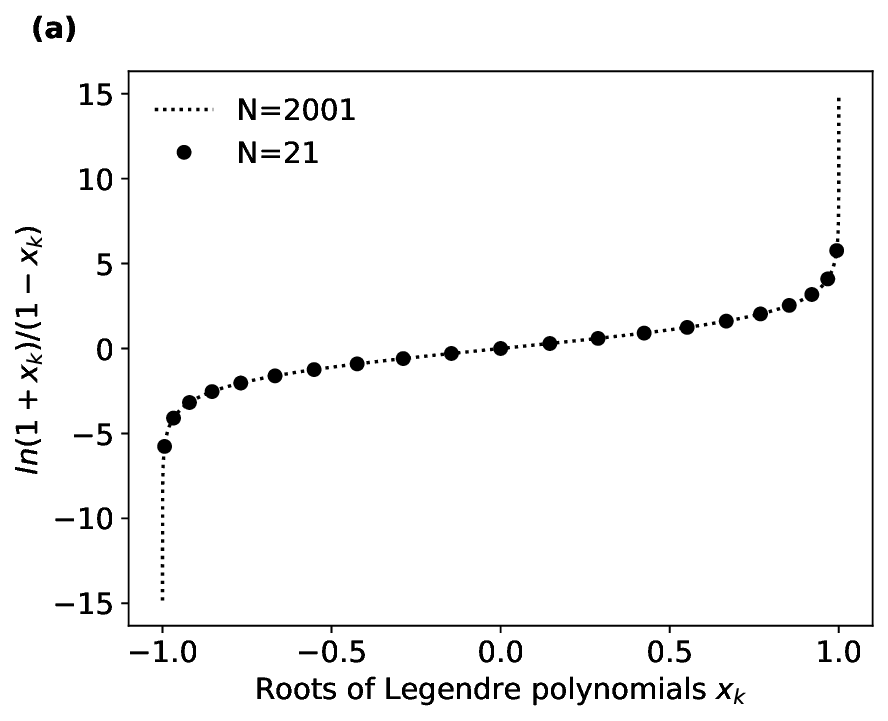}
    \includegraphics[width=0.8\linewidth]{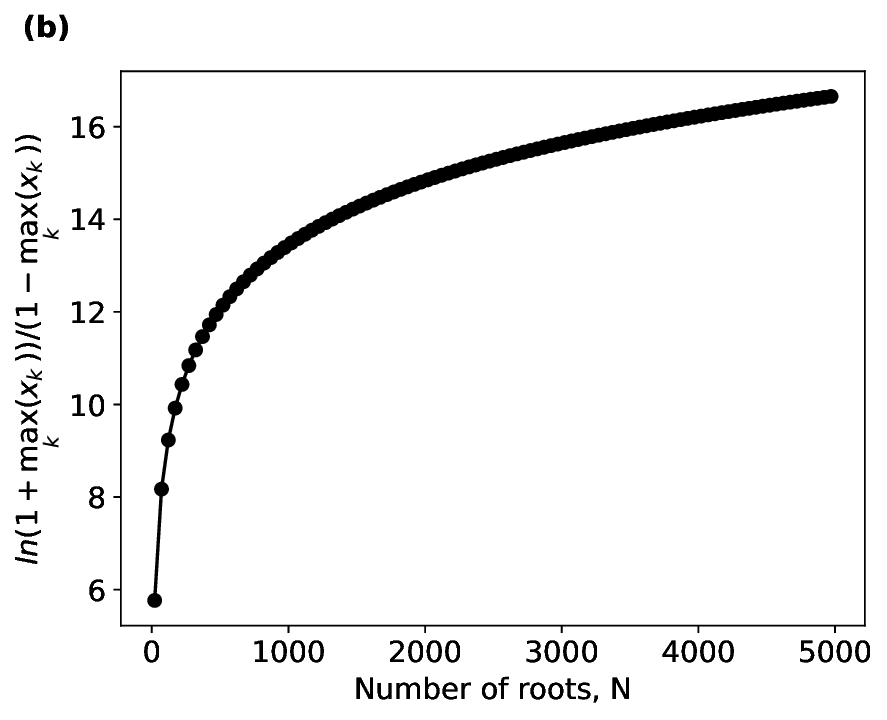}
    \includegraphics[width=0.8\linewidth]{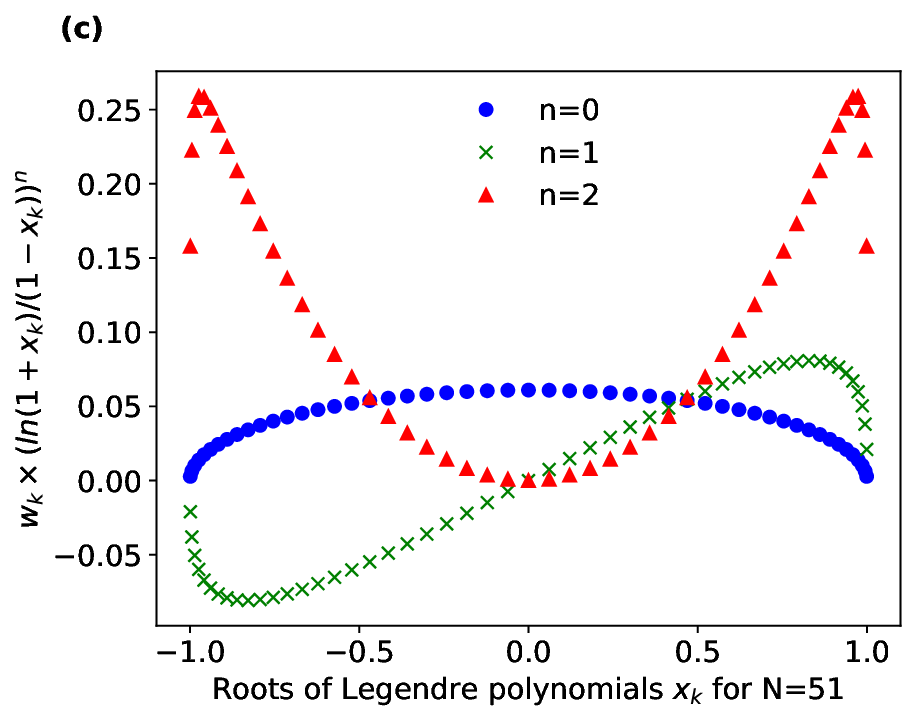}
    \caption{Gauss-Legendre integration grids: (a) $\ln [(x+1)/(x-1)]$ the range of the energy window in units of $k_BT$ around the chemical potential $\mu$, for $N=21$ and $N=2001$, (b) $\ln [(x_N+1)/(x_N-1)]$, with $x_N$ the largest root, as a function of the number of roots $N$, (c) function in the integrand of Eq.~\ref{Lij_x}, $\Big(\ln [(1+x_k)(1-x_k)]\Big)^{n} w_k$, as a function of the $x_k$, for $n=0,1,2$.}
    \label{Gauss-Legendre}
\end{figure}

To compute the Onsager coefficients from Eq.~\ref{eq:Lij} the electronic relaxation times and the electron group velocities are needed.
A naive implementation would require the computation of those relaxation times for all the Kohn-Sham states. However, we can limit the computation of relaxation times to states whose energies lie within the energy range spanned by a Gauss-Legendre integration grid.

Indeed, in Appendix~\ref{GL}, the integral over energy Eq.~\ref{eq:Lij} is replaced by an integration over $x$ using $\epsilon=\mu+ k_BT \ln \frac{1+x}{1-x}$, with $x$ in the range $(-1,1)$. The $x$ axis is discretized using $\{x_k\}_{k=1\dots N}$, the $N$ roots of the Legendre polynomial $P_N(x)$. $N$ is a convergence parameter. It is the number of points in the interval $(1,-1)$ and therefore controls the size of the integration step. However, it also controls the size of the energy window around the chemical potential over which the integration is performed. Indeed, in Fig.~\ref{Gauss-Legendre}~(a), the function $\ln([1+x)/(1-x)]$, which measures the range of the energy window around the chemical potential, in units of $k_BT$, is plotted with respect to the roots $x_k$ for several values of $N$. The roots change with $N$, and in particular, $x_N$, the largest root, approaches $1$ as $N$ increases. In fact, we have the estimate \cite{Bruns1881,szego75} $\cos \pi \frac{1}{N+1/2}<x_N<\cos \pi \frac{1/2}{N+1/2}$. This means that $\ln[(1+x_N)/(1-x_N)]$ may become large for large $N$ (see Fig.~\ref{Gauss-Legendre}~(b)), and therefore the energy range over which the relaxation time needs to be computed may become large as well. Fortunately, to perform the integrals in Eq.~\ref{eq:Lij}, it has been found from the examples we considered that moderate values of $N$, around $N=501$, are sufficient. The energy range around the chemical potential may be restricted because of the fast decay of the derivative of the Fermi-Dirac function as the energy increases, but it can also be restricted because the integration weights $w_k$ decrease as $x_k \to \pm 1$, and as $N$ increases (see Fig.~\ref{Gauss-Legendre}~(c)). This strategy speeds up the calculation while allowing control over the accuracy and computational cost using a single parameter, the number of integration points $N$ of the Gauss-Legendre grid.
 

\begin{figure}
    \centering
    \includegraphics[width=1\linewidth]{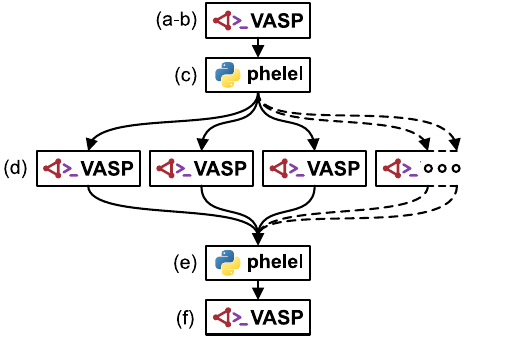}
    \caption{Flow chart for the calculation of electronic transport properties using VASP and phelel. (a) Preparatory steps are performed using the VASP code. (c) Supercell structures are generated with phelel. (d) Each displaced structure is then run as an individual VASP calculation. (e) The results are gathered by phelel to compute the potential derivatives and transferred to the primitive cell. (f) The final VASP calculation in the primitive cell then calculates the electron-phonon interactions and transport properties.}
    \label{fig:flow}
\end{figure}

\subsection{Calculation steps \label{steps}}
The relaxation times given by Eqs.~\ref{tau_SERTA_1}, \ref{tau_MRTA_1} and \ref{tau_ERTA_1} are computed from the electron-phonon interaction $g(\mathbf{k}n,\mathbf{k}'n',\mathbf{q}j)$. Details about the theory are given in Ref.~\citenum{chaput2019} for the PAW method. In a few words, to compute $g(\mathbf{k}n,\mathbf{k}'n',\mathbf{q}j)$, the derivatives of the local and non-local potentials are computed by applying finite displacements to atoms in supercells. The derivatives are obtained from the changes in the potentials using finite differences.

If we go more into the details of such calculation, a series of computations are necessary to obtain $g(\mathbf{k}n,\mathbf{k}'n',\mathbf{q}j)$ and the transport properties. They are all organized using a command line tool of the phelel code\cite{phelel_url}, allowing to prepare the inputs of every VASP calculation, using outputs from previous ones. The general setting of the calculations, such as cut-off energy, Brillouin zone sampling, etc, are obtained from a separate file, defined prior to the start of the calculations, and read by phelel when the inputs of the VASP calculations are being prepared. This approach is convenient because the series of calculations needed to obtain the electron-phonon interactions and the transport properties can be seen as operated by phelel. The series of calculations can then be summarized as a small number of simple phelel commands

The different steps, detailed below for the computations performed in this work, can be categorized as preparatory, and steps related to the computation of the derivatives of the potentials, and transport steps. They are respectively indexed as (a-b), (c-e), and (f) in the following.

The preparatory steps may be optional. In this study, (a) the crystal structure is relaxed. Then, (b) using the relaxed crystal structure, dielectric constants and Born effective charges are computed. This step is required because the compounds we consider in our study have an ionic contribution.

In step (c-e), the potential derivatives are computed. The method is summarized in Fig.~\ref{fig:flow}. At step (c) the supercells with displacements are generated using the phelel code, considering symmetry. Then, (d) a series of VASP calculations is performed to obtain changes in the potentials. In step (e) the potential derivatives are computed from the results of the VASP calculations using the phelel code, and are stored in a file, \texttt{phelel$\_$params.hdf5}. For that step, phelel uses the finufft\cite{finufft} library to perform a nonuniform fast Fourier transform from the grid points in the supercells to the grid points in a primitive cell on which the wavefunctions will later be known.

In a final step, (f), VASP reads the \texttt{phelel$\_$params.hdf5} file to compute $g(\mathbf{k}n,\mathbf{k}'n',\mathbf{q}j)$ and the transport properties. This step is detailed in the next section.

\subsection{Electron-phonon matrix elements}


To obtain $g(\mathbf{k}n,\mathbf{k}'n',\mathbf{q}j)$, the integration in $\langle \varphi_{\mathbf{k}n}|\partial \tilde{h}/ \partial \mathbf{R}| \varphi_{\mathbf{k}'n'}\rangle$ should be performed over a supercell, with $\mathbf{k}$, $\mathbf{k}'$ and $\mathbf{q}$ within the Brillouin zone of a primitive cell\cite{chaput2019}. Indeed, the potential derivatives contained in \texttt{phelel\_params.hdf5}  are known within the supercell. Therefore, to perform the integral, one should  expand the orbitals from the primitive cell, known from the final VASP step in Fig.~\ref{fig:flow}, to the supercell using Bloch's theorem.  However, in practice, we use the opposite strategy to integrate over a primitive cell. Instead of expanding the Bloch functions from the primitive cell to the supercell, the potential derivatives are downfolded from the supercell to the primitive cell. For example, Eq.~A21 in Ref.~\citenum{engel2022} can be written as an integral over $V_0$
\begin{align}
    \int_{V_0} d^3 \mathbf{r}\,
    \tilde{u}_{\mathbf{k}n}^{*} (\mathbf{r})\tilde{v}_{\kappa}(\mathbf{r},\mathbf{k}-\mathbf{k}')
    \tilde{u}_{\mathbf{k}'n'}(\mathbf{r}), \label{int_prim}
\end{align}
with
\begin{align}
\tilde{v}_{\kappa}(\mathbf{r},\mathbf{k}-\mathbf{k}')&=
    \sum_{\mathbf{l}}
    \sum_{\mathbf{L}'} \frac{d \tilde{v}(\mathbf{r}+\mathbf{l})}{d \mathbf{R}_{\mathbf{L}' \kappa}} \nonumber\\
    &\times\frac{1}{|\{ \mathbf{L}\}_{ \kappa,\mathbf{r} }|}
    \sum_{\{ \mathbf{L}\}_{ \kappa, \mathbf{r} }} e^{i(\mathbf{k}'-\mathbf{k})\cdot(\mathbf{r}+\mathbf{l}+\mathbf{L})}.
\end{align}
$\tilde{u}_{\mathbf{k}n}$ are the periodic parts of the pseudo wavefunctions and $\mathbf{l}$ primitive lattice vectors used to define the supercell. The $\mathbf{L}$ vectors are supercell lattices vectors, chosen, according  to the minimal image convention, to minimize the distance between atom $\kappa$ and the point $\mathbf{L}+\mathbf{r}$ in the neighboring supercell. Therefore the exponential in the previous equation is the phase factor needed to bring the primitive cells filling the supercell to the one at the origin. This downfolding produces $\mathbf{q}=\mathbf{k}-\mathbf{k}'(\text{mod } \mathbf{G})$ dependent potential derivatives, but the accumulation over grid points used to compute the integrals is less demanding than over the supercell.  A more detailed description of this downfolding can be found in Ref.~\citenum{engel2022}.


Notice that for polar materials, we remove the long-range part of the electron-phonon potential due to dipole interactions from the supercell before interpolation and then add it in the primitive cell at finite $\mathbf{q}$~\cite{vogl_microscopic_1976,engel2022}. See Appendix A.3 in Ref.~\cite{engel2022}.

The number of electron-phonon matrix elements to be computed may be reduced. Indeed, the relaxation times given by Eqs.~\ref{tau_SERTA_1}, \ref{tau_MRTA_1} and \ref{tau_ERTA_1} involve delta functions. Therefore, to reduce the number of computations, we first evaluate the delta functions between all the selected bands at $\mathbf{k}$ and $\mathbf{k'}$, and only if they are nonzero do we proceed to evaluate the electron-phonon matrix elements. This optimization can only be done because in transport problems it is the relaxation time which is needed. It mimics the imaginary part of a self-energy. To obtain the real part as well, for example, using a Kramers-Kronig transform, the full set of states would be needed.

Once the electron-phonon matrix elements between $\mathbf{k}$ and $\mathbf{k'}$ have been obtained for the chosen states, the relaxation times are computed and stored for different temperatures, doping levels, number of bands or relaxation-time approximations and then the electron-phonon matrix elements are discarded. This way, we compute the transport coefficients for different settings in a single run of the code, where the electron-phonon matrix elements are computed just once.

\begin{widetext}
\begin{center}
\begin{table}[ht]
\centering
\caption{Table of the materials considered. The reported values are obtained using the PBEsol exchange-correlation potential. \label{materials-PBEsol}}
\begin{tabular}[t]{lccccc}
\hline
               & lattice parameter [\AA] & dielectric constant & Born effective charges & Gap [eV] \\
\hline 
PbTe           &  $a=6.441$                    &   $\epsilon_r=30.441$                  & $Z_\text{Pb}=5.876$, $Z_\text{Te}= -5.876$              &   0.65           \\
PbSe           &  $a=6.102$                    &    $\epsilon_r=30.802$                 &  $Z_\text{Pb}=4.601$, $Z_\text{Se}=-4.601$             & 0.17        \\
\hline
Mg$_2$Si       &   $a=6.325$                   &  $\epsilon_r=15.417$                   & $Z_\text{Si}=-3.652$, $Z_\text{Mg}=1.826$             &   0.11    \\
Mg$_2$Ge       &   $a=6.356$                   &   $\epsilon_r=16.260$                  & $Z_\text{Ge}=-3.576$, $Z_\text{Mg}=1.788$             &     0.08            \\
Mg$_2$Sn       &    $a=6.739$                  &     $\epsilon_r=18.775$                &  $Z_\text{Sn}=-3.845$, $Z_\text{Mg}=1.922$              &           \\
\hline
NiTiSn         &  $a=5.865$                    & $\epsilon_r=23.511$                    &   $Z_\text{Ti}= 2.771$, $Z_\text{Sn}=1.179$, $Z_\text{Ni}=-3.950$          &    0.45         \\
NiZrSn         &   $a=6.067$                   &    $\epsilon_r=21.592$                 &  $Z_\text{Zr}=2.593$, $Z_\text{Sn}=1.026$, $Z_\text{Ni}=-3.619$             &   0.50         \\
NiHfSn         &   $a= 6.032$                   & $\epsilon_r= 20.571$                    & $Z_\text{Hf}=2.682$, $Z_\text{Sn}=0.964$, $Z_\text{Ni}=-3.645$            &    0.38      \\
\hline
\end{tabular}
\end{table}
\end{center}  
\end{widetext}

\section{Thermoelectric materials\label{materials}}

\subsection{Computational details}

In the following, we compute the transport properties of the materials listed in Tab. \ref{materials-PBEsol}. We consider several thermoelectric materials, including half-Heusler, rocksalts and antifluorites. The structures of those materials are described in the Supplemental Material\cite{SM}. Transport properties are computed for doped compounds. The carrier concentrations we use are reported in Tab. \ref{roomT}. They are taken from Hall measurements and used to determine the chemical potential according to Eq. \ref{fermi}.

All materials are computed with both the PBEsol\cite{PBEsol} and PBE\cite{pbe} exchange-correlation functionals using the VASP code\cite{VASP1993,VASP1994,VASP1996a,VASP1996b,VASP1999} and the PAW\cite{PAW-Blochl-1994} potentials listed in the Supplemental Material\cite{SM}. A significant challenge arises with materials containing heavy elements, like PbTe and PbSe, where spin-orbit coupling (SOC) may be crucial. However, simply combining SOC with standard PBE or PBEsol functionals may results in an unphysical inverted gap~\cite{hummer_structural_2007}, leading to inaccurate transport properties. Therefore, for simplicity, SOC has been neglected in the present study. Accurately incorporating spin-orbit coupling necessitates the use of an exchange-correlation functional specifically designed for better band gap description. More details about the effects of spin-orbit coupling on transport properties are given in Appendix \ref{SOC}.

To obtain the electron-phonon interactions, different cells, which describe the same crystal structure, may be used to perform the steps listed in Sec. \ref{steps}. We consider primitive, conventional, and supercells. They are generated from each other through simple matrix transformations. The basis vectors of the primitive cells are chosen as $(\frac{1}{2}(\mathbf{b+c}), \frac{1}{2}(\mathbf{c+a}), \frac{1}{2}(\mathbf{a+b}))$, where $(\mathbf{a}, \mathbf{b}, \mathbf{c})$ are the basis vectors of the cubic conventional unit cells. Supercell are generated as multiple of the conventional cells. Details about the size of the supercells can be found in the Supplemental Material\cite{SM}. For each compound, we consider two supercells to be able to check the convergence of the method with respect to the supercell size.

In this works, in step (a), the crystal structure is relaxed in the conventional unit cell and in step (b) dielectric constants and Born effective charges are computed using the primitive cell. In step (c), the atoms in the supercells are displaced using symmetrically irreducible displacements\cite{togo2015} of 0.03 \AA.

For all compounds, the plane-wave cutoff energy is set to 400 eV. The sampling in reciprocal space, for Brillouin zone and supercell shape, are given in the tables of the Supplemental Material\cite{SM}. The mesh number is
roughly estimated by $10 \times |\mathbf{a}^*|$ where $\mathbf{a}^*$ is a reciprocal basis vector of the cell used for the specific calculation, $\mathbf{a}^* \cdot \mathbf{a}=2\pi$.

The long-range part of the electron-phonon potential needs to be considered. It is computed from a sum of $\mathbf{G}$ vectors in reciprocal space (see Appendix A.3 of Ref.~\citenum{engel2022}). We use a cutoff of 50 eV to perform this summation.

The lattice thermal conductivity is computed from phonon-phonon interaction calculations in the supercell approach, using phono3py\cite{phonopy-phono3py-JPCM,phonopy-phono3py-JPSJ}, and solving the phonon Boltzmann equation under the relaxation-time approximation. Random directional displacements of 0.03 A were introduced in 120 supercells to compute anharmonic force constants using the symfc code\cite{symfc}.
In Fig. \ref{fig:conv_k}, it is shown that sampling the Brillouin zone using a $30\times30\times30$ mesh is already sufficient to converge the lattice thermal conductivity.

\begin{widetext}
\begin{center}
\begin{figure}[H]
    \includegraphics[scale=0.45]{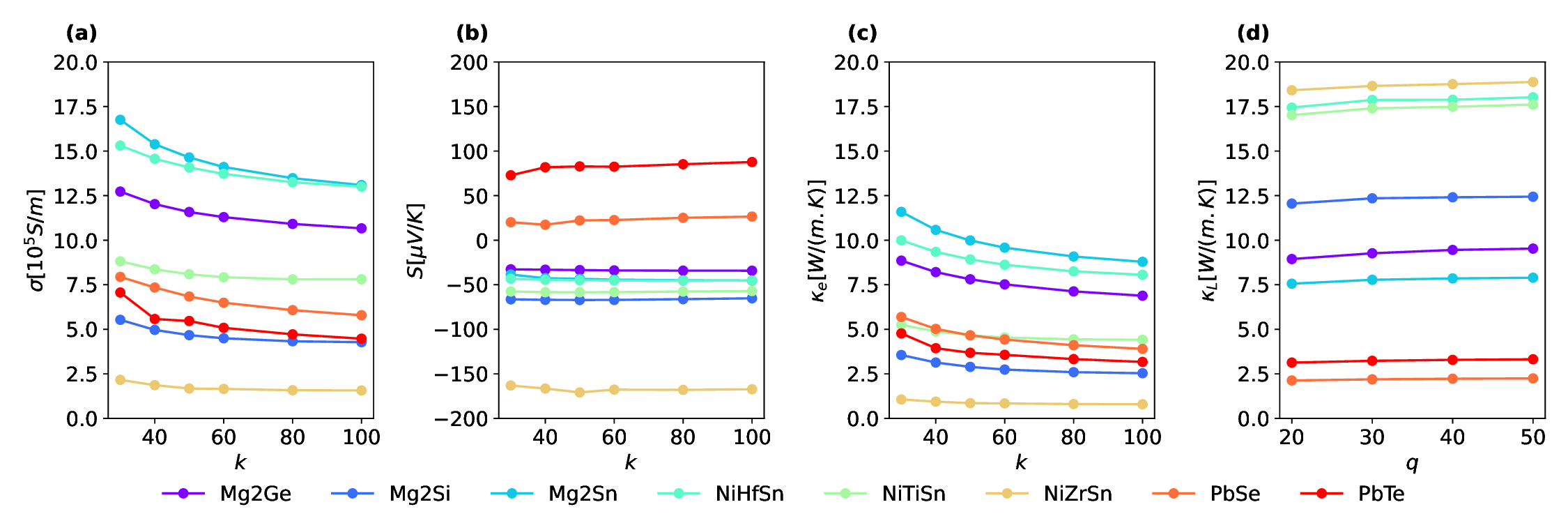}
    \caption{Convergence of the (a) electronic conductivity ($\sigma$),  (b) thermopower ($S$), (c) electronic part of the thermal conductivity ($\kappa_e$) and (d) lattice part of the thermal conductivity ($\kappa_L$), at $300$ K, with respect to the sampling of the first Brillouin zone.}
    \label{fig:conv_k} 
\end{figure}
\end{center}
\end{widetext}

\subsection{Results \label{results}}

The geometries of the materials were optimized using the PBEsol\cite{PBEsol} and PBE\cite{pbe} exchange-correlation functionals. The parameters we obtain are reported in Tab. \ref{materials-PBEsol} for the PBEsol functional, and in Tab.~\ref{materials-PBE}, in Appendix~\ref{TE-mat}, for the PBE functional. For all compounds, we obtain smaller lattice parameters using the PBEsol exchange-correlation functional.

The electronic band structures are plotted in Appendix~\ref{TE-mat}  along the paths given by Seekpath\cite{HINUMA2017140,spglib}. Comparing the results obtained with the PBEsol and PBE exchange-correlation functionals, we do not see major topological changes in the band structure in the vicinity of the Fermi level. However, the values of the band gaps can be very different. For Mg$_2$Si, Mg$_2$Ge, PbTe and PbSe, the band gaps computed with the PBEsol functional are smaller than the ones obtained with PBE.  For example, the PbSe band gap is strongly reduced using the PBEsol functional. This is due to a different parametrization of the exchange-correlation functional which reduces the exchange contribution to the energy thus lowering the value of the band gaps\cite{Borlido}.
The opposite is observed for the half-Heusler, as the PBE gaps are smaller than those obtained from the PBEsol functional. The differences are small, however.

The phonon dispersions are shown in Appendix \ref{TE-mat} for the PBEsol and PBE functionals. They are computed using the small supercells reported in the Supplemental Material\cite{SM}. As already noted, the lattice parameters we obtain are smaller using the PBEsol exchange-correlation functional. Consequently, we obtain higher phonon frequencies. This is what we observe for NiTiSn, NiZrSn, NiHfSn,  Mg$_2$Si, Mg$_2$Ge, PbTe.
The change in the phonon spectrum is not a simple scaling of the frequencies, with larger differences being observed in the optical modes. 

For Mg$_2$Sn  and the PBE functional we find that the real space force-constants computed using symmetric displacements on the small $2\times2\times2$ supercell reported in the Supplemental Material\cite{SM} are insufficient to interpolate the phonon dispersion accurately as shown in Fig.~\ref{phonon_Mg2X} and hence we did not proceed with the calculation of the transport coefficients in this case, reporting only the results for the $4\times4\times4$ supercell in Tab.~\ref{roomT}.

In Fig. \ref{PbX_phonon}, we show that in PbSe the optical modes at the zone center are very different for the PBE and PBEsol approximations. This is due to the different values for the dielectric function $\epsilon_r$ and Born effective charges $Z_i$ as reported in Tab. \ref{materials-PBEsol} and Tab. \ref{materials-PBE}. When using the PBE $\epsilon_r$ and $Z_i$ in the phonon calculation with PBEsol, the phonon dispersion
becomes closer to the PBE one (Fig. 11). The same behavior is observed for the values of the transport properties for which we obtain $\sigma=4.3 \, 10^{5}(\Omega \text{m})^{-1}$, $S=32.2 \, \mu \text{V/K}$, $\kappa_e=2.9\, \text{W/(m $\cdot$ K)}$, closer to the PBE results of Tab.~\ref{roomT}. 


To compute the electronic transport properties, several parameters must be set to ensure the accuracy of $g(\mathbf{k}n,\mathbf{k}'n',\mathbf{q}j)\sim \langle \varphi_{\mathbf{k}n}| \partial \tilde{h}/\partial \mathbf{R}| \varphi_{\mathbf{k}'n'}\rangle \cdot \delta \mathbf{R}^{\mathbf{q}j}$. The plane-wave cutoff energy, $E_\text{cut}$, the number of points used to sample the Brillouin zone of the primitive cell, $N_k$, and the size of the supercell used to compute $g(\mathbf{k}n,\mathbf{k}'n',\mathbf{q}j)$ have to be carefully chosen for those calculations to be both efficient and accurate. 
 
For all materials, to obtain the values of $E_\text{cut}$ and $N_k$, we have converged the transport properties $\sigma$, $S$, and $\kappa_e$ at $T=300$ K using the small supercells listed in the Supplemental Material\cite{SM} and the PBEsol functional. The Brillouin zone of the primitive cell is
 sampled using a uniform $k\times k\times k=N_k$ grid. Fig. \ref{fig:conv_k} shows the result of these calculations with respect to $k$ for $E_\text{cut}=400$ eV. It can be seen that $k=100$ allows to obtain converged values, to 5 $\%$ accuracy,  for all materials.

The derivatives of the Hamiltonian, $\partial \tilde{h}/\partial \mathbf{R}$, and the phonon spectra are obtained from supercell calculations which use $E_\text{cut}=400$ eV. In the computation of $g(\mathbf{k}n,\mathbf{k}'n',\mathbf{q}j)$, this value defines the number of real-space grid points used to compute the integral $\langle \varphi_{\mathbf{k}n}| \partial \tilde{v}_{loc}/\partial \mathbf{R}| \varphi_{\mathbf{k}'n'}\rangle$. 
 The transport properties we obtain are found to be almost insensitive to the value $E_\text{cut}$ for values in the range $300-500$ eV. To rationalize this result, it may be useful to notice that the  largest suggested cutoff energy among the pseudopotentials we use is 310 eV for Ge. 

We have also investigated the effect of the supercell size we use to compute the phonon spectra and $\partial \tilde{h}/\partial \mathbf{R}$. The results are reported in Tab. \ref{roomT} for two supercell sizes for $k=100$ and $E_\text{cut}=400$ eV. It is clear from this table that the electronic transport properties depend very weakly on the size of the supercell.

\begin{figure}
    \centering
    \vspace{1cm}
    \includegraphics[width=1\linewidth]{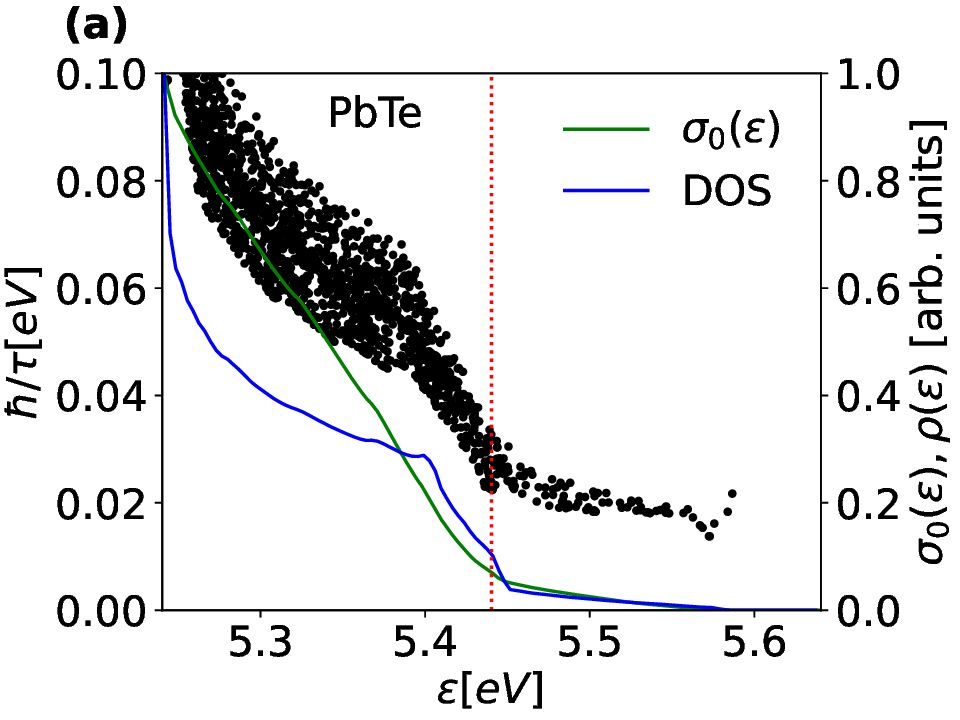}
    \includegraphics[width=1\linewidth]{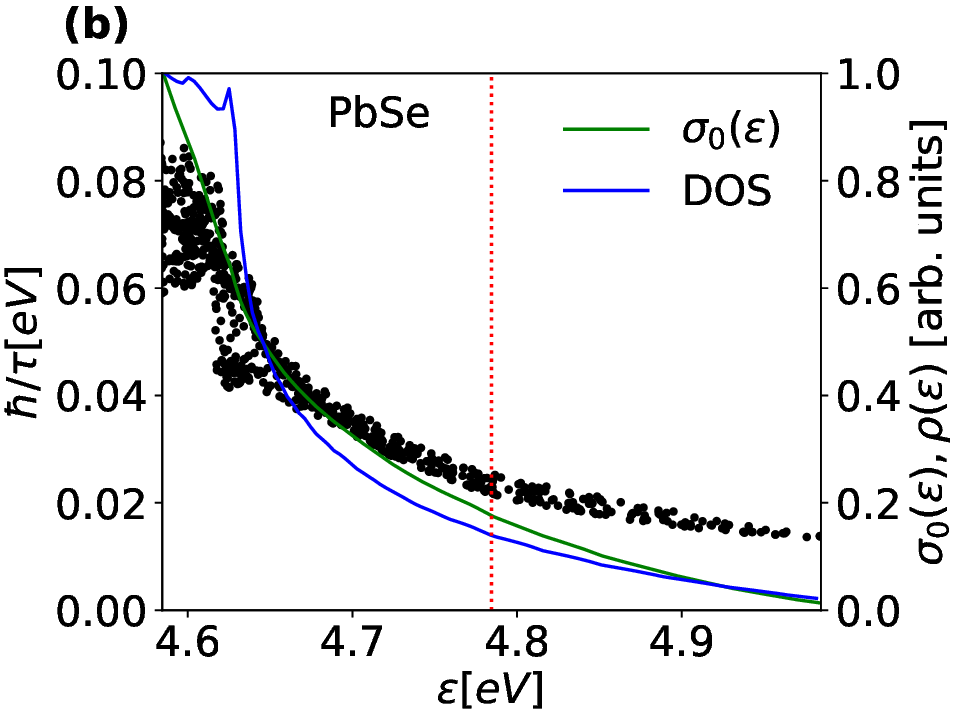}
    \caption{Reciprocal lifetimes, $\hbar/\tau$, at 300 K, as function of the band energies, are shown as black dots for (a) PbTe and (b) PbSe. The PBEsol exchange correlation functional is used. The transport function at constant relaxation time, $\sigma_0(\epsilon)$, and the density of states, $\rho(\epsilon)$, in arbitrary units, are shown as green and blue curves, respectively.
    The chemical potential is shown using a red dotted line. PbTe and PbSe are $p$ doped compounds in our calculations (see Tab. \ref{roomT}), therefore it is the lifetime of hole which is plotted. \label{relax_PbX}}
\end{figure}

The lifetimes of holes or electrons are plotted in Figs. \ref{relax_PbX}, \ref{relax_HH} and \ref{relax_Mg2X} as functions of the Kohn-Sham energies. Two distinct behaviors are observed. In Fig. \ref{relax_PbX} the reciprocal lifetime closely follows the density of states and the transport function. This can easily be understood from Eq. \ref{tau_SERTA_1} if the electron-phonon matrix elements are seen as constants, and the phonon energy is considered to be small. In Figs. \ref{relax_HH} and \ref{relax_Mg2X}, the reciprocal lifetime shows a minimum around the chemical potential. This can be understood from Eq. \ref{tau_SERTA_1} as well. Assuming the electron-phonon matrix elements to be constant, and a single energy $\hbar \omega$ in the phonon spectrum, we obtain
\begin{align*}
\frac{1}{\tau_{\mathbf{k}n}} &\propto [n^0(\hbar \omega)+1-f^0(\epsilon_{\mathbf{k}n}-\hbar \omega)] \rho(\epsilon_{\mathbf{k}n}-\hbar \omega)\nonumber\\
&+
[n^0(\hbar \omega)+f^0(\epsilon_{\mathbf{k}n}+\hbar \omega)]\rho(\epsilon_{\mathbf{k}n}+\hbar \omega).
\end{align*}
If, moreover, the density of states is almost constant around $\epsilon_{\mathbf{k}n}$, then
\begin{align*}
\frac{1}{\tau_{\mathbf{k}n}} &\propto [2n^0(\hbar \omega)+1-f^0(\epsilon_{\mathbf{k}n}-\hbar \omega)+f^0(\epsilon_{\mathbf{k}n}+\hbar \omega)]\\
&\times\rho(\epsilon_{\mathbf{k}n}).
\end{align*}
The factor in the square bracket takes the value $2n^0(\hbar \omega)+1$ for energies far from the chemical potential, but has a pronounced minimum of depth $\tanh (\hbar \omega)/(2 k_B T)$ at the chemical potential. Its width is about $\pm \hbar \omega$ around the chemical potential. This is the behavior observed in Figs. \ref{relax_HH} and \ref{relax_Mg2X}. Fig. \ref{relax_PbX} can also be recovered if the phonon frequency $\hbar \omega$ is assumed to be small. The Fermi functions can be neglected and therefore 
\begin{align*}
\frac{1}{\tau_{\mathbf{k}n}} &\propto \rho(\epsilon_{\mathbf{k}n}).
\end{align*}
Looking at the values of the dips in the lifetimes of Figs.~\ref{relax_HH} and \ref{relax_Mg2X}, we locate the effective phonon frequencies $\omega$ in the acoustic part of the phonon spectra. The values are lower for half-Heusler than for the Mg$_2$X compounds, and even lower for the rocksalt compounds. Therefore, as anticipated, the effective phonon frequency decreases as the mass of the compound increases.

The result in Fig.~\ref{fig:conv_k} suggests that above room temperature we can obtain an acceptable estimation of the thermoelectric figure of merit using a $60 \times 60 \times 60$ mesh in the first Brillouin zone. We use the PBEsol exchange-correlation functional. The results are shown in Fig.~\ref{RS_app}, and in Figs.~\ref{HH_app} and \ref{Mg2X_app} in Appendix~\ref{TE-mat}.

To understand the thermoelectric figure of merit as a function of temperature, we write
\begin{align}
ZT&=(S T)^2 \Big(\frac{\sigma}{\kappa T}\Big), \\
&= (S T)^2 \Big(\frac{\sigma T}{\kappa_e} \Big) \Big(\frac{1}{1+\frac{\kappa_L}{\kappa_e}} \frac{1}{T^2} \Big). \label{ZT_factors}
\end{align}

\begin{figure}
    \centering
    \includegraphics[width=1\linewidth]{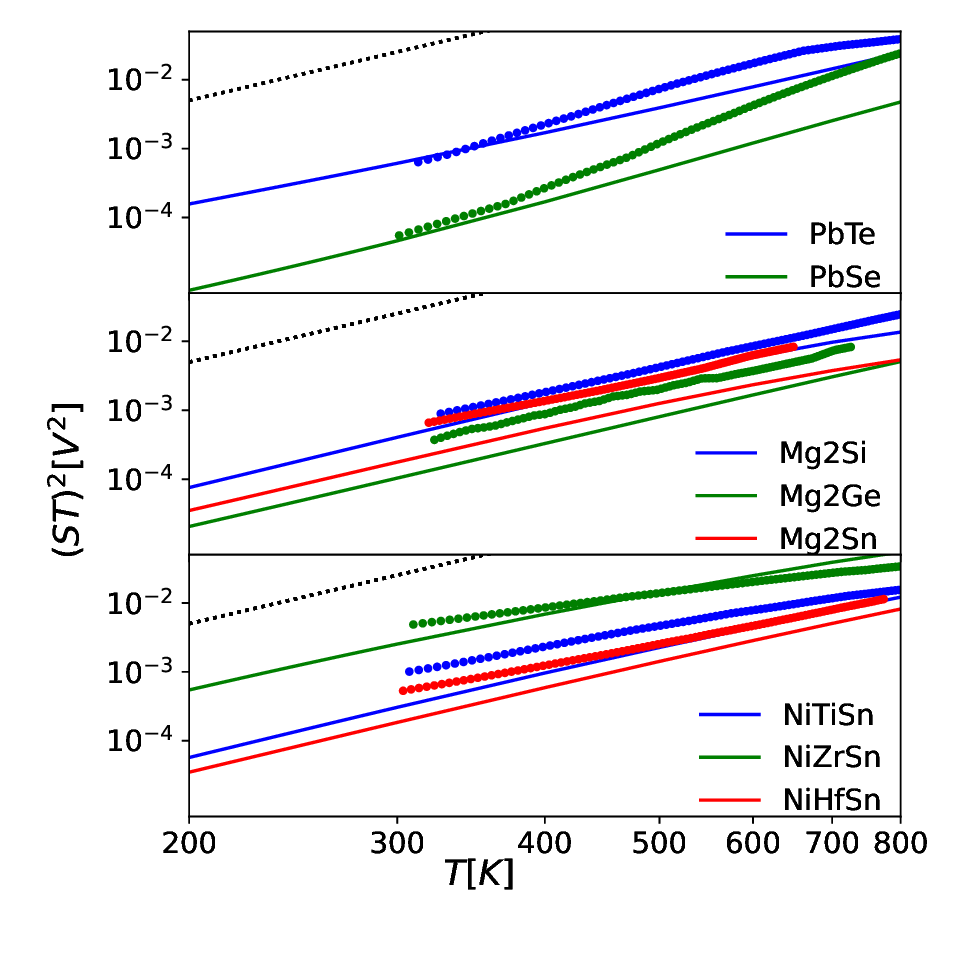}
    \caption{$(ST)^{2}$ as a function of temperature. The calculations are shown as continuous lines and experimental measurements\cite{Jood2020,Zhang2012,TANI2007,GAO201533,Saito2020,ren,xie_2014,yu2009} as filled circles. The $T^4$ law is shown as a black dotted line. }
    \label{ST2}
\end{figure}

As shown in Fig.~\ref{ST2}, the first factor in the above formula is a strongly increasing function of the temperature, 
roughly as $T^4$ because of Mott's law.

\begin{figure}
    \centering
    \includegraphics[width=0.9\linewidth]{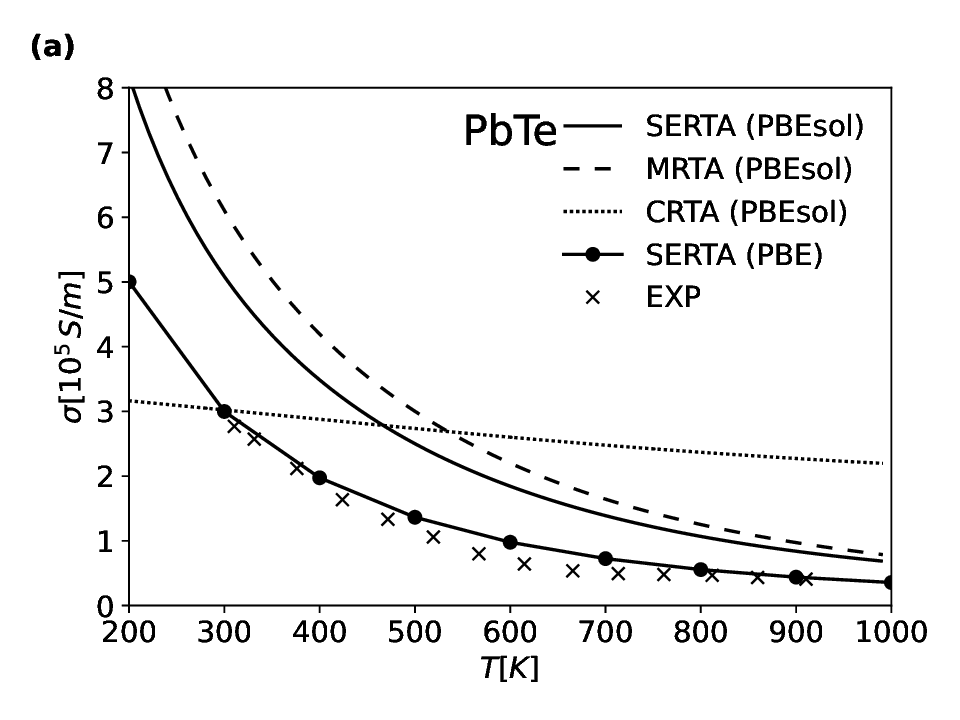}
    \includegraphics[width=0.9\linewidth]{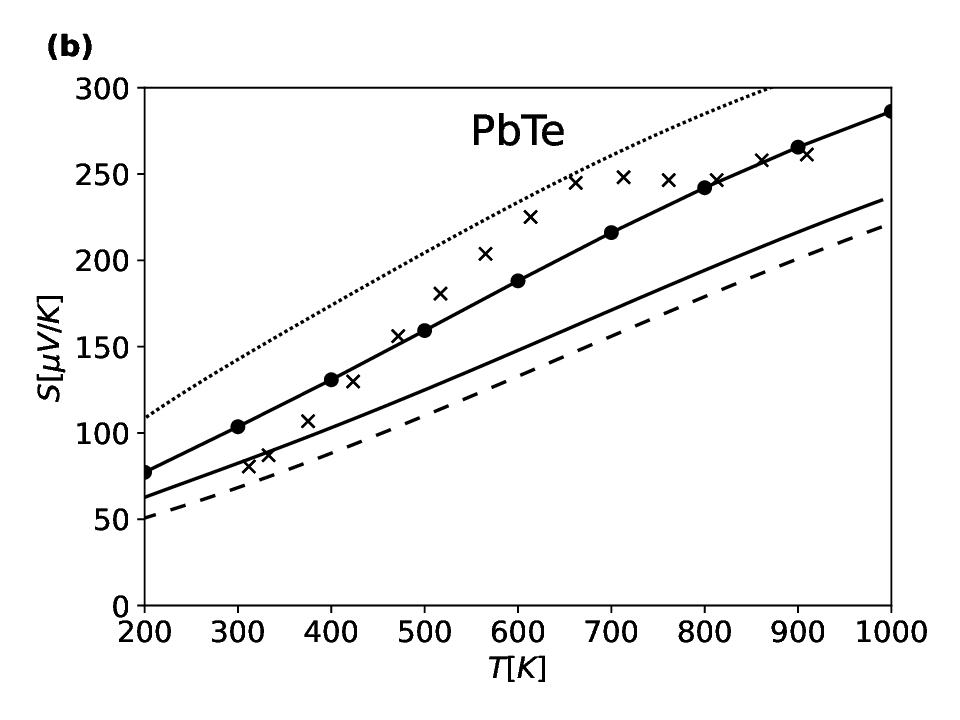}
    \includegraphics[width=0.9\linewidth]{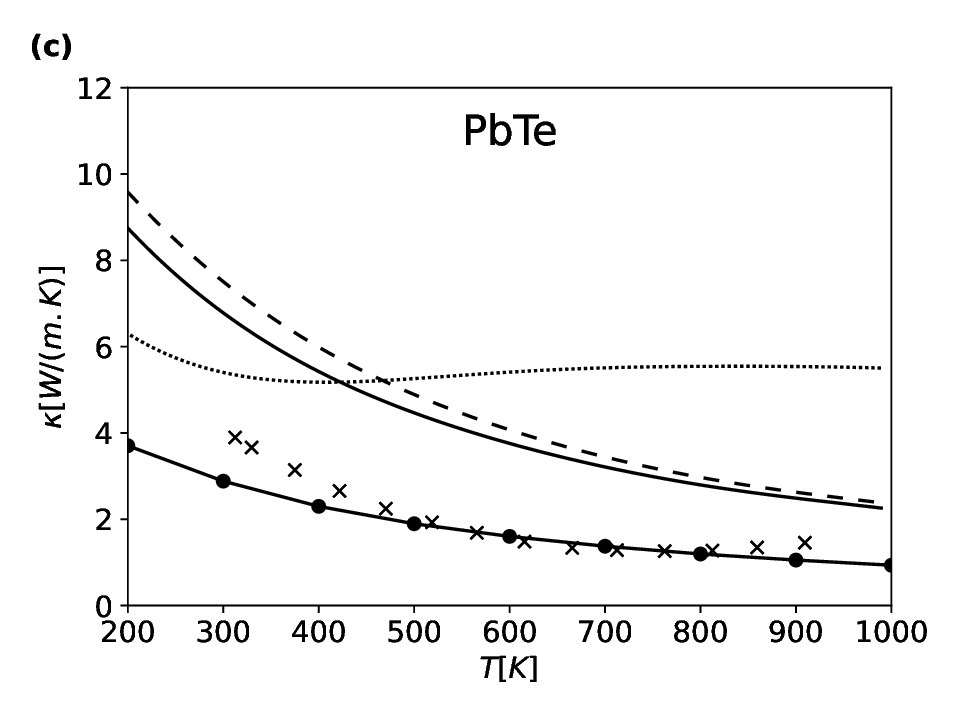}
    \includegraphics[width=0.9\linewidth]{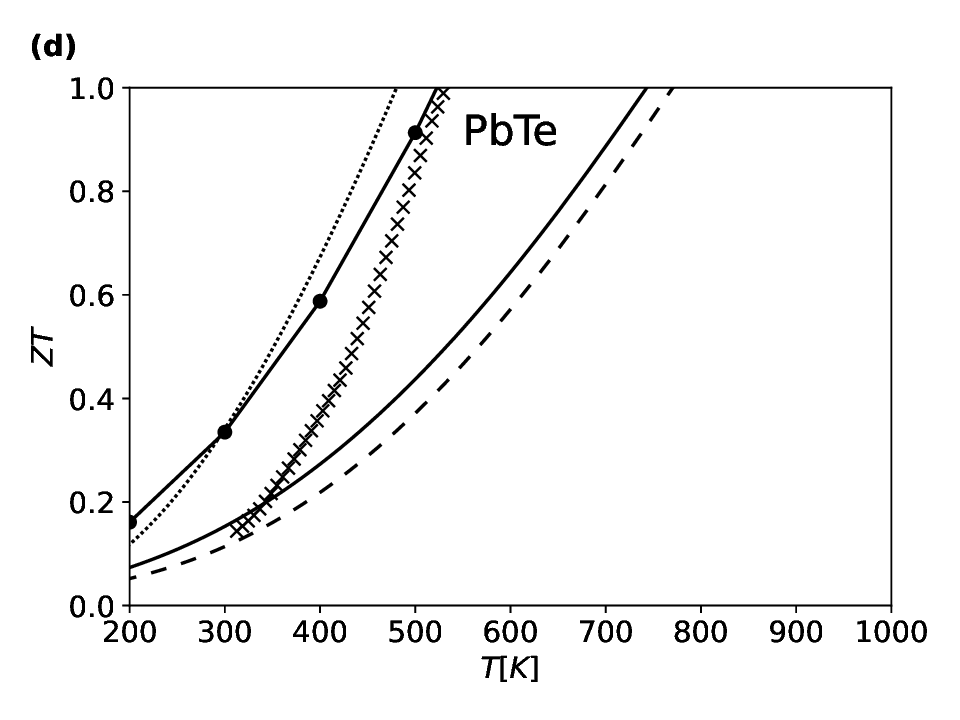}
    \caption{Transport properties of PbTe computed with PBEsol and $k=60$.  (a) electronic conductivity $\sigma$, (b) thermopower $S$, (c) thermal conductivity $\kappa$ and (d) ZT thermoelectric figure of merit. Experimental measurements\cite{Jood2020} are shown using crosses. The results of computations are shown using continuous, dashed and dotted lines for the SERTA, MRTA and CRTA approximations, respectively. For the CRTA, $\tau=10^{-14 }\, \text{s}$ is used. To compute the lattice thermal conductivity $q=27$ is used.\label{RS_app}}
\end{figure} 

\begin{figure}
    \centering
    \includegraphics[width=0.9\linewidth]{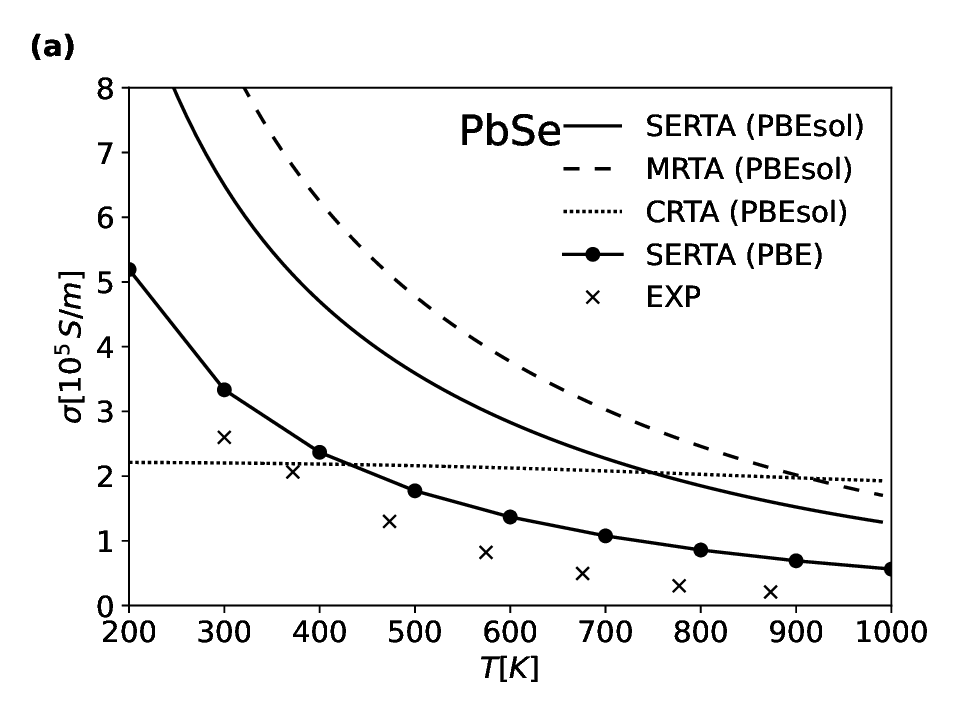}
    \includegraphics[width=0.9\linewidth]{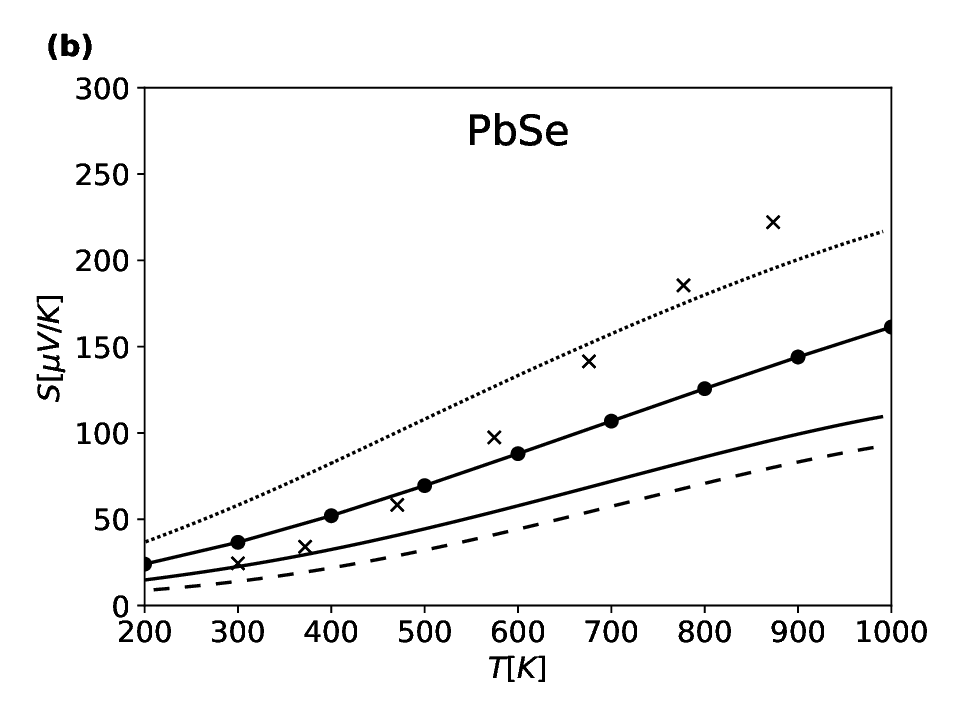}
    \includegraphics[width=0.9\linewidth]{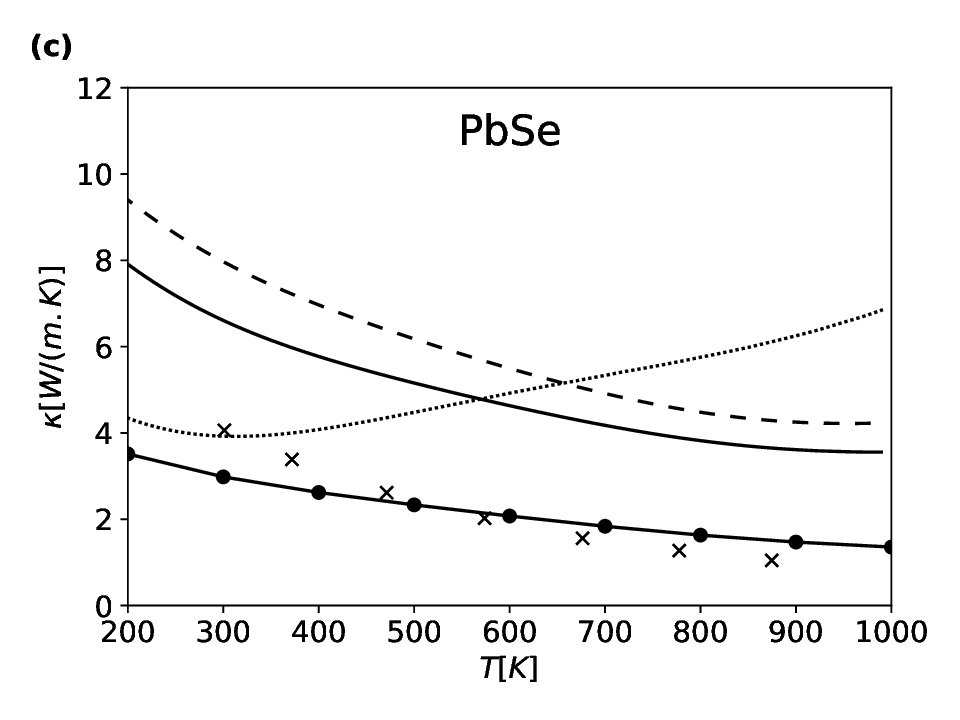}
    \includegraphics[width=0.9\linewidth]{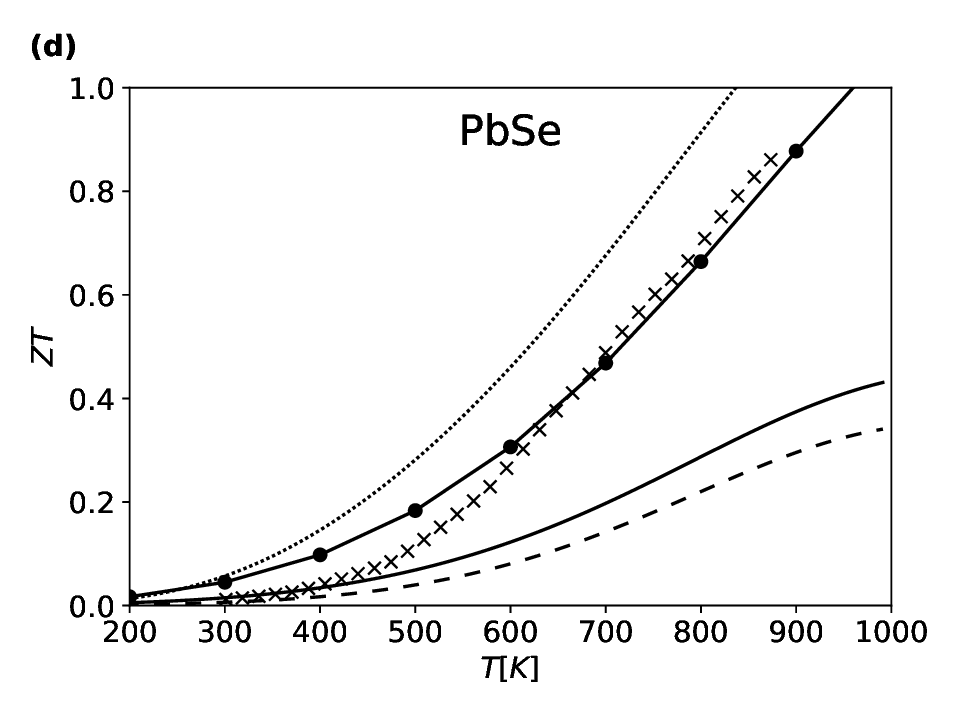}
    \caption{Transport properties of PbSe computed with PBEsol and $k=60$.  (a) electronic conductivity $\sigma$, (b) thermopower $S$, (c) thermal conductivity $\kappa$ and (d) ZT thermoelectric figure of merit. Experimental measurements\cite{Zhang2012} are shown using crosses. The results of computations are shown using continuous, dashed and dotted lines for the SERTA, MRTA and CRTA approximations, respectively. For the CRTA, $\tau=10^{-14 }\, \text{s}$ is used. To compute the lattice thermal conductivity $q=28$ is used.\label{RS_app}}
\end{figure}

In Figs.~\ref{ST2},  \ref{RS_app}, \ref{HH_app} and \ref{Mg2X_app}, we observe that the agreement between theory and experiment is roughly correct for the thermopower. 
The compounds we consider, and thermoelectric compounds in general, are highly doped. Therefore,
to compute $\sigma$, $S$ and $\kappa_e$, one should add the scattering rate of electrons on
impurities to the electron-phonon scattering rate to obtain the total relaxation time. One could for example, add those two scattering mechanisms using Matthiesen rule. 
However, the thermopower $S$ is known to depend weakly on the relaxation time. Indeed, it is the 
ratio of two Onsager coefficients (Eq.~\ref{seebeck}), and some cancellations between the numerator and the denominator occur.
Therefore, the moderate agreement observed between theory and experiment in Figs.~\ref{ST2},  \ref{RS_app}, \ref{HH_app} and \ref{Mg2X_app}  for the thermopower cannot entirely be
explained by a too approximate relaxation time. The doping is so strong in those compounds that the electronic
structure itself may be  modified around the chemical potential and  impact the thermopower.
For example, in the FeVSb half-Heusler, it is well known that, around the chemical potential, the electronic structure is strongly modified by doping \cite{jodin}. 

\begin{figure}
    \centering
    \includegraphics[width=1\linewidth]{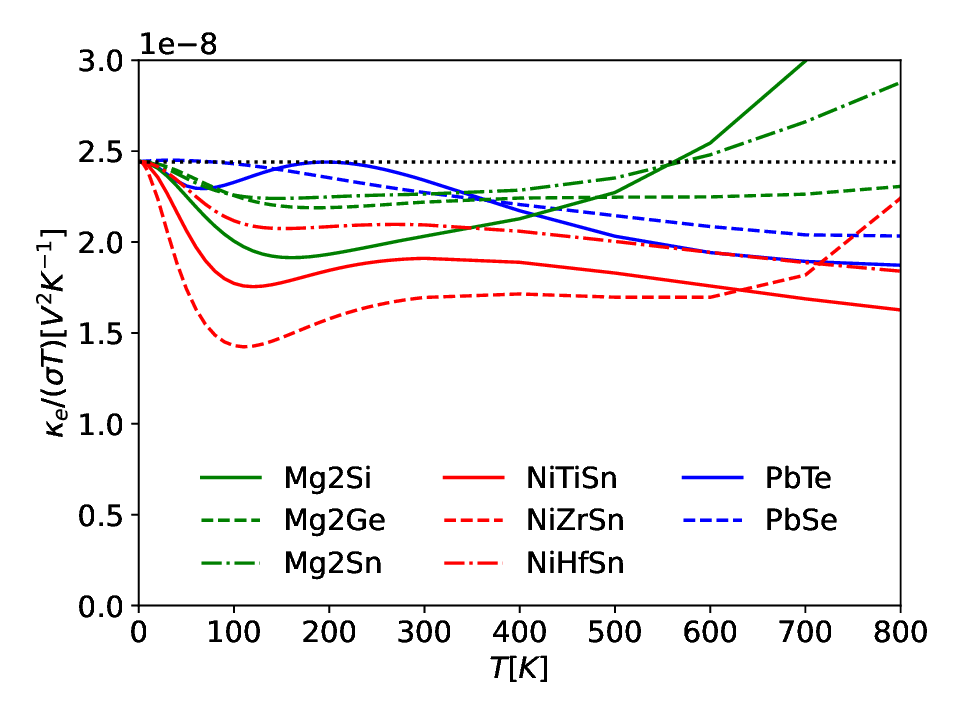}
    \caption{The Lorenz function $\kappa_e/(\sigma T)$ as a function of temperature.}
    \label{Lorenz}
\end{figure}

The second factor in Eq.~\ref{ZT_factors}, $\sigma T/\kappa_e$ is approximately independent of temperature 
and is given by the inverse of the Lorenz number, $L_0= 2.44 \, 10^{-8} \, \text{V}^2 \text{K}^{-2}$, if the metallic regime is reached and the temperature is low enough.
This factor is shown in Fig.~\ref{Lorenz}. We observe that at $T=0$ K it reaches the theoretical limit $L_0$, but in the operating temperature range
$200-800$ K, on average it takes a lower value.  In models with parabolic bands and a power-law relaxation
time \cite{espen_lorenz,kim_lorenz,wang_lorenz} such low values can be obtained when the density of state is rapidly
changing which coincides with large absolute values of the Seebeck coefficient. The rise of the Lorenz number above the Wiedemann-Franz value, $L_0$, at high temperature, indicates multiband effects\cite{wang_lorenz,thesberg_lorenz}. Notice that the convergence with respect to $k$ of the reported transport properties has been studied at $300$ K. Therefore, the values below that temperature are only indicative, the main purpose of reporting them is to show that at $T=0$ K the Wiedemann-Franz value $L_0$ is reached, as a measure of the quality of our numerical scheme.

The last factor in Eq.~\ref{ZT_factors} measures the ratio of the lattice to electronic thermal conductivity. If we use the high-temperature limit of the electron-phonon lifetime, 
$\hbar/\tau_{ep}=2\pi \lambda k_B T$, with $\lambda$ the electron-phonon coupling constant\cite{grimvall1981,chaput2019}, then the lifetime is inversely proportional to
the temperature, and the electronic thermal conductivity becomes independent of temperature. It is obviously a gross approximation, because, for example, 
at high enough temperature the electronic thermal conductivity increases due to multiband effects\cite{zhang}, as shown in Fig.~\ref{Mg2X_app}.
Nevertheless, if we use this approximation and because $\kappa_L \approx C/T^{\delta}$ with $\delta \approx 1$, then we obtain that the last factor in 
Eq.~\ref{ZT_factors} decreases with temperature as $1/(T^2+cT)$, and therefore much slower than $(ST)^2$.

\begin{figure}
    \centering
    \includegraphics[width=1\linewidth]{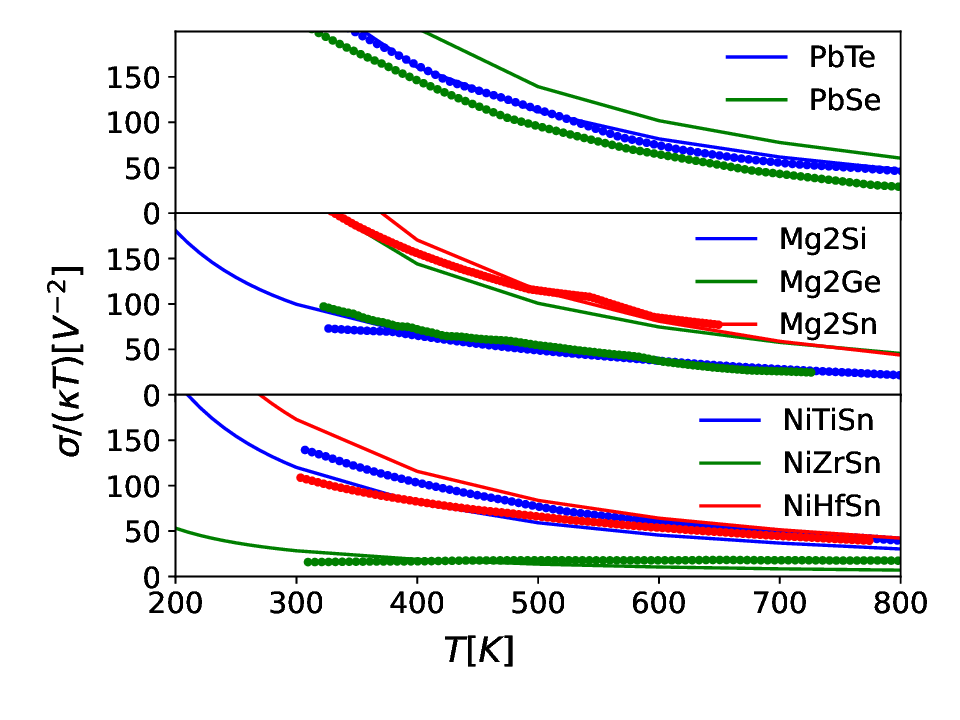}
    \caption{$\sigma/(\kappa T)$ as a function of temperature. The calculations are shown as continuous lines and experimental measurements\cite{Jood2020,Zhang2012,TANI2007,GAO201533,Saito2020,ren,xie_2014,yu2009} as filled circles.}
    \label{skt}
\end{figure}

\begin{figure}
    \centering
    \includegraphics[width=1\linewidth]{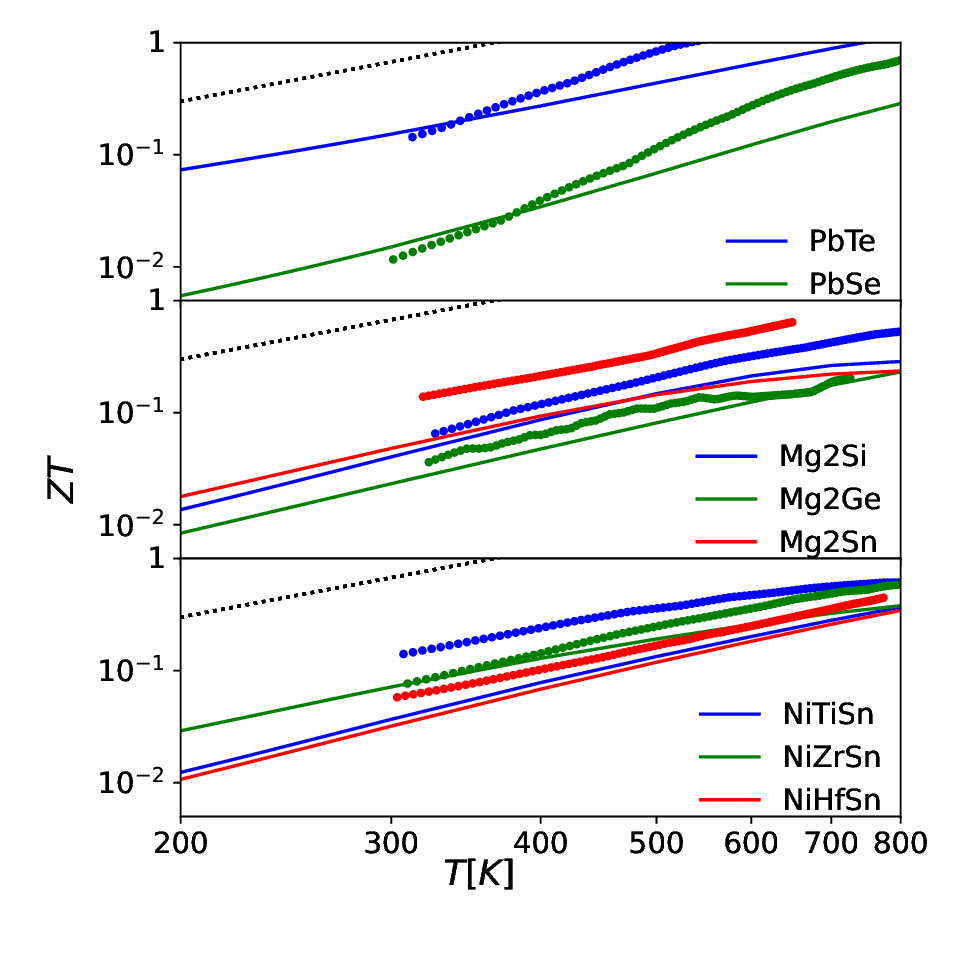}
    \caption{$ZT$ as a function of temperature. The calculations are shown as continuous lines and experimental measurements\cite{Jood2020,Zhang2012,TANI2007,GAO201533,Saito2020,ren,xie_2014,yu2009} as filled circles. The $T^2$ law is shown as a black dotted line. }
    \label{ZT_fig}
\end{figure}

Because experimentally it is difficult to split $\kappa$ into $\kappa_e$ and $\kappa_L$, to compare with theory, in Fig.~\ref{skt} we show the product of the last two factors of Eq.~\ref{ZT_factors}, 
$\sigma/(\kappa T)$. According to the above analysis, it behaves as $1/(T^2+cT)$ with temperature. As in the case of $(ST)^2$, the agreement
with experiment is roughly correct. The results of Fig. \ref{ST2} and Fig.~\ref{skt} can be combined, in Fig \ref{ZT_fig},  to obtain the figure of merit as a function of temperature. The results are also replotted individually for each compound family in Figs.~\ref{RS_app}, \ref{HH_app} and \ref{Mg2X_app}. It is observed that the $ZT$ follow approximately a $T^2$ law, as obtained from the above analysis.

This reasonable agreement between theory
and experiment should be considered with caution. If we look more into the details, in Tab.~\ref{roomT}, but also in Figs.~\ref{RS_app}, \ref{HH_app} and \ref{Mg2X_app}, where $\sigma$ and $\kappa$ are plotted separately, then
we see that the agreement with experiment indicates room for improvement. As previously mentioned, we may attribute, at least partially, the difference between theory and experiments to impurities. 
In $\kappa_L$, the impurities would manifest themselves through a relaxation time due to the scattering of the phonons on mass fluctuations\cite{tamura}, and in 
$\kappa_e$ through the scattering of the electrons on the change in the potential due to doping. It seems that those two effects partially cancel in the ratio $\kappa_L/\kappa_e$, 
which allows us to obtain a rough estimate of the thermoelectric figure of merit, in Fig.~\ref{ZT_fig}, considering only the electron-phonon interaction.

In Figs.~\ref{RS_app}, \ref{HH_app} and \ref{Mg2X_app}, we considered the different relaxation-time approximations given by Eq.~\ref{tau_SERTA_1} and Eq.~\ref{tau_MRTA_1}. 
The trends we observe are the same and therefore the above conclusions still apply.  Moreover, it can be noted that for the conductivity we have MRTA $>$ SERTA.

From Tab.~\ref{roomT} and Fig.~\ref{RS_app} we observe that the agreement of electrical conductivity with experiment is better for PbTe than for other compounds, and in fact even improved if we use the 
PBE exchange-correlation functional instead of PBEsol (See Fig.~\ref{RS_app}). This, however, may be accidental since impurity scattering is neglected.

Finally, we mention a peculiar behavior in Fig.~\ref{ZT_fig}. The increase of $ZT$ with temperature for PbTe and PbSe, given by the experiments in Ref.~\citenum{Jood2020} and \citenum{Zhang2012} seems to be faster than those given by the calculations and experiments on other materials. This behavior is also observed in the thermopower in Fig.~\ref{ST2}, and therefore is likely due to an electronic effect. It could be due to the temperature dependence of the electronic structure, as mentioned in Ref.~\cite{yanzhong}, or due to changes in the electronic structure due to doping.

\subsection{Discussion }
In this paper we have considered the transport properties of thermoelectric compounds. Those compounds are heavily doped compounds containing heavy elements, therefore it is interesting to discuss how impurity scattering and spin orbit coupling may impact the results.

\begin{figure}
\begin{center}
    \includegraphics[scale=0.25]{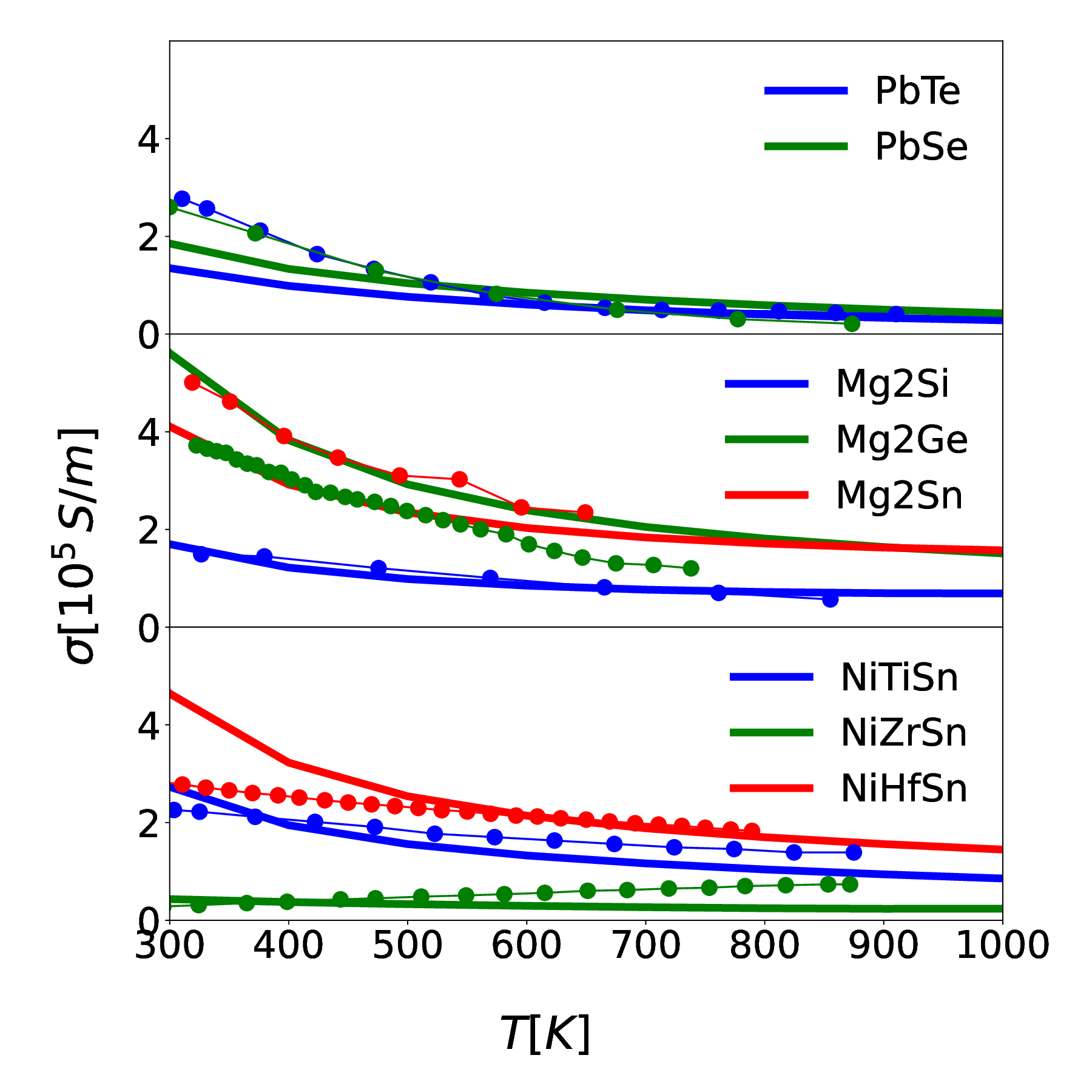}
    \caption{Electrical conductivity of the thermoelectric compounds considered in this work.  The results of the calculations are shown as continuous lines, and the experimental values as dots.  The calculations include impurity scattering as given by the  Debye and Conwell formula.  NiXSn and Mg$_2$X have been computed using the PBEsol exchange correlation functional using parameters discussed explicitly in the previous sections while the PBE functional is used for PbX compounds. Experimental values are obtain from the references given in  Tab. \ref{roomT}. 
     \label{cond_I} 
   }
\end{center}
\end{figure}

The scattering of electrons on impurities is not yet implemented in our approach, therefore direct comparison with experiments is difficult. 
Indeed, in the previous sections we have presented the results without accounting for this contribution.
However, using the model of Debye and Conwell\cite{debye} it is possible to get an estimate of how strongly impurity scattering impact the electrical conductivity at a given impurity concentration. In Appendix \ref{Si} we reproduce the results of \cite{ponce} and \cite{protik} for $n$-type silicon and describe the model. The agreement with experiments is excellent as long as impurity scattering is included using the above mentioned model. Figs. \ref{mob_n} and \ref{mob_T} evidence that for heavy doping ($n \sim  10^{19}\, cm^{-3}$)  impurity scattering is essential to consider, even at room temperature. This is in this concentration range that thermoelectric compounds usually operate, as seen in Tab. \ref{roomT}.
Therefore we applied the same empirical model to the compounds considered in this study. The results are shown in Fig. \ref{cond_I}. According to Eq.  \ref{mix}, the electrical conductivity we obtain from our \textit{ab initio} calculations ($\sigma_{ep}$) is multiplied by the  Debye and Conwell\cite{debye} factor $\Big(1+X^2 [\text{ci}(X)\cos(X)+\text{si}(X)\sin(X)]\Big)$ to account for impurity scattering.  Notice that in this case to compute $X=\sqrt{6 \sigma_{ep}/\sigma_{ei}}$ we use the Brooks-Herring formula\cite{RevModPhys.53.745} for the electrical conductivity $\sigma_{ei}$ corresponding to impurity scattering only, since an empirical formula equivalent to Eq. \ref{mu_i} for Si is not available for the compounds we consider. We see on Fig. \ref{cond_I} that we obtain a good agreement with experiments. One exception is Mg$_2$Sn, but it has been noted already that the PBEsol and PBE functionals give a negative band gap, therefore the discrepancy with experiments is likely to be due to the too approximate electronic structure we use.

As mentioned already, spin-orbit coupling (SOC) was not considered in those calculations because at the level of semi-local functionals it may not lead to a systematic improvement of the transport properties. Indeed, it is shown in \cite{hummer_structural_2007} that for band gaps, at the level of semi-local functionals, including SOC is not always beneficial.  This is also discussed in \cite{souza} for the effective mass. In Appendix \ref{SOC} we confirm this behaviour for the electrical conductivity considering PbSe as a special case.

\begin{figure}
\begin{center}
    \includegraphics[scale=0.25]{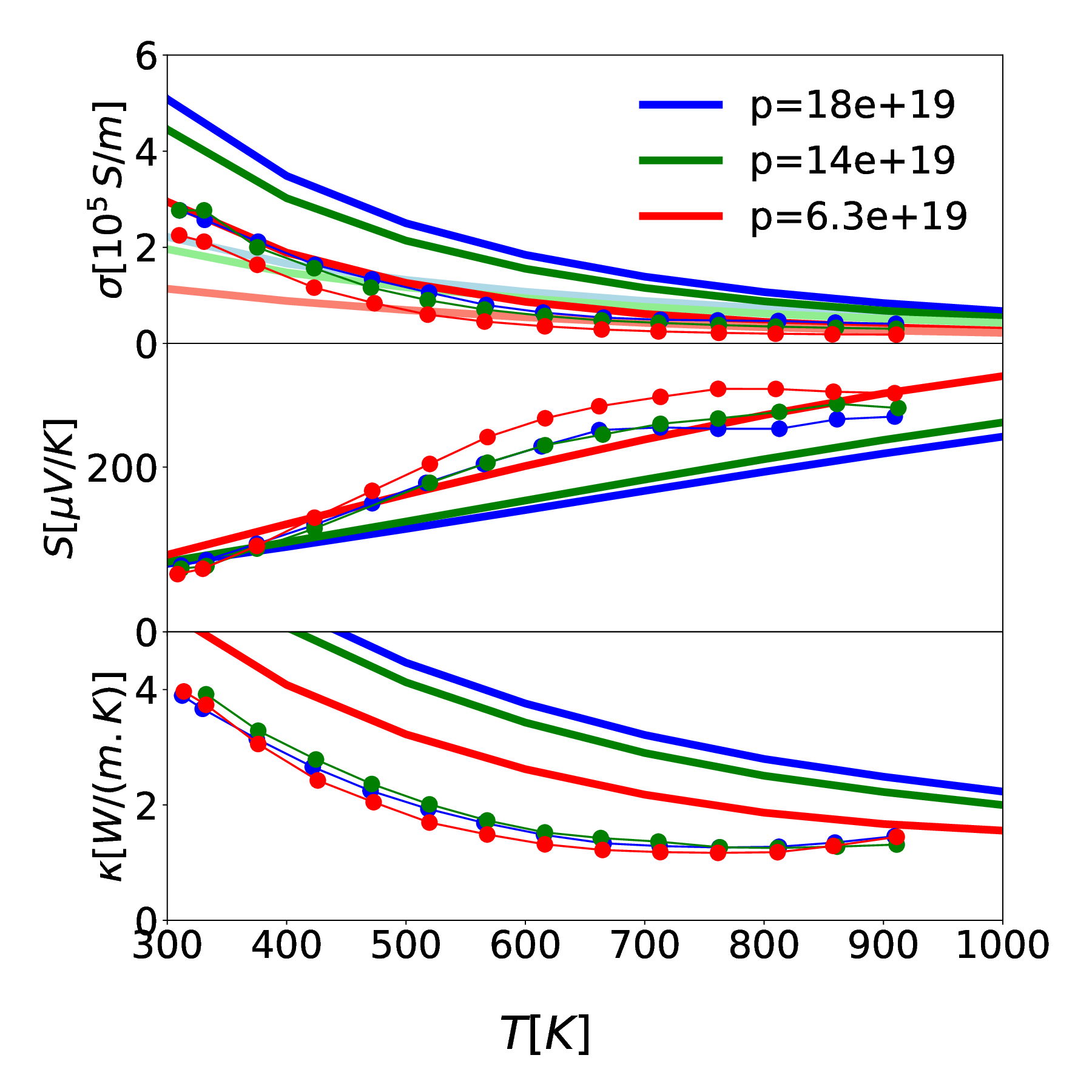}
    \caption{ Transport properties of PbTe for different carrier concentrations. Calculations are shown as continuous lines and experiments as dots. Calculations are performed using the PBEsol functional in the SERTA approximation. Experiments are from \cite{Jood2020}. The light blue green and red lines for the conductivity show the results of the calculations corrected for impurity scattering using the Debye and Conwell\cite{debye} formula.  \label{PbTe_new_c}  }
\end{center}
\end{figure}

We end this section discussing the impact of carrier concentration on our results. Our approach to compute the transport coefficients assumes the doping is weak enough for bands to exists and the Boltzmann equation to hold. The Boltzmann equation, Eq. \ref{boltz}, is written using the electron band structure $\epsilon_{\mathbf{k}n}$ of the perfect crystal. Therefore to obtain the heat flux, $ (\epsilon_{\mathbf{k}n}-\mu)\mathbf{v}_{\mathbf{k}n}f_{\mathbf{k}n}$, the chemical potential is computed from Eq. \ref{fermi} with $\rho(\epsilon)$ the density of states of the perfect crystal. To obtain the value of $\mu$ at a given temperature Eq. \ref{fermi} is solved considering the carrier concentration $n_0$ as an input parameter. In this work we have assumed that $n_0$ can be obtained from Hall measurements and is equal to the Hall number $n_H$ given in Tab. \ref{roomT}. This is the approach used in most works, but it is an approximation. At first $n_H$ equals the carrier concentration only for a spherical Fermi surface\cite{Ziman_1972}. But, in fact, $n_0$ does not even need to equal the carrier concentration. It should be regarded as an effective parameter used to locate the chemical potential which properly selects the states which carry the currents. In Eq. \ref{fermi} if $\rho(\epsilon)$ is the density of the real crystal, then obviously $n_0$ is the carrier concentration. But in our approach $\rho(\epsilon)$ is the density of the perfect crystal, therefore $n_0$ contains implicit information about the change of the density of states due to the doping and the microstructure of the sample.
We do not have a procedure to choose $n_0$ accurately, therefore we regard $n_0\approx n_H$ as a practical approximation. 
In Fig. \ref{PbTe_new_c} we computed the transport properties of PbTe for the different Hall concentrations reported in \cite{Jood2020}. The change of the transport properties with carrier concentration reproduce well the experiments, and once the conductivity is corrected for impurity scattering the agreement is excellent. 
This is an interesting result because it can be seen from the band structure and from\cite{PhysRevB.94.195141}  that, despite a complicated, non-spherical Fermi surface, the deviation between $n_0$ and $n_H$ still allows for a reasonable calculation of the transport properties. As reported in\cite{Jood2020}, at large doping concentration, $18 \times 10^{19} \text{ cm}^{-3}$, $n$-type intrinsic defects are created when the Na doping atoms are entering the lattice. Therefore, the $n_H$ value obtained experimentally characterizes a complicated microstructure. At a given temperature, the same $n_H$ value can be obtained for different microstructures; our approach does not allow us to distinguish between them. Therefore, whether or not the electronic states selected by the chemical potential approximately describe the electronic states of the actual sample is assessed by comparing the computed transport properties to the experiments. In Fig. 12, for $p = 18 \times 10^{19} \text{ cm}^{-3}$, we observe that despite the rather complicated microstructure, and the known convergence of the L and $\Sigma$ bands due to doping\cite{Jood2020}, the agreement with experiment is acceptable when impurity scattering is taken into account.

\section{Conclusions}

We propose a computational approach to the thermoelectric figure of merit based on the calculation of the electron-phonon and phonon-phonon interactions from first-principles. Numerical strategies have been developed to make these calculations efficient.
In particular, we show that the transport coefficients can be computed without any interpolation of the electron-phonon matrix elements, other than the minimal image convention \cite{Parlinski,Brunin,chaput2019,engel2022}. Avoiding the Wannierization procedure, which often requires human intervention, facilitates the direct application of this approach to large sets of materials under different physical conditions such as strain or using different exchange-correlation functionals.

The transport coefficients we obtain reproduce the experiments, even if the agreement is not perfect.
A better description of the electronic structure such as band gaps and effective masses is required to reduce the discrepancies. This, in turn, is also expected to be important in the description of the long-range part of the electron phonon potential since the value of the band gap also affects the value of the dielectric tensor and Born effective charges and we have seen in Sec. \ref{results} that those quantities strongly impact the transport properties. But of course a better description of the band gaps should not come at the expense of a good description of the effective masses~\cite{kim_towards_2010},  since those are also crucial for the computation of the transport coefficients.

For the heavily doped conditions of the systems under study, we have seen that the inclusion of impurity scattering and the band structure relaxation should have a non-negligible impact on the final results.
Additional work should also be done to investigate the role of the lattice thermal expansion as well as the effect of the temperature dependence of the band-structure in the electronic transport coefficients.


\begin{acknowledgments}
This work was supported by JSPS KAKENHI Grant Numbers JP24K08021,
JP24H00190, JP25H01246, and 25H01252. Part of the calculations in this
study were performed on the Numerical Materials Simulator at NIMS.
\end{acknowledgments}

\newpage
\begin{widetext}

{\small
\begin{table}[H]
\centering
\caption{Table of thermoelectric transport properties at room temperature. Negative $n$ values mean $p$ doping. The units are  $n[10^{19}\text{cm}^{-3}]$, $\sigma[10^5(\Omega \cdot \text{m})^{-1}]$, $S[10^{-6}\text{V}/\text{K}]$, $\kappa[\text{W}/(\text{m} \cdot \text{K})]$. A $100\times100\times100$ mesh is used to sample the Brillouin zone to compute the electronic transport properties. To compute the total thermal conductivity ($\kappa$) of large supercells, we used the value of the lattice thermal conductivities of the small supercells. The mesh sizes used to sample the Brillouin zone to compute the lattice thermal conductivities are given in Figs. \ref{RS_app}, \ref{HH_app} and \ref{Mg2X_app}. The \textit{information} column contains the experimental nominal composition, or the supercell size, of the conventional cells, used for the calculation.\label{roomT}}
\begin{tabular}[t]{lccccccccccc}
\hline
 &$n$ & $\sigma$ & $S$ & $\kappa$ & $\kappa_e$ & $\kappa_l$ &  $ZT$ & method & information \\
\hline 
PbTe   & -18 & 2.9 & 76 & 4 &  & & 0.12 & exp & Pb$_{0.96}$TeNa$_{0.04}$\cite{Jood2020}  \\
  &  -18 &  4.5&  87.7 & 6.4 & 3.2 & 3.2& 0.16 &PBEsol & $2\times2\times2$ \\
  &  -18 & 4.4 &  86.8 & 6.3 & 3.1  & & 0.16 &PBEsol & $4\times4\times4$  \\
  &  -18 & 2.6 & 109.8  & 2.6 & 1.7  &0.9 & 0.36 &PBE & $2\times2\times2$  \\  
  &  -18 & 2.6 & 109.5  & 2.6 & 1.7 & & 0.36 &PBE & $4\times4\times4$  \\  
PbSe   & -16 & 2.6 & 24.3 & 2.4 &  & & 0.02 & exp & Pb$_{0.985}$SeK$_{0.015}$\cite{Zhang2012} \\
  & -16 & 5.8 & 26.5 & 6.1 & 3.9 & 2.2 & 0.02 & PBEsol & $2\times2\times2$  \\
  & -16 & 5.6 & 26.5& 6.0 &3.8  & & 0.02 & PBEsol & $4\times4\times4$   \\
  &  -16 & 2.9 & 40.4  & 2.7 & 1.9 &0.8 & 0.05 &PBE & $2\times2\times2$  \\  
  &  -16 & 2.9 & 40.8  & 2.7 & 1.9 & & 0.05 &PBE & $4\times4\times4$  \\  
\hline
Mg$_2$Si   & 15 & 1.5 & -88 & 6.7 &  & & 0.05 & exp & Mg2SiSb$_{0.02}$ \cite{TANI2007}  \\
           & 15 &   4.3     &  -65.2     & 14.8 & 2.5      & 12.3 &  0.04  & PBEsol & $2\times2\times2$ \\
           & 15 &   4.1     &   -65.0   & 14.8 &  2.5    &  &  0.04  & PBEsol & $4\times4\times4$ \\
  &  15 & 3.7 & -79.5  & 14.2 & 2.3  &11.9 & 0.05 &PBE & $2\times2\times2$  \\  
  &  15 & 3.6 & -79.4  & 14.1 & 2.2 & &0.05 & PBE & $4\times4\times4$  \\  
Mg$_2$Ge   & 46 & 4 & -59 & 12 &  & & 0.03&  exp & Mg$_{2.2}$Ge$_{0.995}$Sb$_{0.005}$ \cite{GAO201533}\\
          & 46 & 10.6  & -34.2 & 16.1 & 6.9 & 9.2 &0.02 &  PBEsol & $2\times2\times2$ \\
          & 46 & 10.3  & -34.1 & 15.8  & 6.6 & & 0.02 & PBEsol & $4\times4\times4$ \\
  &  46 & 9.6 & -40.1  & 14.7 & 6.2 & 8.5& 0.03 & PBE & $2\times2\times2$  \\  
  &  46 & 9.3 & -40.0  & 14.5 & 6.0 & & 0.03& PBE & $4\times4\times4$  \\  
Mg$_2$Sn   & 44.5 & 4.9 & -78 & 7.4 &  & &0.12 &  exp & Mg2Sn$_{0.98}$Sb$_{0.02}$ \cite{Saito2020} \\
          & 44.5 & 13.1 & -45.4 & 16.5 & 8.8 & 7.7 & 0.05&  PBEsol & $2\times2\times2$ \\  
          & 44.5 & 12.7 & -45.4 & 16.2 & 8.5 & &0.05 &  PBEsol & $4\times4\times4$ \\  
  &   44.5 &   &   &  &  & 6.0& & PBE & $2\times2\times2$  \\  
  &   44.5 & 15.2 & -35.4  & 15.7 &9.7   & & 0.04& PBE & $4\times4\times4$  \\  
\hline
NiTiSn   & 72 & 2.2 & -102 & 5.3 &  & & 0.13&  exp &   Ti$_{0.975}$Ta$_{0.025}$NiSn\cite{ren} \\ 
         & 72 &  7.8   &  -57.1    &  21.8   & 4.4  & 17.4 & 0.03 &   PBEsol & $2\times2\times2$   \\
         & 72 &   7.8 &  -57.1    & 21.8    &  4.4 &  & 0.03   & PBEsol & $4\times4\times4$   \\
  &   72 & 7.0 & -60.3 &18.8 & 4.0 &14.8   &0.04 & PBE & $2\times2\times2$  \\  
  &   72 & 7.0 &  -60.3&18.8 & 4.0 &   &0.04 & PBE & $4\times4\times4$  \\ 
NiZrSn   & 5.9 & 0.28 & -224 & 6.1 &  & & 0.07&  exp & ZrNiSn\cite{xie_2014}\\ 
         & 5.9 &  1.6   &   -167.3   & 19.4  &0.8     & 18.6 & 0.07 &   PBEsol & $2\times2\times2$   \\
         & 5.9 &   1.6 &   -167.3  & 19.4  &  0.8   &  & 0.07 &   PBEsol & $4\times4\times4$   \\-
  &   5.9 &1.4  & -174.6  &17.4  &0.7  & 16.7&0.07 & PBE & $2\times2\times2$  \\  
  &   5.9 & 1.4 &  -174.6 & 17.4 & 0.7 & & 0.07 &PBE & $4\times4\times4$  \\ 
NiHfSn   & 72.3 & 2.83 & -74 & 8.4 &  & &0.05 &  exp & HfNiSn$_{0.98}$Sb$_{0.02}$\cite{yu2009} \\ 
         & 72.3 &  13.0   &   -45.3   &  25.9   &  8.1 & 17.8 & 0.03 &   PBEsol & $2\times2\times2$   \\
         & 72.3 &  12.9   &  -45.2    &  25.8   & 8.0  &  & 0.03 &   PBEsol & $4\times4\times4$   \\
  &   72.3 & 12.0 & -47.4  &  22.6&  7.4&15.2 &0.04 & PBE & $2\times2\times2$  \\  
  &   72.3 & 11.9 &  -47.4 & 22.6 & 7.4 & &0.04 & PBE & $4\times4\times4$  \\
\hline
\end{tabular}
\end{table}
}
\newpage



\appendix

\section{Thermoelectric materials \label{TE-mat}}
The materials properties we obtained are summarized in Tab. \ref{materials-PBEsol} when the PBEsol exchange correlation potential is used, and in Tab. \ref{materials-PBE} for PBE.

\begin{center}
\begin{table}[H]
\centering
\caption{Table of the materials considered. The reported values are obtained using the PBE exchange-correlation potential. \label{materials-PBE}}
\begin{tabular}[t]{lccccc}
\hline
               & lattice parameter [\AA] & dielectric constant & Born effective charges & Gap [eV] \\
\hline 
PbTe           &  $a=6.566$                    &   $\epsilon_r=26.400$                  & $Z_\text{Pb}=5.730$, $Z_\text{Te}=-5.730 $              &   0.83         \\
PbSe           &  $a=6.216$                    &    $\epsilon_r=20.213$                 &  $Z_\text{Pb}=4.797$, $Z_\text{Se}=-4.797$         &      0.44           \\
\hline
Mg$_2$Si       &   $a=6.360$                   &  $\epsilon_r=15.365$                   & $Z_\text{Si}=-3.702$, $Z_\text{Mg}=1.851$            &    0.21    \\
Mg$_2$Ge       &   $a=6.411$                   &   $\epsilon_r=16.851$                  & $Z_\text{Ge}=-3.650$, $Z_\text{Mg}=1.825$             &   0.16             \\
Mg$_2$Sn       &    $a=6.808$                  &     $\epsilon_r=20.552$                &  $Z_\text{Sn}=-3.938$, $Z_\text{Mg}=1.969$            &              \\
\hline
NiTiSn         &  $a=5.945$                    & $\epsilon_r=24.382$                    &   $Z_\text{Ti}=2.814 $, $Z_\text{Sn}=1.122$, $Z_\text{Ni}=-3.936$       &    0.44           \\
NiZrSn         &   $a=6.147$                   &    $\epsilon_r=22.047$                 &  $Z_\text{Zr}=2.653$, $Z_\text{Sn}=0.955$, $Z_\text{Ni}=-3.608$         &     0.48          \\
NiHfSn         &   $a=6.113 $                   & $\epsilon_r=21.067 $                    & $Z_\text{Hf}=2.735$, $Z_\text{Sn}=0.892$, $Z_\text{Ni}=-3.627$        &     0.37           \\
\hline
\end{tabular}
\end{table}
\end{center}

\subsection{Rocksalt}

\begin{figure}[H]
    \centering
    \includegraphics[scale=0.36]{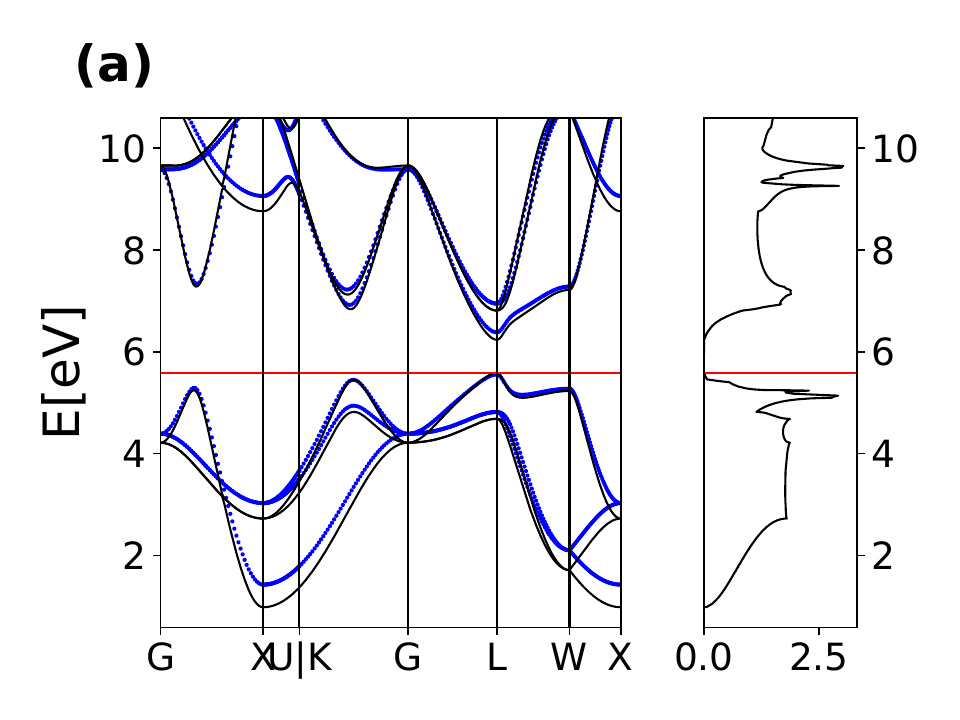}
    \includegraphics[scale=0.36]{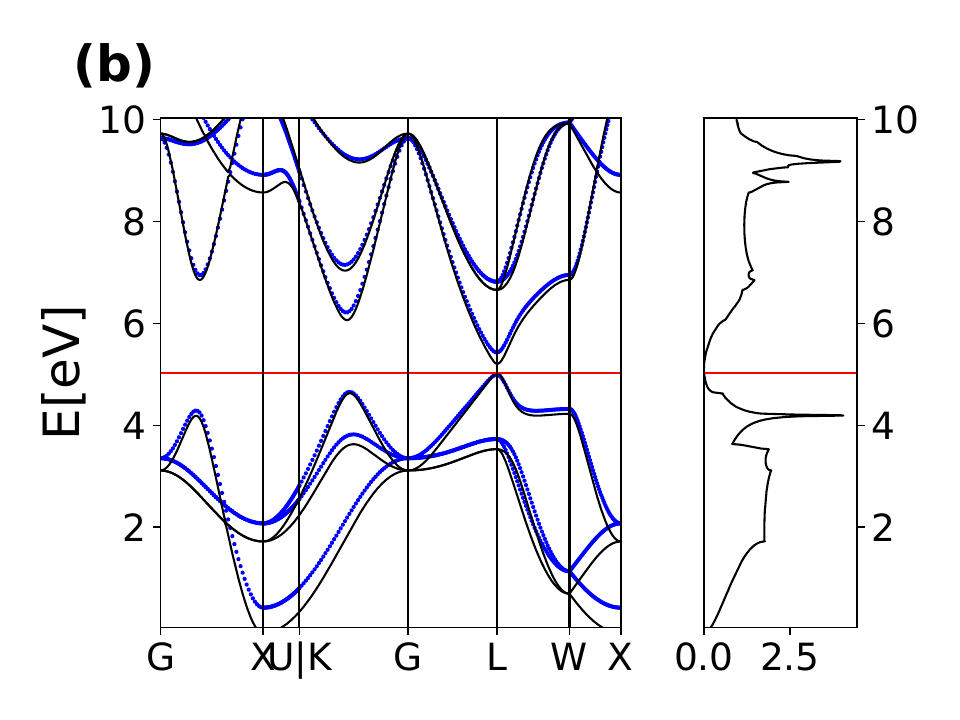}
    \caption{Electron band structures of (a) PbTe and (b) PbSe using the PBEsol exchange-correlation functional, in black and PBE in blue.}
\end{figure}

\begin{figure}[H]
    \centering
    \includegraphics[scale=0.36]{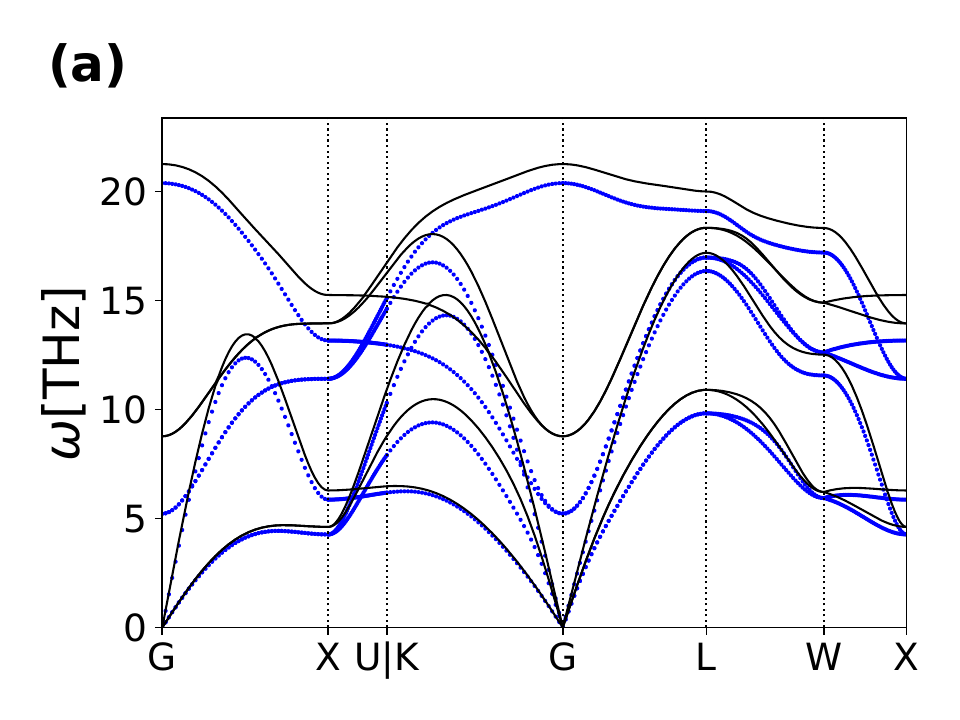}
    \includegraphics[scale=0.36]{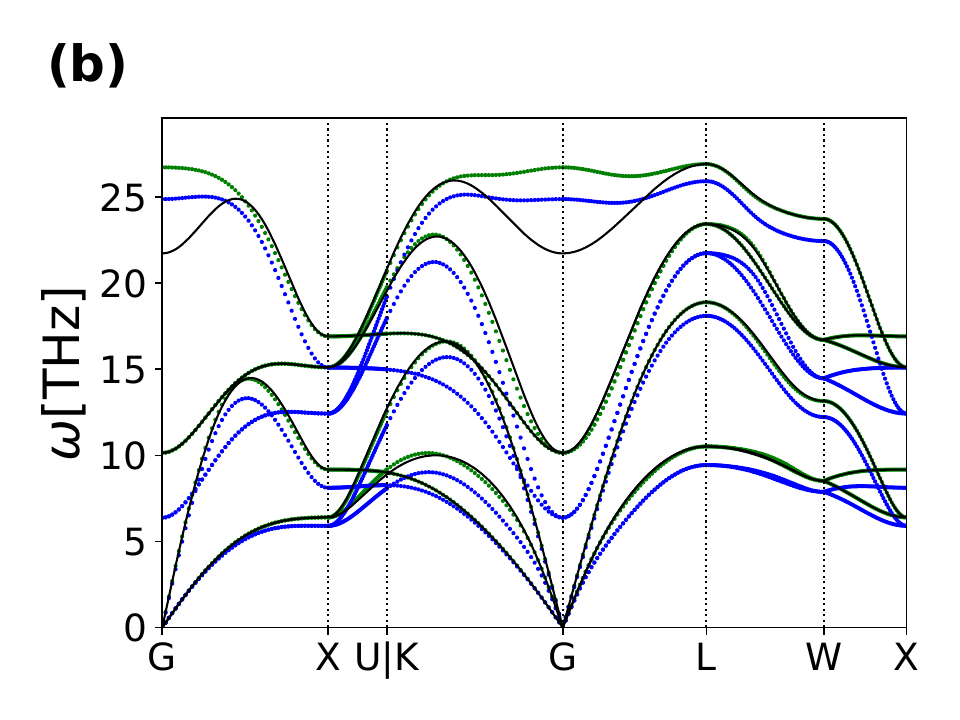}
    \caption{Phonon dispersion of (a) PbTe and (b) PbSe using the PBEsol exchange-correlation functional, in black and PBE in blue. In (b) the PBEsol dispersion obtained using the PBE dielectric function and Born effective charges is also shown in green (see main text). \label{PbX_phonon}}
\end{figure}

\subsection{Half Heusler}
\begin{figure}[H]
    \centering
    \includegraphics[scale=0.36]{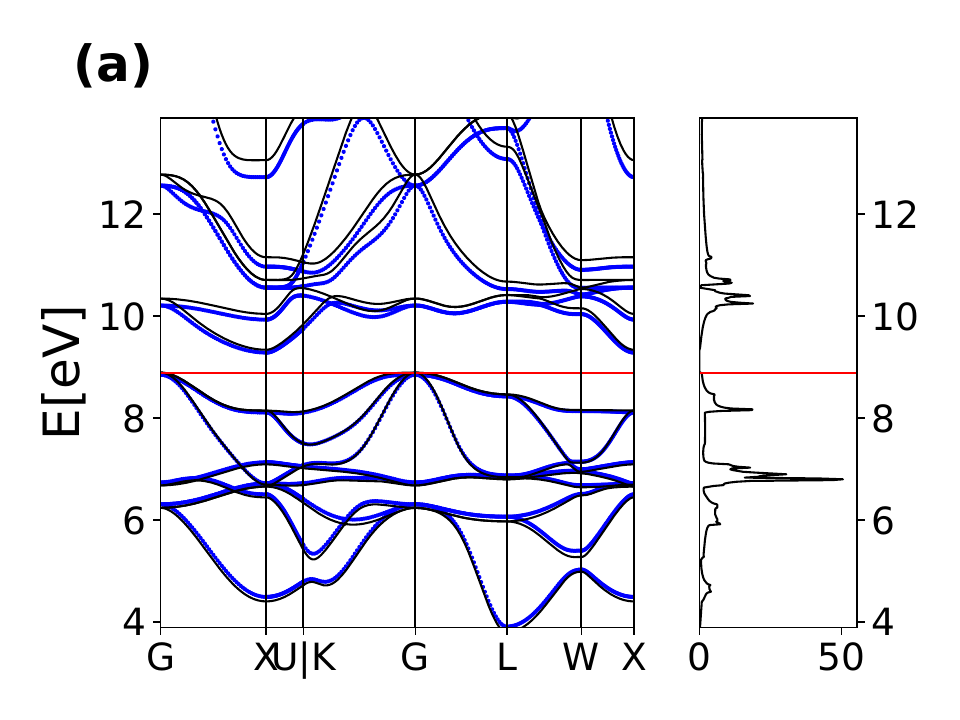}
    \includegraphics[scale=0.36]{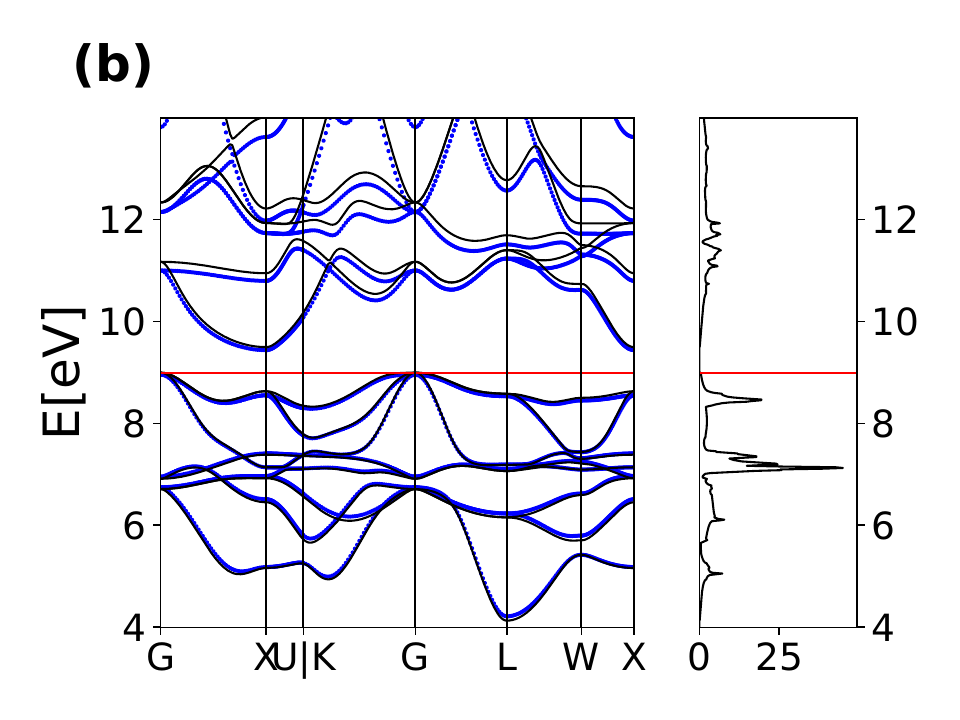}
    \includegraphics[scale=0.36]{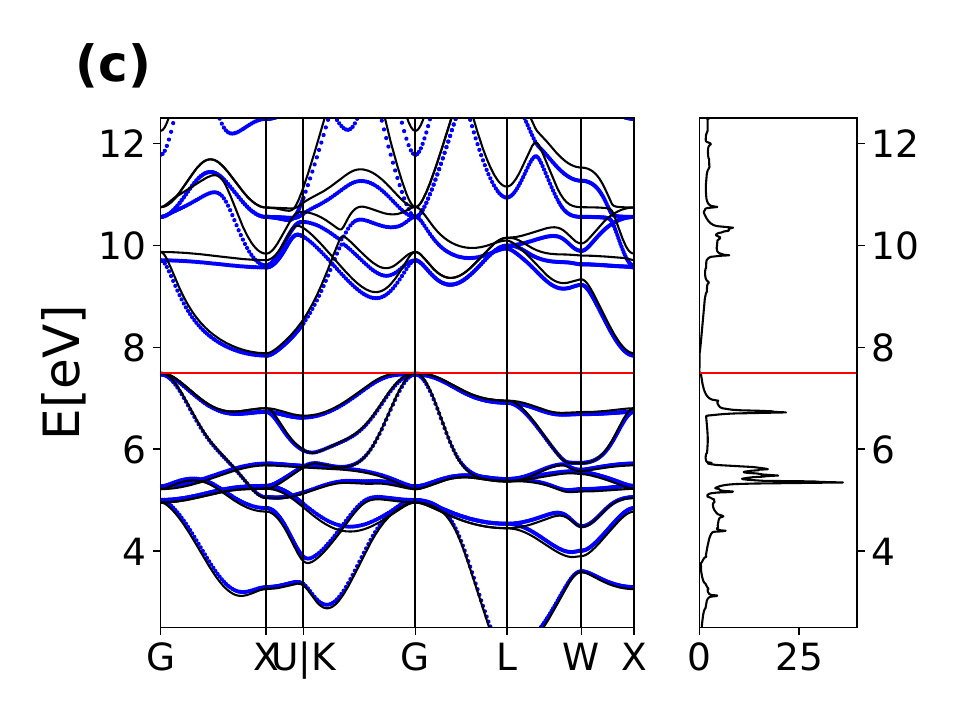}
    \caption{Electron band structures of (a) NiTiSn, (b) NiZrSn and (c) NiHfSn using the PBEsol exchange-correlation functional, in black and PBE in blue.}
\end{figure}
\begin{figure}[H]
    \centering
    \includegraphics[scale=0.36]{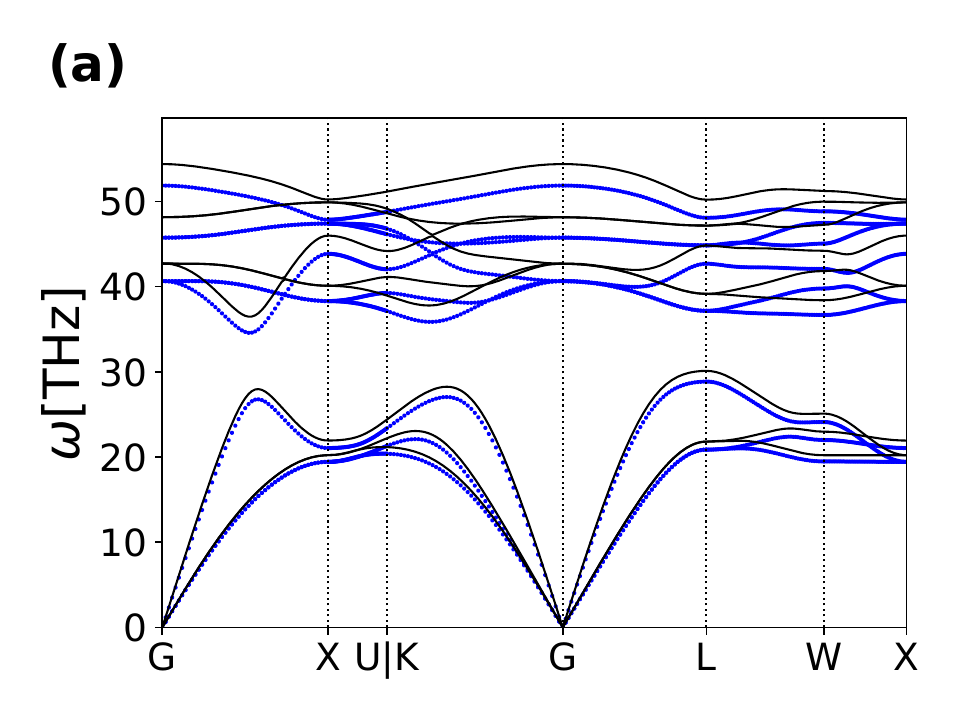}
    \includegraphics[scale=0.36]{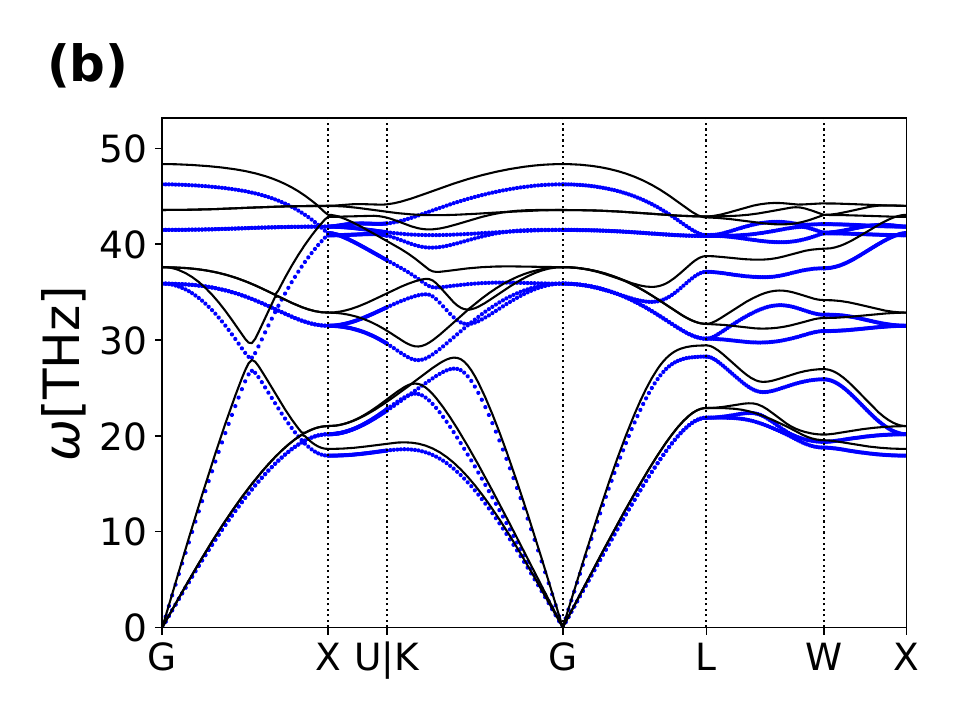}
    \includegraphics[scale=0.36]{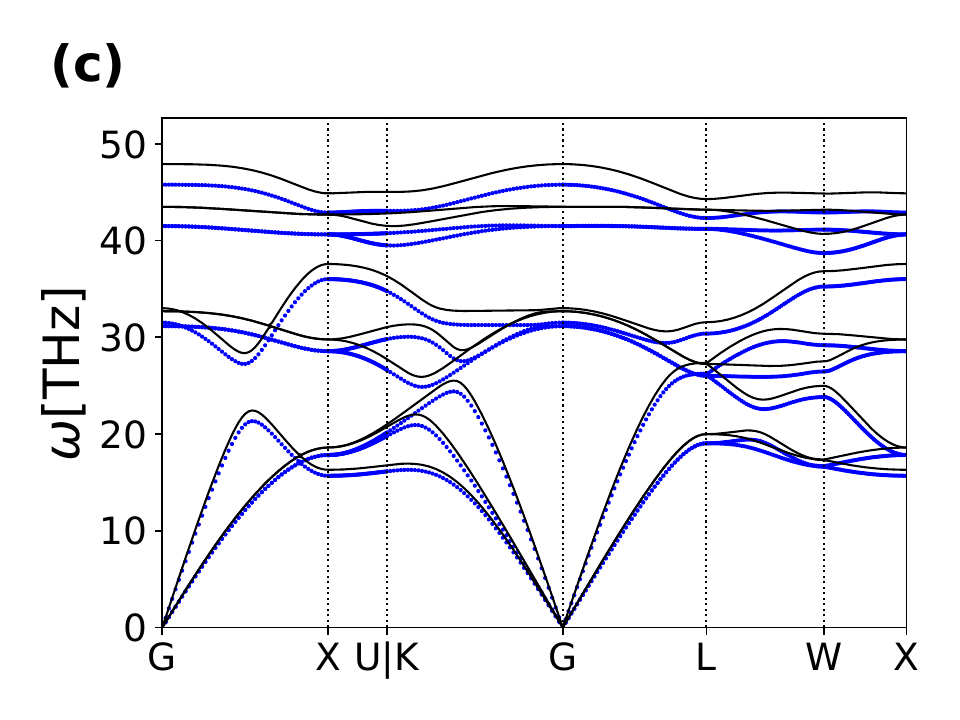}
    \caption{Phonon dispersion of (a) NiTiSn, (b) NiZrSn and (c) NiHfSn using the PBEsol exchange-correlation functional, in black and PBE in blue.}
\end{figure}
\begin{figure}[H]
    \centering
    \includegraphics[width=0.32\linewidth]{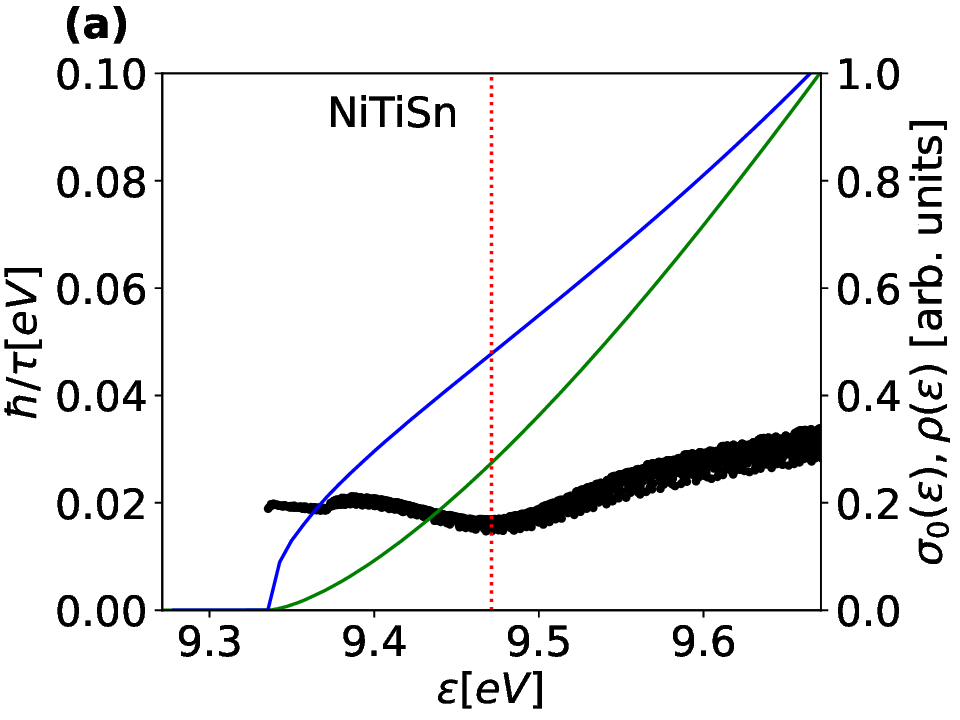}
    \includegraphics[width=0.32\linewidth]{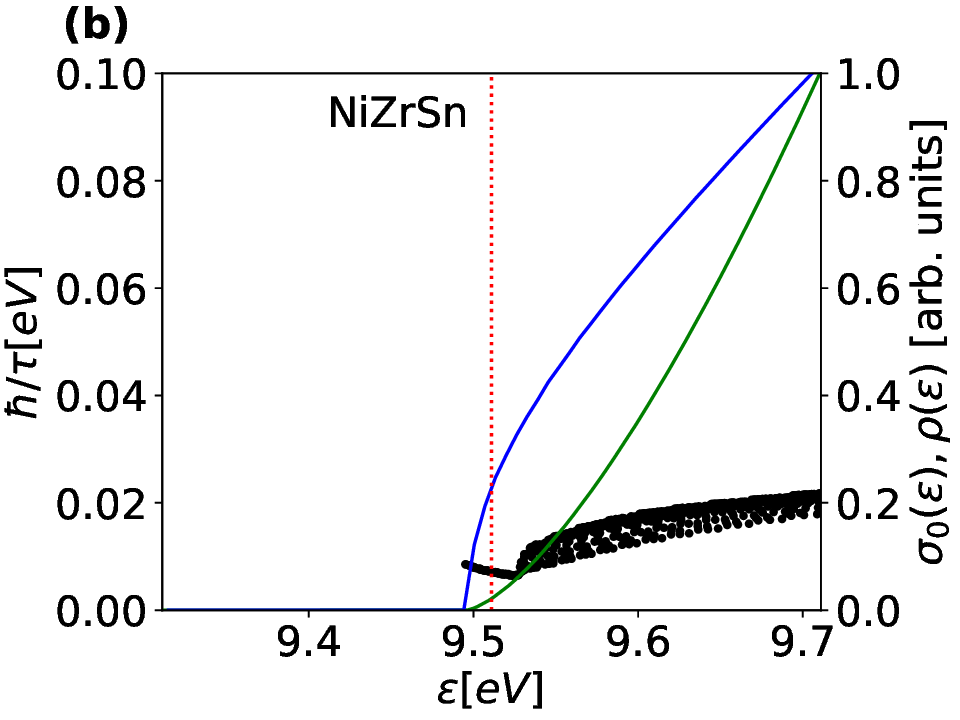}
    \includegraphics[width=0.32\linewidth]{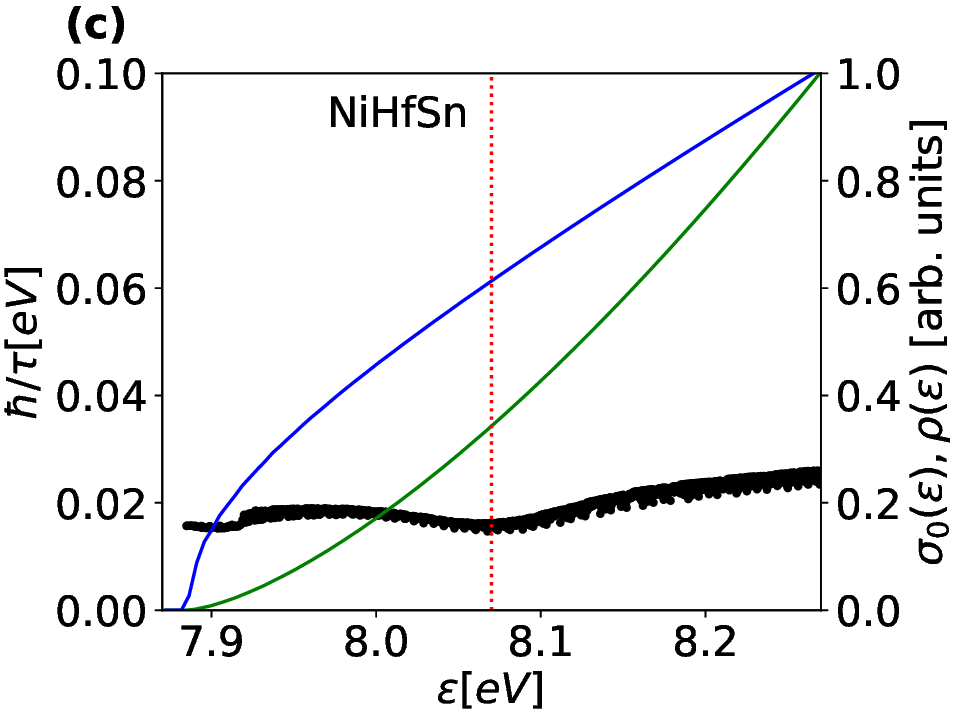}
    \caption{ Reciprocal lifetimes,  $\hbar/\tau$, at 300 K, as function of the band energies, are shown as black dots for (a) NiTiSn, (b) NiZrSn and (c) NiHfSn. The PBEsol exchange correlation functional is used. 
   The transport function at constant relaxation time, $\sigma_0(\epsilon)$ is shown in green, the density of states, $\rho(\epsilon)$, in blue. Arbitrary units are used. The chemical potential $\mu$ is shown using a vertical red dotted line.\label{relax_HH}}
\end{figure}
\begin{figure}[H]
    \centering
    \includegraphics[scale=0.4]{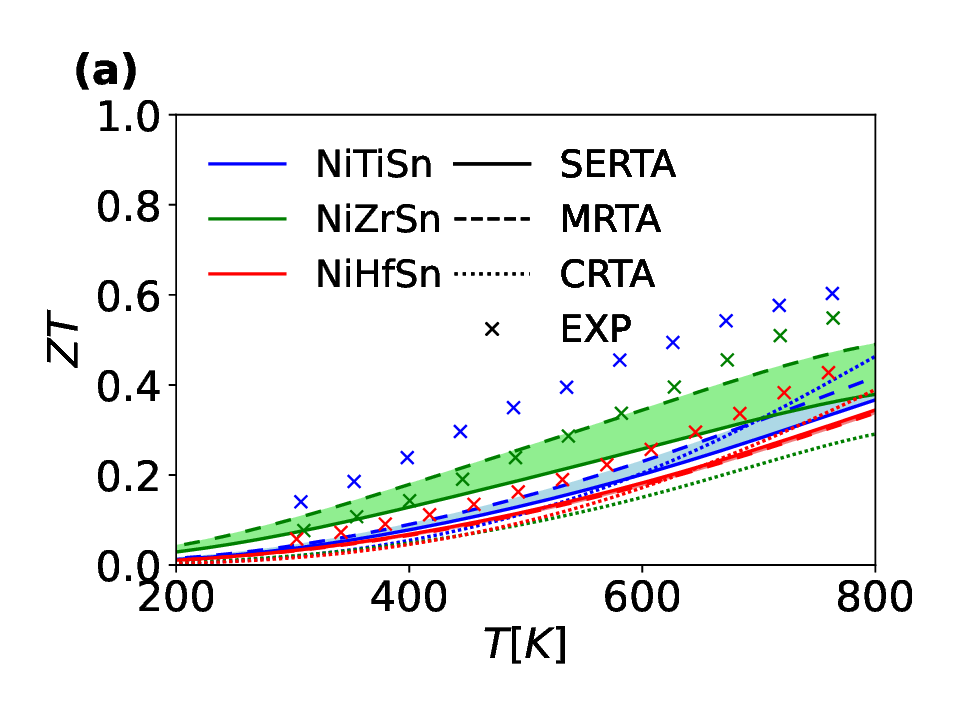}
    \includegraphics[scale=0.4]{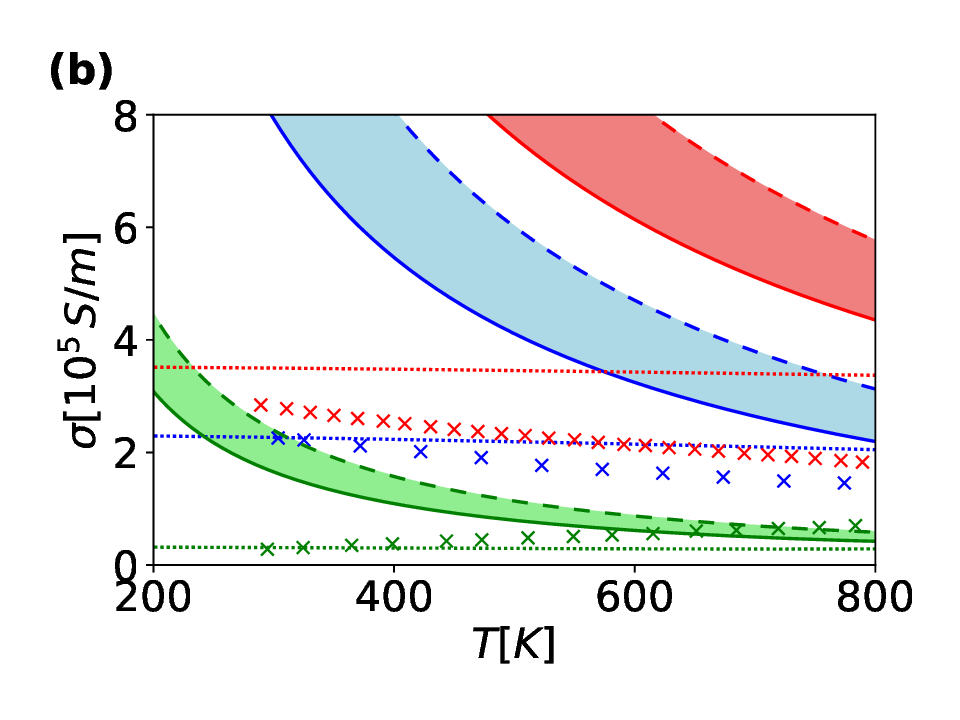}
    \includegraphics[scale=0.4]{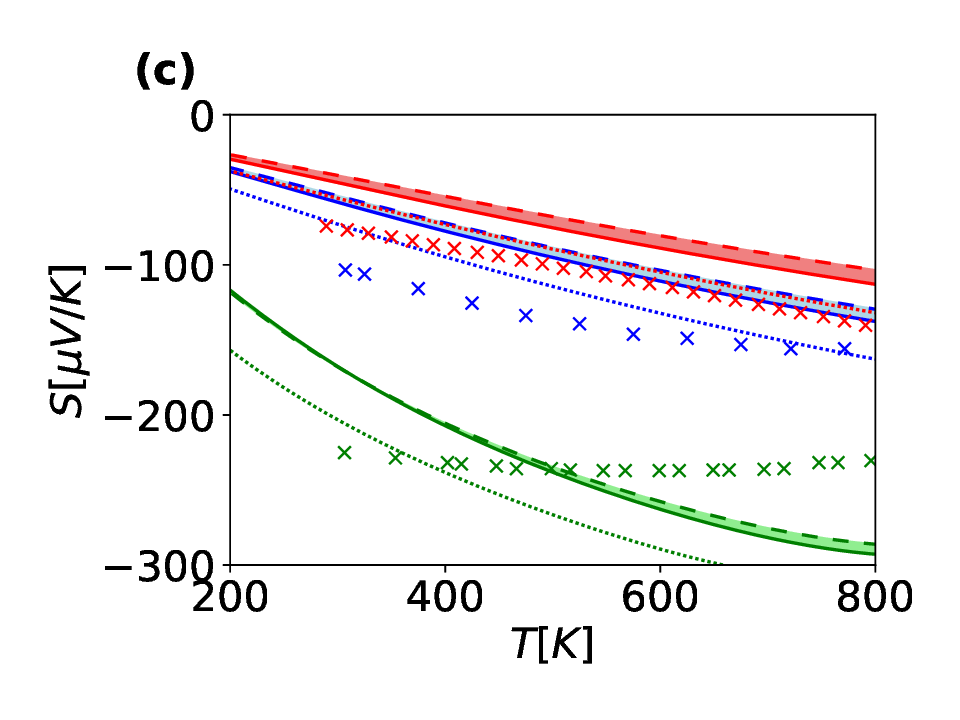}
    \includegraphics[scale=0.4]{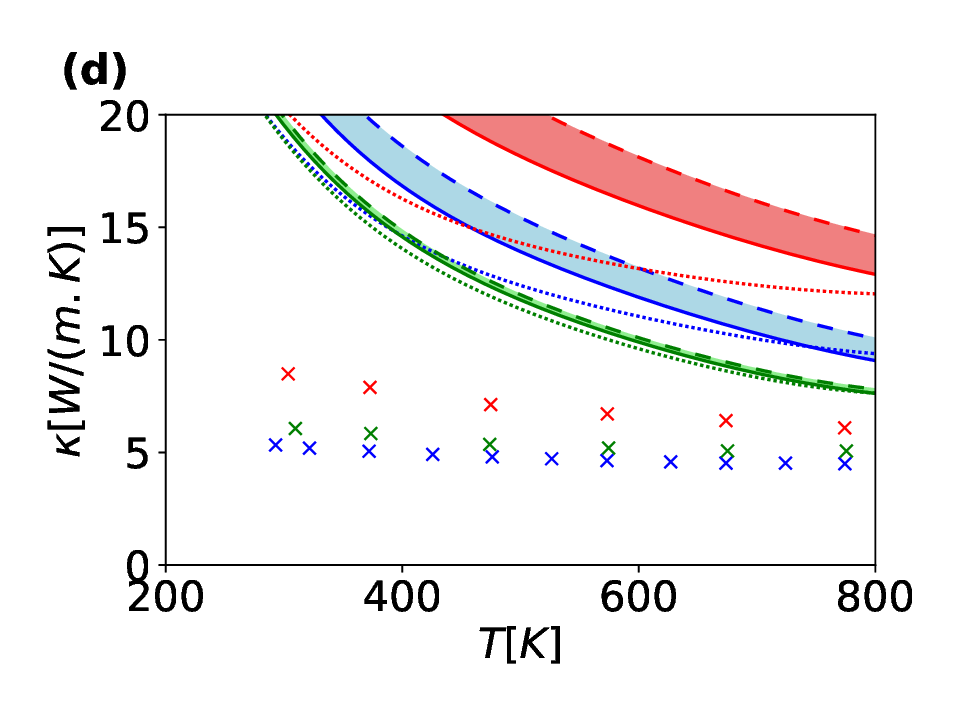}
    \caption{Transport properties of half Heusler compounds computed with PBEsol. (a) ZT thermoelectric figure of merit, (b) electronic conductivity $\sigma$, (c) thermopower $S$, and (d) thermal conductivity $\kappa$. A $k=60$ mesh sampling is used. Experimental measurements\cite{ren, xie_2014, yu2009} are shown using crosses. The results of computations are shown using continuous, dashed and dotted lines for the SERTA, MRTA and CRTA approximations, respectively. For the CRTA, $\tau=10^{-14 }\, \text{s}$ is used.  To compute the lattice thermal conductivity, a sampling of $q=30$, $q=29$ and $q=29$ are used for NiTiSn, NiZrSn and NiHfSn respectively.  For each compound, the differences between SERTA and MRTA approximations are shown using shaded areas.\label{HH_app}}
\end{figure} 

\newpage
\subsection{Mg$_2$Si, Mg$_2$Ge and Mg$_2$Sn}

\begin{figure}[H]
    \centering
    \includegraphics[scale=0.36]{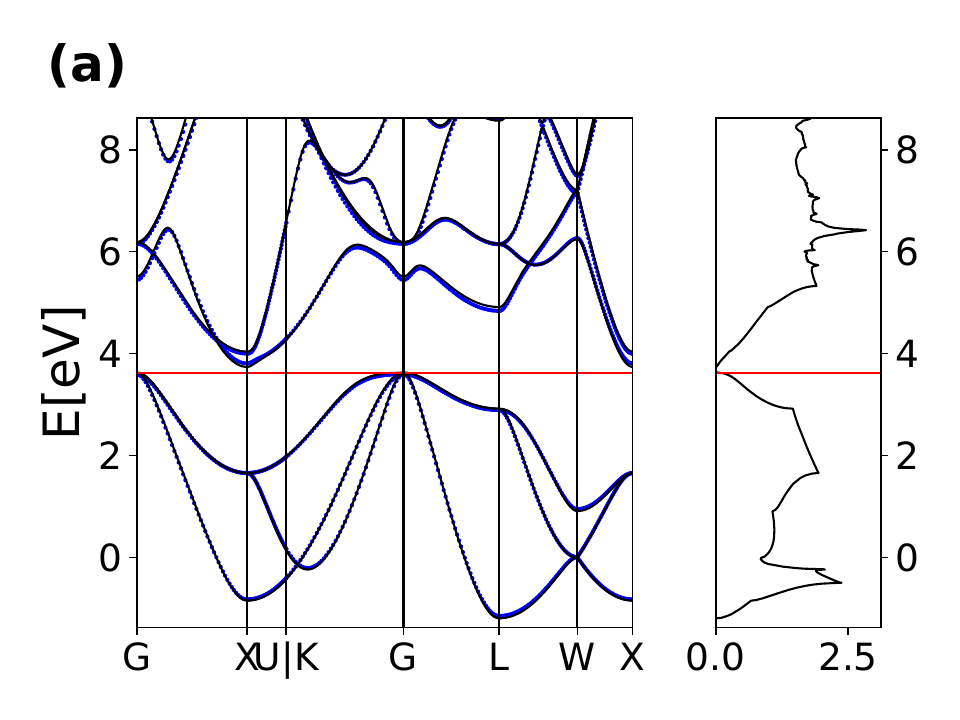}
    \includegraphics[scale=0.36]{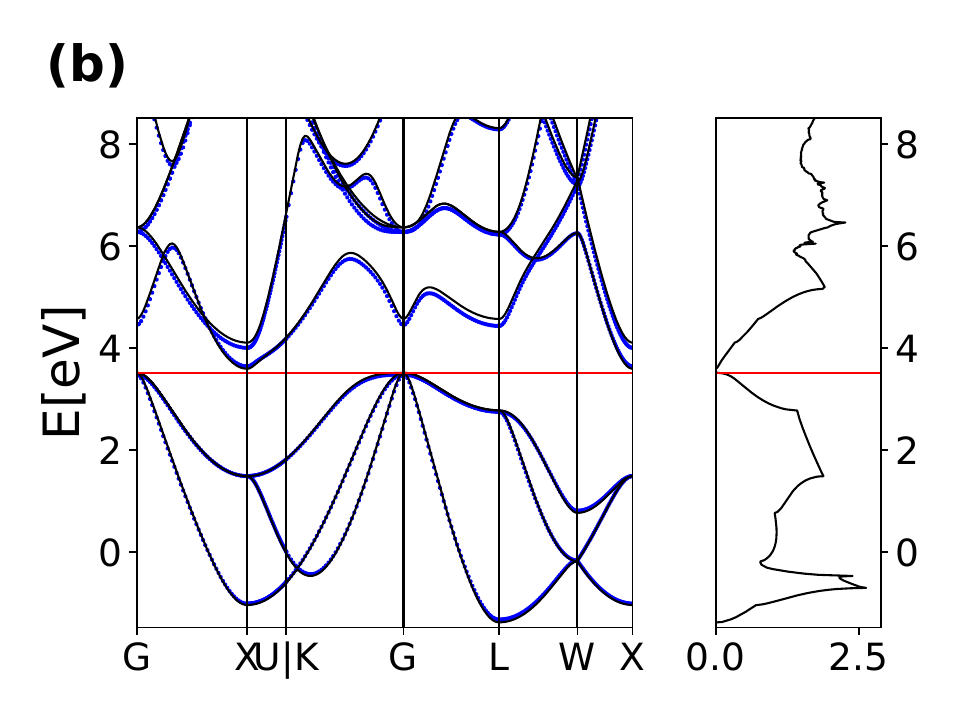}
    \includegraphics[scale=0.36]{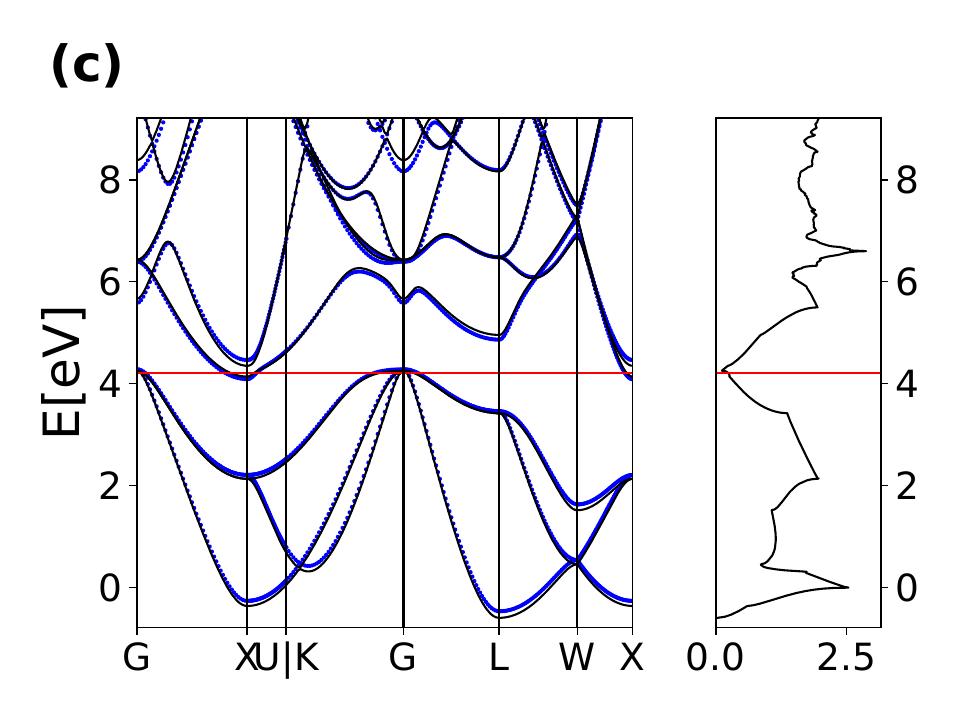}
    \caption{Electron band structures of (a) Mg$_2$Si, (b) Mg$_2$Ge and (c) Mg$_2$Sn using the PBEsol exchange-correlation functional, in black and PBE in blue.}
\end{figure}

\begin{figure}[H]
    \centering
    \includegraphics[scale=0.36]{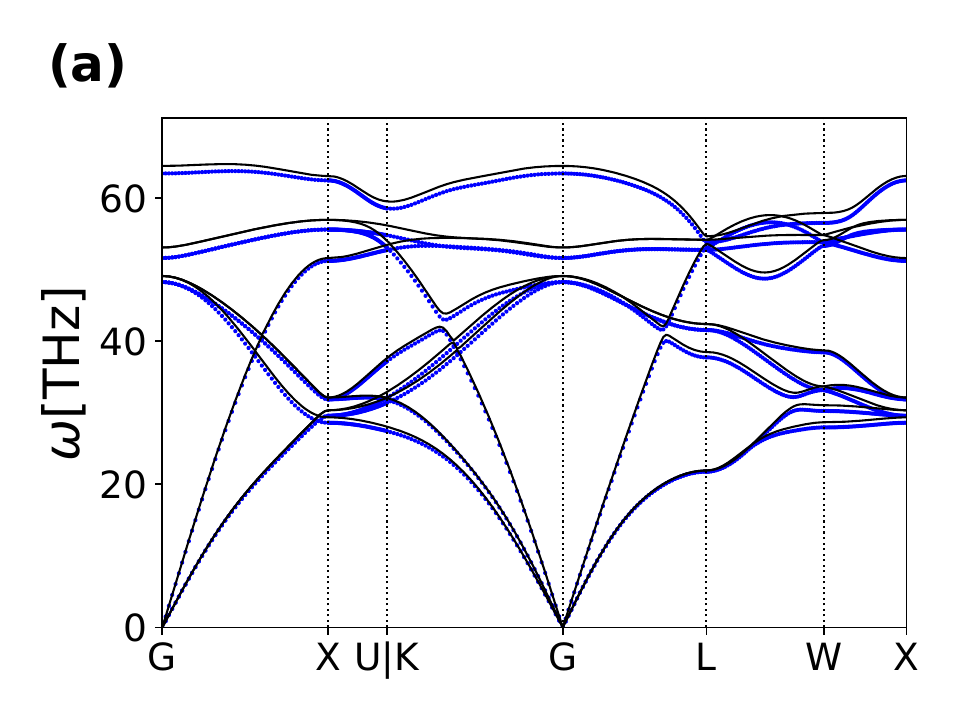}
    \includegraphics[scale=0.36]{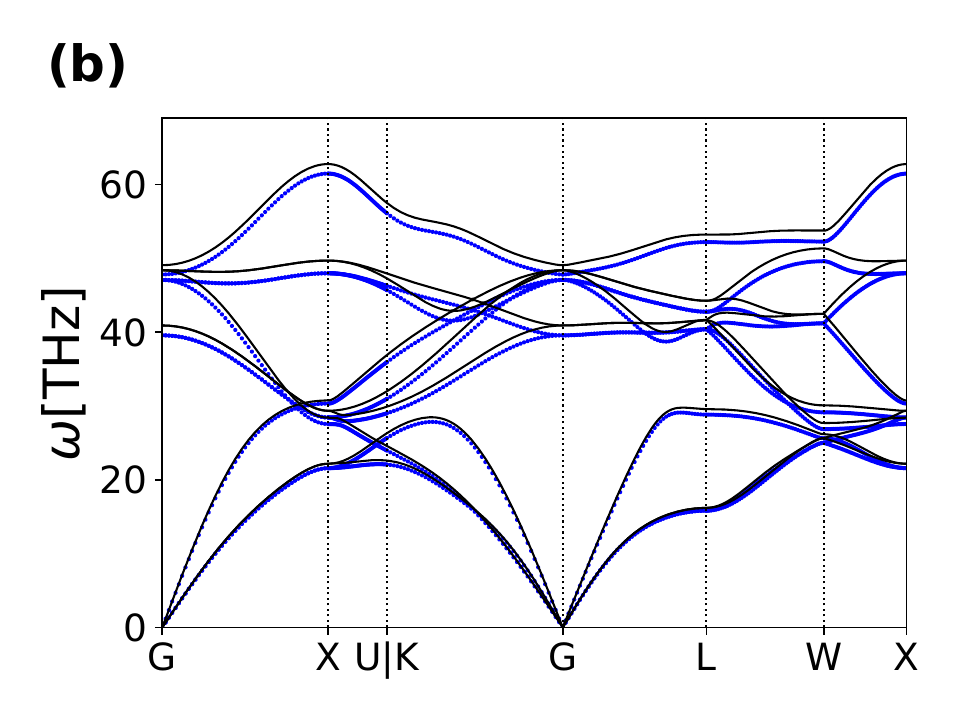}
    \includegraphics[scale=0.36]{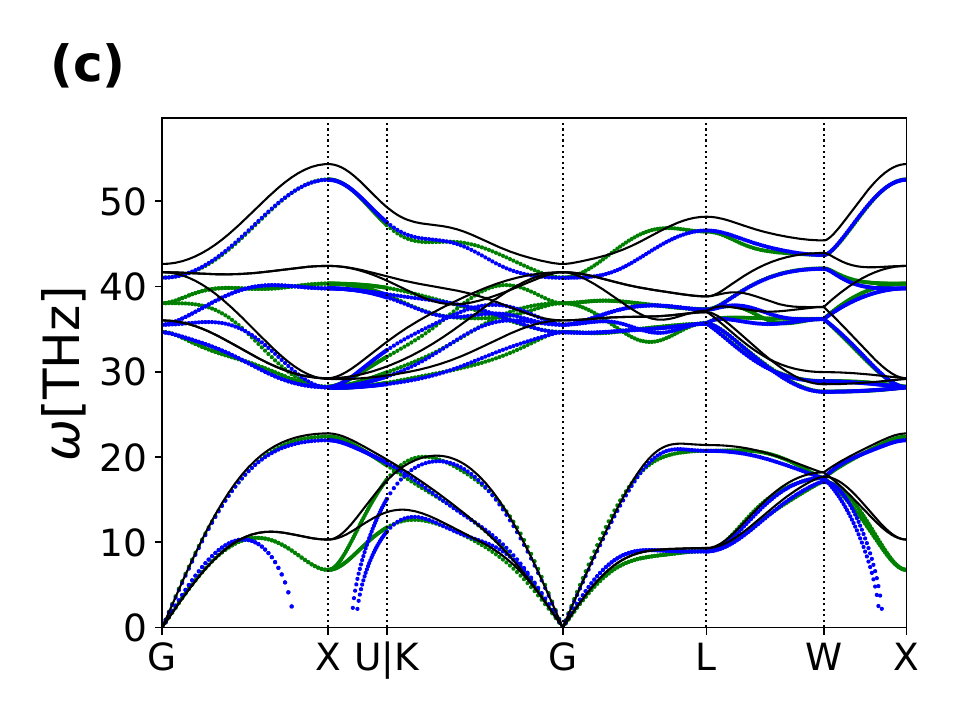}
    \caption{Phonon dispersion of (a) Mg$_2$Si, (b) Mg$_2$Ge and (c) Mg$_2$Sn using the PBEsol exchange-correlation functional in black and PBE in blue. In (c) the PBE dispersion obtained using a $4\times4\times4$ supercell is also shown in green. \label{phonon_Mg2X}}
\end{figure}

\begin{figure}[H]
    \centering
    \includegraphics[width=0.32\linewidth]{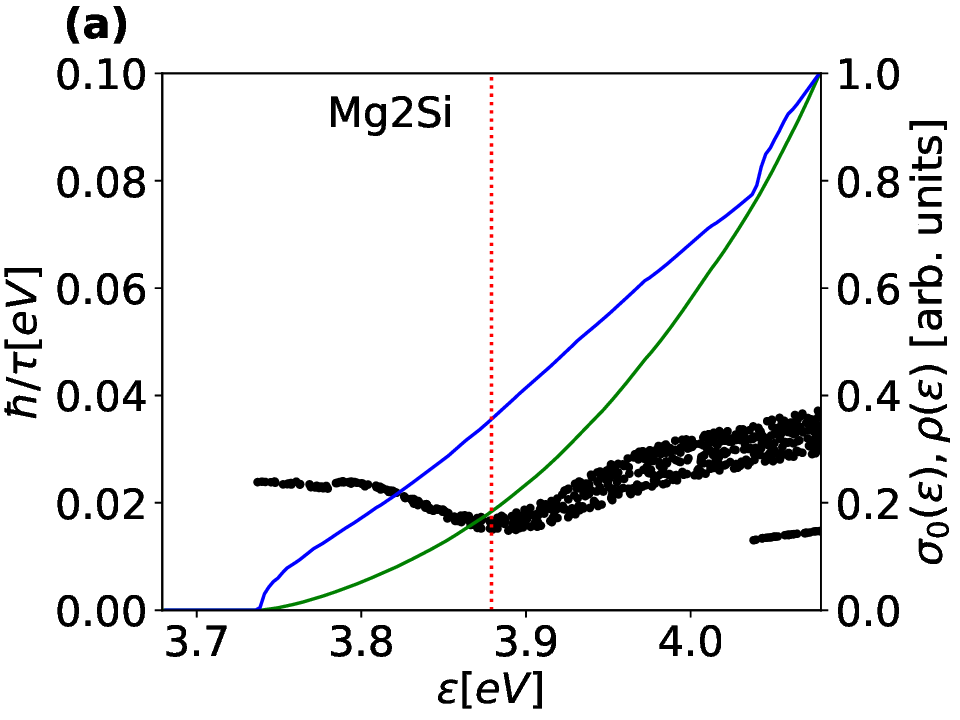}
    \includegraphics[width=0.32\linewidth]{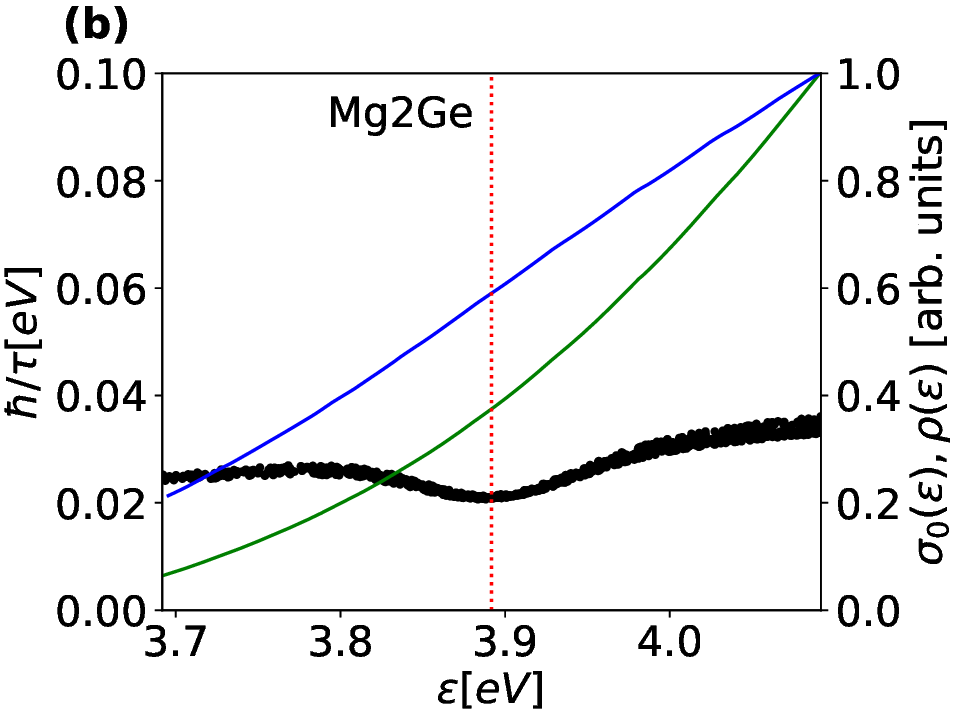}
    \includegraphics[width=0.32\linewidth]{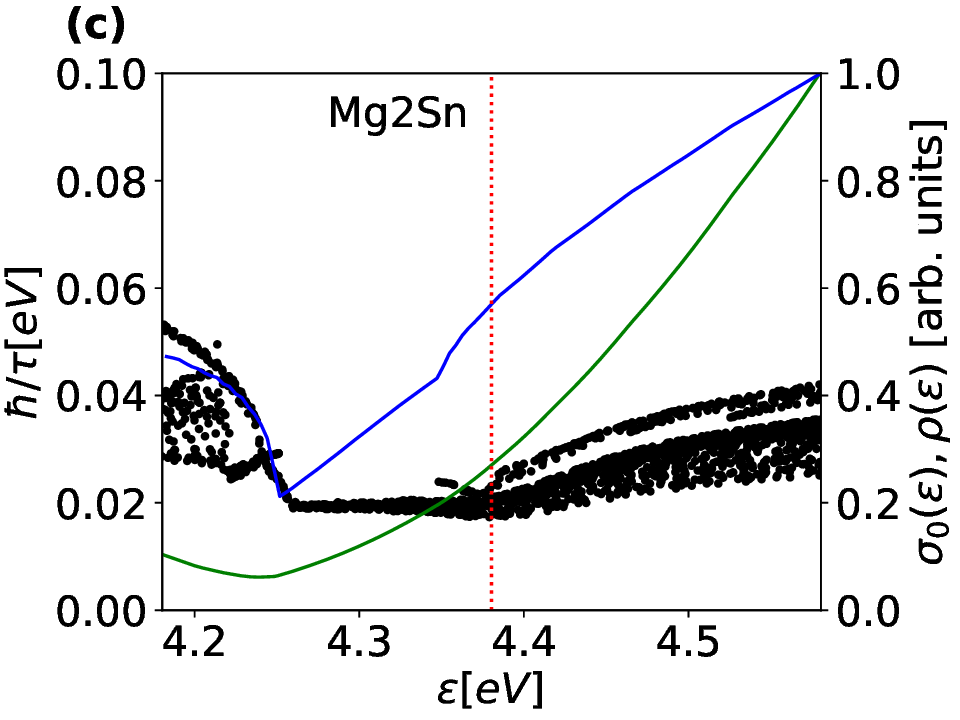}
    \caption{ Reciprocal lifetimes, $\hbar/\tau$, at 300 K, as function of the band energies, are shown as black dots for (a) Mg$_2$Si, (b) Mg$_2$Ge and (c) Mg$_2$Sn . The PBEsol exchange correlation functional is used. 
    The transport function at constant relaxation time, $\sigma_0(\epsilon)$ is shown in green, the density of states, $\rho(\epsilon)$, in blue. Arbitrary units are used. The chemical potential $\mu$ is shown using a vertical red dotted line. \label{relax_Mg2X}}
\end{figure}

\begin{figure}[H]
    \centering
    \includegraphics[scale=0.4]{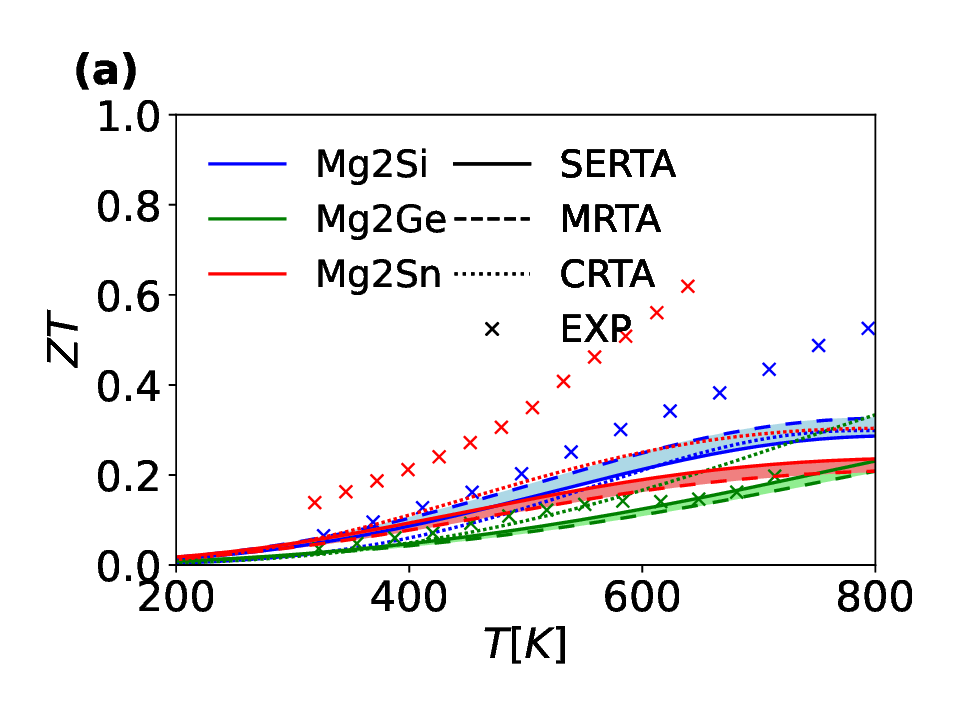}
    \includegraphics[scale=0.4]{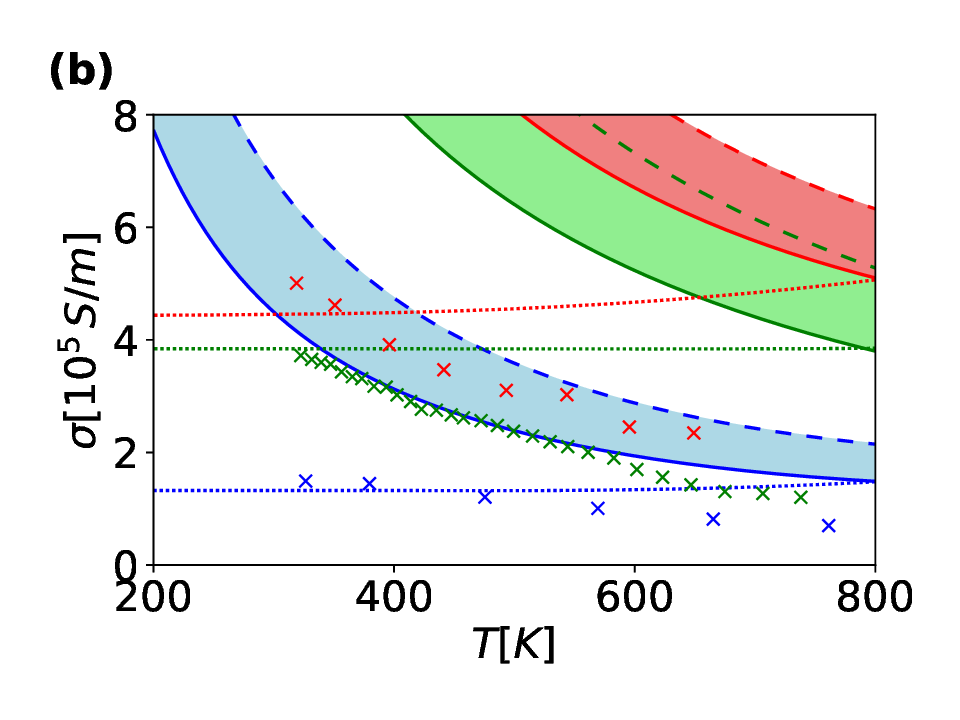}
    \includegraphics[scale=0.4]{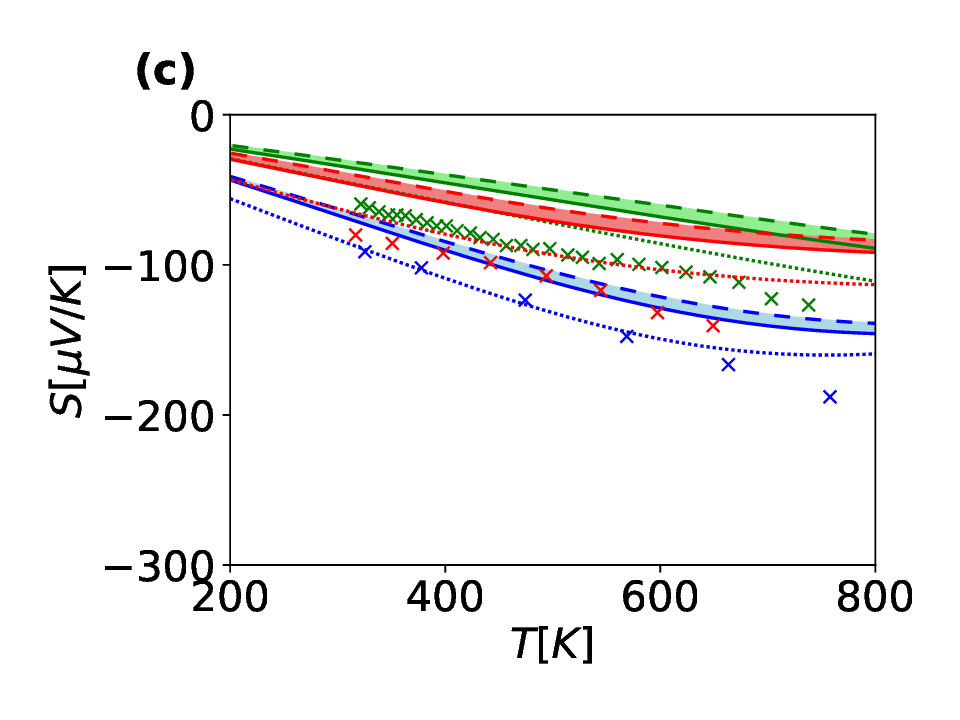}
    \includegraphics[scale=0.4]{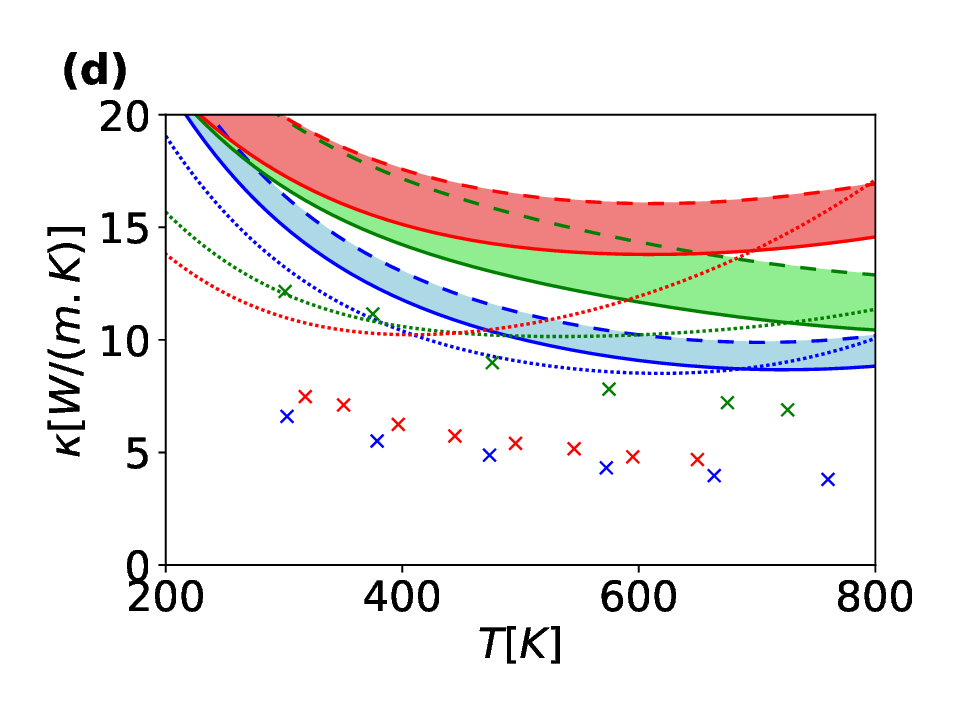}
    \caption{Transport properties of Mg$_2$Si, Mg$_2$Ge and Mg$_2$S compounds computed with PBEsol. (a) ZT thermoelectric figure of merit, (b) electronic conductivity $\sigma$, (c) thermopower $S$, and (d) thermal conductivity $\kappa$. A $k=60$ mesh sampling is used. Experimental measurements\cite{TANI2007, GAO201533, Saito2020} are shown using crosses. The results of computations are shown using continuous, dashed and dotted lines for the SERTA, MRTA and CRTA approximations, respectively.  For the CRTA, $\tau=10^{-14 }\, \text{s}$ is used. To compute the lattice thermal conductivity, a sampling of $q=27$, $q=27$ and $q=26$ are used for  Mg$_2$Si, Mg$_2$Ge and Mg$_2$Sn respectively. For each compound, the differences between SERTA and MRTA approximations are shown using shaded areas.\label{Mg2X_app}}
\end{figure}

\section{ERTA and MRTA \label{XRTA}}

Assuming the phonon system to be at equilibrium, the collision integral, Eq. \ref{col}, can be written
\begin{align}
\frac{\partial f_{\mathbf{k}n} }{\partial t} \Big|_{col}
   &=\sum_{\mathbf{k}'n'} 
    \Big(\frac{(1-f_{\mathbf{k}n})f_{\mathbf{k}'n'}}{(1-f^0_{\mathbf{k}n})f^0_{\mathbf{k}'n'}}-\frac{f_{\mathbf{k}n}(1-f_{\mathbf{k}'n'})}{f^0_{\mathbf{k}n}(1-f^0_{\mathbf{k}'n'})} \Big)(1-f^0_{\mathbf{k}n})f^0_{\mathbf{k}'n'}\nonumber\\
    &\times \frac{2\pi}{\hbar}\sum_{\mathbf{q}j}|g(\mathbf{k}n,\mathbf{k}'n',\mathbf{q}j) |^2\Big(n^0_{\mathbf{q}j}\delta(\epsilon_{\mathbf{k}n}-\epsilon_{\mathbf{k}'n'}-\hbar \omega_{\mathbf{q}j})
    +(1+n^0_{-\mathbf{q}j})\delta(\epsilon_{\mathbf{k}n}-\epsilon_{\mathbf{k}'n'}+\hbar \omega_{-\mathbf{q}j}) \Big)\\
   &\equiv\sum_{\mathbf{k}'n'} 
    \Big(\frac{(1-f_{\mathbf{k}n})f_{\mathbf{k}'n'}}{(1-f^0_{\mathbf{k}n})f^0_{\mathbf{k}'n'}}-\frac{f_{\mathbf{k}n}(1-f_{\mathbf{k}'n'})}{f^0_{\mathbf{k}n}(1-f^0_{\mathbf{k}'n'})} \Big)(1-f^0_{\mathbf{k}n})f^0_{\mathbf{k}'n'} S(\mathbf{k}'n',\mathbf{k}n).
\end{align}
It is linearized as
\begin{align}
\frac{\partial f_{\mathbf{k}n} }{\partial t} \Big|_{lin}
   &=\sum_{\mathbf{k}'n'} 
    \Big(\frac{ 1-f^0_{\mathbf{k}n}}{1-f^0_{\mathbf{k}'n'}}\delta f_{\mathbf{k}'n'}
    -\frac{ f^0_{\mathbf{k}'n'}}{f^0_{\mathbf{k}n}}\delta f_{\mathbf{k}n} \Big) S(\mathbf{k}'n',\mathbf{k}n).
\end{align}

\noindent If a relaxation time is defined by the relation
\begin{align*}
\frac{\partial f_{\mathbf{k}n} }{\partial t} \Big|_{lin}
   =-\frac{\delta f_{\mathbf{k}n}}{\tau_{\mathbf{k}n}},
\end{align*}
then, using  $\Big(-\frac{\partial f^0_{\mathbf{k}n} }{\partial \epsilon_{\mathbf{k}n}}\Big)=\frac{1}{k_B T}f^0_{\mathbf{k}n}(1-f^0_{\mathbf{k}n})$, 
the Boltzmann equation can be rewritten as an equation for $\tau_{\mathbf{k}n}$, 
\begin{align}
\frac{1}{\tau_{\mathbf{k}n}}= \sum_{\mathbf{k}'n'} \frac{f^0_{\mathbf{k}'n'}}{f^0_{\mathbf{k}n}} \Big( 1- \frac{\tau_{\mathbf{k}'n'} \mathbf{G}_{\mathbf{k}'n'}\cdot \mathbf{v}_{\mathbf{k}'n'} }{\tau_{\mathbf{k}n} \mathbf{G}_{\mathbf{k}n}\cdot \mathbf{v}_{\mathbf{k}n} } \Big) S(\mathbf{k}'n',\mathbf{k}n) \label{eq:tau}
\end{align}
where we have defined the vector of applied fields, 
\begin{align}
 \mathbf{G}_{\mathbf{k}n}=(\epsilon_{\mathbf{k}n}-\mu)\frac{\nabla T}{T}+e\bm{\mathcal{E}}.
\end{align}
The above equation means that the relaxation time depends on the applied fields. Two different approximations are obtained when we assume the $\mathbf{G}_{\mathbf{k}n}$ vector to be a pure electric field or a pure temperature gradient. 

We have to consider those two cases because then the dot product in Eq. \ref{eq:tau} can be performed along a vector independent of the wavevector. Therefore in Eq. \ref{eq:tau}, the group velocities of the states $\mathbf{k}n$ and $\mathbf{k}'n'$ are projected in the same direction. Denoting this vector $\mathbf{G}$, this allows to use the law of cosines to write\cite{callaway1974}
\begin{align}
    \cos \Big( \mathbf{G}, \mathbf{v}_{\mathbf{k}'n'} \Big) = \cos \Big( \mathbf{v}_{\mathbf{k}n}, \mathbf{v}_{\mathbf{k}'n'} \Big) \cos \Big( \mathbf{G}, \mathbf{v}_{\mathbf{k}n} \Big) + \sin \Big( \mathbf{v}_{\mathbf{k}n}, \mathbf{v}_{\mathbf{k}'n'} \Big) \sin \Big( \mathbf{G}, \mathbf{v}_{\mathbf{k}n} \Big) \cos \phi'
\end{align}
where $\phi'$ is the angle between the planes $ \mathbf{v}_{\mathbf{k}n}, \mathbf{v}_{\mathbf{k}'n'}$ and $\mathbf{G}, \mathbf{v}_{\mathbf{k}n}$.
If nothing would depend on $\phi'$ in the integrand of Eq. \ref{eq:tau}, $\cos \phi'$ would average to zero in the integration over $\mathbf{k}'$, in which case we could use
\begin{align}
   \frac{ \cos \Big( \mathbf{G}, \mathbf{v}_{\mathbf{k}'n'} \Big)}{\cos \Big( \mathbf{G}, \mathbf{v}_{\mathbf{k}n}\Big)} = \cos \Big( \mathbf{v}_{\mathbf{k}n}, \mathbf{v}_{\mathbf{k}'n'} \Big)  = \frac{\mathbf{v}_{\mathbf{k}n}\cdot \mathbf{v}_{\mathbf{k}'n'}}{|\mathbf{v}_{\mathbf{k}n}|| \mathbf{v}_{\mathbf{k}'n'}|} 
\end{align}
and 
\begin{align}
\frac{1}{\tau_{\mathbf{k}n}}= \sum_{\mathbf{k}'n'} \frac{f^0_{\mathbf{k}'n'}}{f^0_{\mathbf{k}n}} \Big( 1- \frac{\tau_{\mathbf{k}'n'}|\mathbf{v}_{\mathbf{k}'n'}|   }{\tau_{\mathbf{k}n} |\mathbf{v}_{\mathbf{k}n}|  }
\frac{\mathbf{G}_{\mathbf{k}'n'}\cdot \mathbf{G}}{\mathbf{G}_{\mathbf{k}n}\cdot \mathbf{G}}
\frac{\mathbf{v}_{\mathbf{k}n}\cdot \mathbf{v}_{\mathbf{k}'n'}}{|\mathbf{v}_{\mathbf{k}n}|| \mathbf{v}_{\mathbf{k}'n'}|} \Big) S(\mathbf{k}'n',\mathbf{k}n).
\end{align}

To eliminate the relaxation time from the right hand side, one further assumption has to be made. Often in the literature\cite{gunst_2016} it is assumed that $\tau_{\mathbf{k}'n'}\approx \tau_{\mathbf{k}n}$ and we get
\begin{align}
\frac{1}{\tau_{\mathbf{k}n}}= \sum_{\mathbf{k}'n'} \frac{f^0_{\mathbf{k}'n'}}{f^0_{\mathbf{k}n}} \Big( 1- 
\frac{\mathbf{G}_{\mathbf{k}'n'}\cdot \mathbf{G}}{\mathbf{G}_{\mathbf{k}n}\cdot \mathbf{G}}
\frac{\mathbf{v}_{\mathbf{k}n}\cdot \mathbf{v}_{\mathbf{k}'n'}}{|\mathbf{v}_{\mathbf{k}n}|^2} \Big) S(\mathbf{k}'n',\mathbf{k}n). 
\end{align}
However, with this approximation the relaxation could become negative\cite{gunst_2016}. Therefore, instead of assuming  the relaxation to be a weak function of $\mathbf{k}$, we assume that the mean free path is such a weak function of $\mathbf{k}$,
$\tau_{\mathbf{k}'n'}|\mathbf{v}_{\mathbf{k}'n'}|\approx \tau_{\mathbf{k}n}|\mathbf{v}_{\mathbf{k}n}|$. Then we get
\begin{align}
\frac{1}{\tau_{\mathbf{k}n}}= \sum_{\mathbf{k}'n'} \frac{f^0_{\mathbf{k}'n'}}{f^0_{\mathbf{k}n}} \Big( 1- 
\frac{\mathbf{G}_{\mathbf{k}'n'}\cdot \mathbf{G}}{\mathbf{G}_{\mathbf{k}n}\cdot \mathbf{G}}
\frac{\mathbf{v}_{\mathbf{k}n}\cdot \mathbf{v}_{\mathbf{k}'n'}}{|\mathbf{v}_{\mathbf{k}n}|| \mathbf{v}_{\mathbf{k}'n'}|} \Big) S(\mathbf{k}'n',\mathbf{k}n). 
\end{align}

For pure electric field, $\mathbf{G}_{\mathbf{k}n}=e\bm{\mathcal{E}}$, and $\mathbf{G}=\bm{\mathcal{E}}$, this gives the momentum relaxation time approximation (MRTA)
\begin{align}
\frac{1}{\tau^\text{MRTA}_{\mathbf{k}n}}&= \sum_{\mathbf{k}'n'} \frac{f^0_{\mathbf{k}'n'}}{f^0_{\mathbf{k}n}} \Big( 1- 
\frac{\mathbf{v}_{\mathbf{k}n}\cdot \mathbf{v}_{\mathbf{k}'n'}}{|\mathbf{v}_{\mathbf{k}n}|| \mathbf{v}_{\mathbf{k}'n'}|} \Big) S(\mathbf{k}'n',\mathbf{k}n) \\
&=\frac{2\pi}{\hbar}\sum_{\mathbf{k}'n'} \sum_{\mathbf{q}j}|g(\mathbf{k}n,\mathbf{k}'n',\mathbf{q}j) |^2 \frac{f^0_{\mathbf{k}'n'}}{f^0_{\mathbf{k}n}}  \Big(n^0_{\mathbf{q}j}\delta(\epsilon_{\mathbf{k}n}-\epsilon_{\mathbf{k}'n'}-\hbar \omega_{\mathbf{q}j})
    +(1+n^0_{-\mathbf{q}j})\delta(\epsilon_{\mathbf{k}n}-\epsilon_{\mathbf{k}'n'}+\hbar \omega_{-\mathbf{q}j}) \Big)\nonumber\\
&\times\Big( 1- 
\frac{\mathbf{v}_{\mathbf{k}n}\cdot \mathbf{v}_{\mathbf{k}'n'}}{|\mathbf{v}_{\mathbf{k}n}|| \mathbf{v}_{\mathbf{k}'n'}|} \Big)\\
&=\frac{2\pi}{\hbar}\sum_{\mathbf{k}'n'} \sum_{\mathbf{q}j}|g(\mathbf{k}n,\mathbf{k}'n',\mathbf{q}j) |^2   \Big((1+n^0_{\mathbf{q}j}-f^0_{\mathbf{k}'n'})\delta(\epsilon_{\mathbf{k}n}-\epsilon_{\mathbf{k}'n'}-\hbar \omega_{\mathbf{q}j})
    +(f^0_{\mathbf{k}'n'}+n^0_{-\mathbf{q}j})\delta(\epsilon_{\mathbf{k}n}-\epsilon_{\mathbf{k}'n'}+\hbar \omega_{-\mathbf{q}j}) \Big)\nonumber\\
&\times\Big( 1- 
\frac{\mathbf{v}_{\mathbf{k}n}\cdot \mathbf{v}_{\mathbf{k}'n'}}{|\mathbf{v}_{\mathbf{k}n}|| \mathbf{v}_{\mathbf{k}'n'}|} \Big)
\end{align}
Therefore the computation of the momentum relaxation time is the same than for the SERTA approximation, except for the last factor which gives more weight to back scattering events.

For pure thermal gradient, $\mathbf{G}_{\mathbf{k}n}=(\epsilon_{\mathbf{k}n}-\mu)\frac{\nabla T}{T}$,  and $\mathbf{G}=\nabla T$, this gives the energy relaxation time approximation (ERTA),
\begin{align}
\frac{1}{\tau^\text{ERTA}_{\mathbf{k}n}}&= \sum_{\mathbf{k}'n'} \frac{f^0_{\mathbf{k}'n'}}{f^0_{\mathbf{k}n}} \Big( 1- 
\frac{\epsilon_{\mathbf{k}'n'}-\mu}{\epsilon_{\mathbf{k}n}-\mu}
\frac{\mathbf{v}_{\mathbf{k}n}\cdot \mathbf{v}_{\mathbf{k}'n'}}{|\mathbf{v}_{\mathbf{k}n}|| \mathbf{v}_{\mathbf{k}'n'}|} \Big) S(\mathbf{k}'n',\mathbf{k}n)\\
&=\frac{2\pi}{\hbar}\sum_{\mathbf{k}'n'} \sum_{\mathbf{q}j}|g(\mathbf{k}n,\mathbf{k}'n',\mathbf{q}j) |^2   \Big((1+n^0_{\mathbf{q}j}-f^0_{\mathbf{k}'n'})\delta(\epsilon_{\mathbf{k}n}-\epsilon_{\mathbf{k}'n'}-\hbar \omega_{\mathbf{q}j})
    +(f^0_{\mathbf{k}'n'}+n^0_{-\mathbf{q}j})\delta(\epsilon_{\mathbf{k}n}-\epsilon_{\mathbf{k}'n'}+\hbar \omega_{-\mathbf{q}j}) \Big)\nonumber\\
    &\times\Big( 1- 
\frac{\epsilon_{\mathbf{k}'n'}-\mu}{\epsilon_{\mathbf{k}n}-\mu}
\frac{\mathbf{v}_{\mathbf{k}n}\cdot \mathbf{v}_{\mathbf{k}'n'}}{|\mathbf{v}_{\mathbf{k}n}|| \mathbf{v}_{\mathbf{k}'n'}|} \Big). 
\end{align}

\section{Gauss-Legendre integration \label{GL}}
In the integral Eq. \ref{eq:Lij}, we can make the substitution $x=1-2f^0$. $x(\epsilon)$ is a monotonic function, going from $-1$ to $1$ as $\epsilon$ goes from $-\infty$ to $\infty$. Moreover, $x(\mu)=0$. Using $dx= 2\frac{\partial f}{\partial \mu}d\epsilon$ and $\epsilon=\mu+ k_BT \ln \frac{1+x}{1-x}$, we obtain
\begin{align}
    \mathcal{L}_{ij}=\frac{1}{2}\Big( \frac{k_B T}{-e}\Big)^{i+j-2} \int_{-1}^{+1} d x \,  \Big(\ln \frac{1+x}{1-x}\Big)^{i+j-2} \sigma(\mu+ k_BT \ln \frac{1+x}{1-x}). \label{Lij_x}
\end{align}
The transport function $\sigma(\epsilon)$ is assumed to be zero outside the interval $(\epsilon_{\text{min}},\epsilon_{\text{max}})$, therefore the integrand of the above interval is assumed to be zero outside $(x(\epsilon_{\text{min}}),x(\epsilon_{\text{max}}))$. This way $x$ is never reaching $-1$ or $+1$ and therefore $\ln \frac{1+x}{1-x}$ is never becoming infinite. Under those conditions, the above integral can be approximated using the Gauss-Legendre quadrature
\begin{align}
    \mathcal{L}_{ij}\approx\frac{1}{2}\Big( \frac{k_B T}{-e}\Big)^{i+j-2} \sum_{k=1}^N\  \Big(\ln \frac{1+x_k}{1-x_k}\Big)^{i+j-2} \sigma(\mu+ k_BT \ln \frac{1+x_k}{1-x_k}) w_k \label{Lij_GL},
\end{align}
where $x_k$ are the roots of the Legendre polynomial of order $N$, $P_{N}(x)$, and the weight $w_k$ are 
\begin{align}
    w_k=\frac{2(1-x_k^2)}{[(N+1) P_{N+1}(x_k)]^2}, 
\end{align}
with $P_0(x)=1$, $P_{1}(x)=x$, $P_{2}(x)=1/2(3x^2-1)$, and $P_n(x)=(2n-1)/n x P_{n-1}(x)-(n-2)/n P_{n-2}(x)$ for $n>2$.

This integration method is valuable for the calculation of transport properties. Indeed, the integrand is approximated by an Hermite interpolating polynomial. It means that the interpolating polynomial matches the values of the integrand at the points $x_k$, but also matches the values of its derivative. This provides an increased smoothness for the interpolation. Finally, the Gauss-Legendre method is obtained by choosing the $x_k$ such that the part of the interpolation involving the derivatives of the function to be integrated does not contribute to the integral. This choice ensures that the quadrature can take the form $\sum_k f(x_k)w_k$ rather than 
 $\sum_k f(x_k)w_k+f'(x_k)u_k$, which would be inconvenient in our case since the derivative of the transport function is not known. Finally, the $x_k$ being the roots of the Legendre polynomials guarantees they are symmetric around $x=0$ (the chemical potential), and that the weights are maximum around that point.

\section{Electron's velocities\label{velocity}}
The expectation value of the velocity operator is obtained from the derivative of the band structure, 
\begin{align}
    \mathbf{v}_{\mathbf{k}n}=\langle \varphi_{\mathbf{k}n} | \mathbf{v} | \varphi_{\mathbf{k}n} \rangle = \langle \varphi_{\mathbf{k}n} |\frac{i}{\hbar}[H,\mathbf{r}]| \varphi_{\mathbf{k}n} \rangle  = \frac{1}{\hbar}\frac{\partial \epsilon_{\mathbf{k}n}}{\partial \mathbf{k}}.
\end{align}
This relation is valid for a general one particle Hamiltonian $H$, even if including spin orbit coupling or a Hubbard U correction.

We would like to perform the above derivative with respect to $\mathbf{k}$ within the framework of the PAW method\cite{Gajdos-2006}. To do this, we proceed along the line which is used in the Heisenberg representation to take time derivative. We define $\mathbf{k}$ dependent operators as
\begin{align}
    A_{\mathbf{k}}= e^{-i \mathbf{k}\cdot \mathbf{r}} A e^{i \mathbf{k}\cdot \mathbf{r}}
\end{align}
and obtain their derivative, 
\begin{align}
    \frac{\partial A_{\mathbf{k}}}{\partial \mathbf{k}}= i[ A_{\mathbf{k}},\mathbf{r}]+  e^{-i \mathbf{k}\cdot \mathbf{r}}  \frac{\partial A}{\partial \mathbf{k}} e^{i \mathbf{k}\cdot \mathbf{r}}. \label{V3}
\end{align}
The second term is zero if the operator does not have an intrinsic  $\mathbf{k}$ dependence.

In the PAW method, we have the equations
\begin{align}
    &\tilde{h}|\tilde{\varphi}_{\mathbf{k}n}\rangle = \epsilon_{\mathbf{k}n}\tilde{o}|\tilde{\varphi}_{\mathbf{k}n}\rangle \\
    &\langle \tilde{\varphi}_{\mathbf{k}n} |\tilde{o}|\tilde{\varphi}_{\mathbf{k}n'}\rangle= \delta_{nn'} \\
    &\tilde{h}=-\frac{\hbar^2}{2m}\nabla^2+v_l+\sum_{l \tau}\sum_{LL'}|\tilde{p}_{L}(l \tau)\rangle D_{LL'}(l \tau)\langle \tilde{p}_{L'}(l \tau)|\\
    &\tilde{o}=1+\sum_{l \tau}\sum_{LL'}|\tilde{p}_{L}(l \tau)\rangle Q_{LL'}(l \tau)\langle \tilde{p}_{L'}(l \tau)|
\end{align}
where $l$ and $\tau$ label the lattice vectors and the atoms in a primitive unit cell, and $L$ and $L'$ quantum numbers used to label atomic orbitals. 

Using the periodic part of the Bloch wavefunctions, $\tilde{u}_{\mathbf{k}n}$ , 
\begin{align}
   \tilde{\varphi}_{\mathbf{k}n}(\mathbf{r})= e^{i \mathbf{k}\cdot \mathbf{r}} \tilde{u}_{\mathbf{k}n}(\mathbf{r}),
\end{align}
the above equations become 
\begin{align}
    &\tilde{h}_{\mathbf{k}}|\tilde{u}_{\mathbf{k}n}\rangle = \epsilon_{\mathbf{k}n}\tilde{o}_{\mathbf{k}}|\tilde{u}_{\mathbf{k}n}\rangle  \label{V1}\\
    &\langle \tilde{u}_{\mathbf{k}n} |\tilde{o}_{\mathbf{k}}|\tilde{u}_{\mathbf{k}n'}\rangle= \delta_{nn'} \label{V2} \\
    &\tilde{h}_{\mathbf{k}}= e^{-i \mathbf{k}\cdot \mathbf{r}} \tilde{h} e^{i \mathbf{k}\cdot \mathbf{r}}=-\frac{\hbar^2}{2m}(\nabla+i\mathbf{k})^2+v_l+\sum_{l \tau}\sum_{LL'} e^{-i \mathbf{k}\cdot \mathbf{r}}|\tilde{p}_{L}(l \tau)\rangle D_{LL'}(l \tau)\langle \tilde{p}_{L'}(l \tau)|e^{i \mathbf{k}\cdot \mathbf{r}}  \\
    &\tilde{o}_{\mathbf{k}}= e^{-i \mathbf{k}\cdot \mathbf{r}} \tilde{o} e^{i \mathbf{k}\cdot \mathbf{r}} = 1+\sum_{l \tau}\sum_{LL'} e^{-i \mathbf{k}\cdot \mathbf{r}} |\tilde{p}_{L}(l \tau)\rangle Q_{LL'}(l \tau)\langle \tilde{p}_{L'}(l \tau)|e^{i \mathbf{k}\cdot \mathbf{r}} 
\end{align}

Eq. \ref{V1} gives
\begin{align}
    \langle \tilde{u}_{\mathbf{k}n} |  \tilde{h}_{\mathbf{k}}|\tilde{u}_{\mathbf{k}n}\rangle = \epsilon_{\mathbf{k}n} \langle \tilde{u}_{\mathbf{k}n} |   \tilde{o}_{\mathbf{k}}|\tilde{u}_{\mathbf{k}n}\rangle. 
\end{align}
As a standard result of perturbation theory, the first order change of the eigenvalues can be obtained differentiating the above equation, and using Eqs. \ref{V1} and \ref{V2}. We obtain
\begin{align}
\frac{\partial \epsilon_{\mathbf{k}n}}{\partial \mathbf{k}} & =\langle \tilde{u}_{\mathbf{k}n} |  \frac{\partial \tilde{h}_{\mathbf{k}}}{\partial \mathbf{k}}-\epsilon_{\mathbf{k}n}\frac{\partial \tilde{o}_{\mathbf{k}}}{\partial \mathbf{k}}|\tilde{u}_{\mathbf{k}n}\rangle.
\end{align}


The derivatives with respect to $\mathbf{k}$ in the above equation are now computed from the commutator with the position operator, as in Eq. \ref{V3}, assuming the Hamiltonian and overlap operators does not have an intrinsic $\mathbf{k}$ dependence.
\begin{align}
\frac{\partial \epsilon_{\mathbf{k}n}}{\partial \mathbf{k}} &=i\langle \tilde{u}_{\mathbf{k}n} |  [\tilde{h}_{\mathbf{k}},\mathbf{r}]-\epsilon_{\mathbf{k}n}[\tilde{o}_{\mathbf{k}},\mathbf{r}]|\tilde{u}_{\mathbf{k}n}\rangle \\
&=\langle \tilde{u}_{\mathbf{k}n} |\frac{\hbar}{m}(-i\hbar \nabla+\hbar \mathbf{k})|\tilde{u}_{\mathbf{k}n}\rangle \nonumber\\
&+i\sum_{l \tau}\sum_{LL'}\langle \tilde{u}_{\mathbf{k}n} |\Big[ e^{-i \mathbf{k}\cdot \mathbf{r}}|\tilde{p}_{L}(l \tau)\rangle (D_{LL'}(l \tau)-\epsilon_{\mathbf{k}n}Q_{LL'}(l \tau))\langle \tilde{p}_{L'}(l \tau)|e^{i \mathbf{k}\cdot \mathbf{r}}  , \mathbf{r}\Big]|\tilde{u}_{\mathbf{k}n}\rangle \\
&=\langle \tilde{u}_{\mathbf{k}n} |\frac{\hbar}{m}(-i\hbar \nabla+\hbar \mathbf{k})|\tilde{u}_{\mathbf{k}n}\rangle \nonumber\\
&+i\sum_{l \tau}\sum_{LL'}\langle \tilde{u}_{\mathbf{k}n} |\Big[ e^{-i \mathbf{k}\cdot (\mathbf{r}-\mathbf{R}_{l\tau})}|\tilde{p}_{L}(l \tau)\rangle (D_{LL'}(l \tau)-\epsilon_{\mathbf{k}n}Q_{LL'}(l \tau))\langle \tilde{p}_{L'}(l \tau)|e^{i \mathbf{k}\cdot (\mathbf{r}-\mathbf{R}_{l\tau})}  , \mathbf{r}-\mathbf{R}_{l\tau}\Big]|\tilde{u}_{\mathbf{k}n}\rangle 
\end{align}

Using Bloch wavefunctions the final results can also be written as

\begin{align}
\frac{\partial \epsilon_{\mathbf{k}n}}{\partial \mathbf{k}} 
&=\frac{1}{m}\langle \tilde{\varphi}_{\mathbf{k}n} |-i\hbar \nabla|\tilde{\varphi}_{\mathbf{k}n}\rangle +i\sum_{l \tau}\sum_{LL'}(D_{LL'}(l \tau)-\epsilon_{\mathbf{k}n}Q_{LL'}(l \tau))\nonumber\\
&\times\Big[
\langle \tilde{\varphi}_{\mathbf{k}n} |\tilde{p}_{L}(l \tau)\rangle \langle \tilde{p}_{L'}(l \tau)|   \mathbf{r}-\mathbf{R}_{l\tau}|\tilde{\varphi}_{\mathbf{k}n}\rangle -\langle \tilde{\varphi}_{\mathbf{k}n} |\mathbf{r}-\mathbf{R}_{l\tau} |\tilde{p}_{L}(l \tau)\rangle\langle \tilde{p}_{L'}(l \tau)||\tilde{\varphi}_{\mathbf{k}n}\rangle \Big].
\end{align}

\end{widetext}

\section{Example case: Si\label{Si}}

In this appendix we consider cubic silicon. The \textit{ab initio} calculations were performed using a cut off energy of $400$ eV and for transport calculations Brillouin zone integrals are performed using a $100 \times 100 \times 100$ uniform grid. Because we consider low carrier concentration, the energy integrals of the Onsager coefficients are performed using $5001$ points.

The electron and hole mobilities are defined as the ratio between the band-resolved electrical conductivity and the corresponding carrier density, 
\begin{align}
\mu^e_{ep} = \frac{\sigma_{n \in \mathrm{CB}}}{n_e}, \quad
n_e = \frac{1}{V_0 N} \sum_{\mathbf{k}n \in \mathrm{CB}} f^0(\varepsilon_{\mathbf{k}n}, T, \mu),
\end{align}
\begin{align}
\mu^h_{ep} = \frac{\sigma_{n \in \mathrm{VB}}}{n_h}, \quad 
n_h = \frac{1}{V_0  N} \sum_{\mathbf{k}n \in \mathrm{VB}} [1 - f^0(\varepsilon_{\mathbf{k}n}, T, \mu)],
\end{align}
where \( \sigma_{n \in \mathrm{CB}} \) and  \( \sigma_{n \in \mathrm{VB}} \)  are the conductivites from conduction-band (CB) and valence-band (VB) states.

The mobility of $n$-type silicon, computed using the MRTA approximation, at $300$ K, is shown as a function of the impurity donors concentration in Fig. \ref{mob_n}. Black lines are used to show the results using the PBE, PBEsol, and LDA exchange correlation functionals. Experimental measurements  from \cite{JACOBONI1977}, similar to those used in \cite{ponce}, are shown using black dots. We observe that above $n \approx 10^{16}\,\text{cm}^{-3}$, the difference between calculations and experiments becomes significant. The calculated electron–phonon-limited mobilities ($\mu_{\mathrm{ep}}$) for silicon at $300\,\mathrm{K}$ are $1388$ (PBE), $1489$ (PBEsol), and $1534$ (LDA)~\(\mathrm{cm^2\,V^{-1}\,s^{-1}}\), in good agreement with previous theoretical results. Indeed, although the details of the electronic structure calculations and Boltzmann equation solvers may differ, previous works report mobilities of $1343$ and $1296\,\text{cm}^2/(\text{V}\cdot\text{s})$ for PBE~\cite{ponce,lu}, and $1509$, $1555$, $1385$, and $1970\,\text{cm}^2/(\text{V}\cdot\text{s})$ for LDA~\cite{Brunin,ponce,protik,restrepo}.

Most of the discrepancy between theory and experiment comes from the neglect of impurity scattering. In \cite{ponce} the formula derived by Debye and Conwell\cite{debye}, under the assumptions of free electrons and power scattering laws for phonons and impurities, was used to take into account the effect of impurity scattering. The above mentioned formula may be written as
\begin{align}
\mu=\mu_{ep} \Big(1+X^2 [\text{ci}(X)\cos(X)+\text{si}(X)\sin(X)]\Big), \label{mix}
\end{align}
with $X=\sqrt{6\mu_{ep}/\mu_{ei}}$, $\mu_{ei}$ the mobility due to electron-impurity scattering and $\mu_{ep}$ the mobility due to electron-phonon scattering which is computed in the present work. For $n$-type silicon, Li et al. \cite{LI1977609} proposed the empirical formula 
\begin{align}
\mu_{ei}= \frac{7.3 \times 10^{17} \, T^{3/2}}{n G(b)}, \label{mu_i}
\end{align}
with $G(b)=\ln(b+1)-b/(b+1)$, $b=24\epsilon m^* (k_BT)^2/(\hbar^2 e^2 n)$, $\epsilon=11.9 \epsilon_0$ and  $m^*=1.08m$.

When it is used in Eq. \ref{mix}, we obtain the gray curves in Fig. \ref{mob_n} which reconcile theory and experiment. 

The effect of impurity scattering is also clearly visible with temperature. In Fig. \ref{mob_T} the mobility is plotted as a function of temperature, as a dark blue curve for low doping concentration ($n=2.75\times 10^{14}\, \text{cm}^{-3}$), and as a dark green curve for higher doping  ($n=2\times 10^{19}\, \text{cm}^{-3}$). The same experimental values than in \cite{protik} are used. The data shown as blue dots, for low concentrations, are taken from \cite{canali}. For higher concentration, the data shown as green dots, diamonds and stars are taken from \cite{Yamanouchi}. For low doping concentration the agreement with experiments is excellent, but at higher doping the effect of impurities needs to be considered. Eq. \ref{mix} is used to obtain the light blue and green curves, denoted as PBE+I. Once again this empirical formula {yields better agreement with experiment.

The result of this section evidences that for highly doped compounds, the effect of impurities cannot be neglected. In the case of silicon, such an effect can be described using a simple model, as long as an empirical formula like Eq. \ref{mu_i} is available.

\begin{figure}[H]
    \centering
    \includegraphics[scale=0.2]{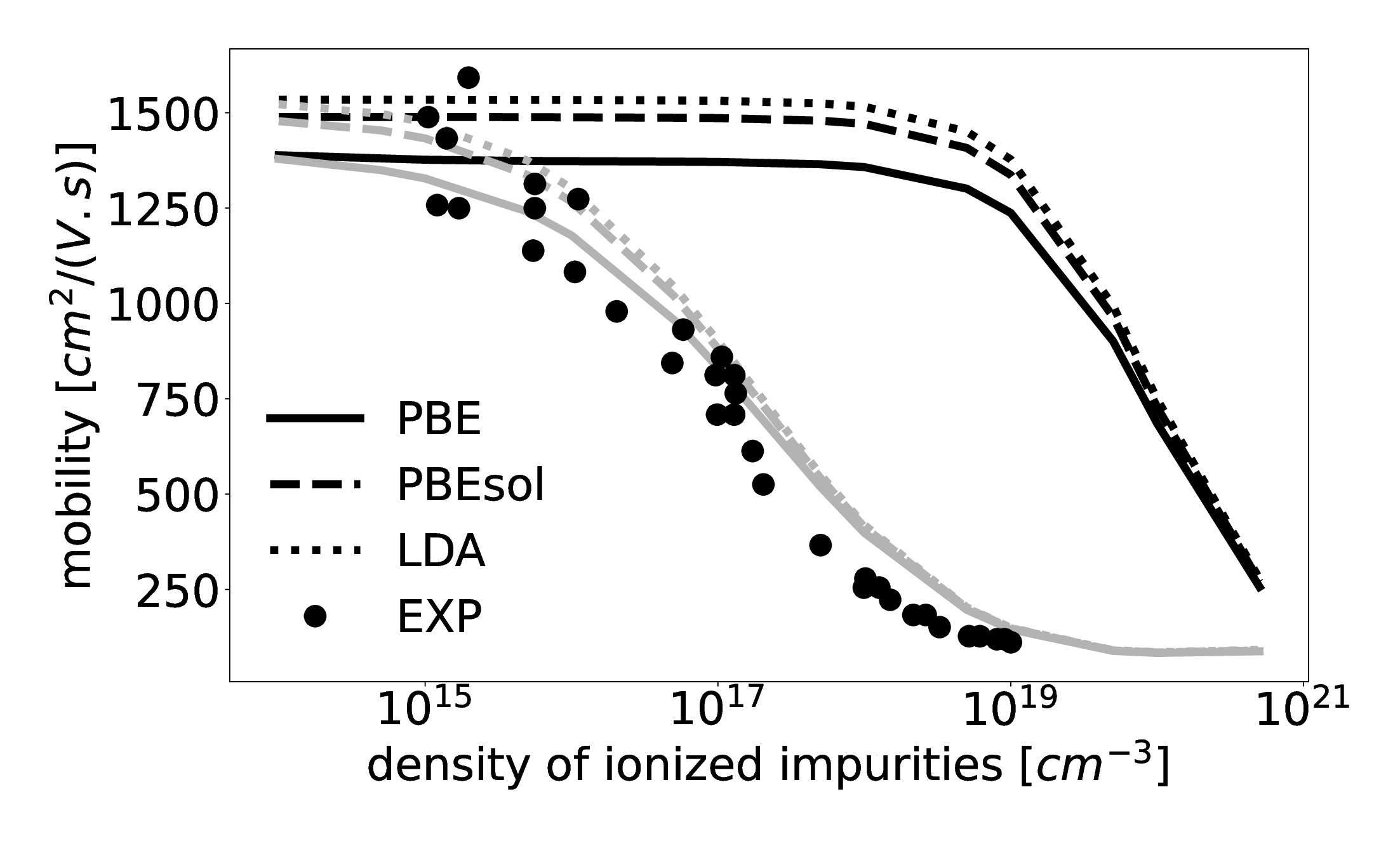}
    \caption{Mobility of silicon as a function of the impurity donors concentration at $300$ K. The lines show the results obtained for the PBE, PBEsol and LDA exchange correlation functionals, while the gray lines correspond to the same calculations but accounting for impurity scattering according to the Debye and Conwell\cite{debye} formula. \label{mob_n}}
\end{figure}
\begin{figure}[H]
    \centering
    \includegraphics[scale=0.2]{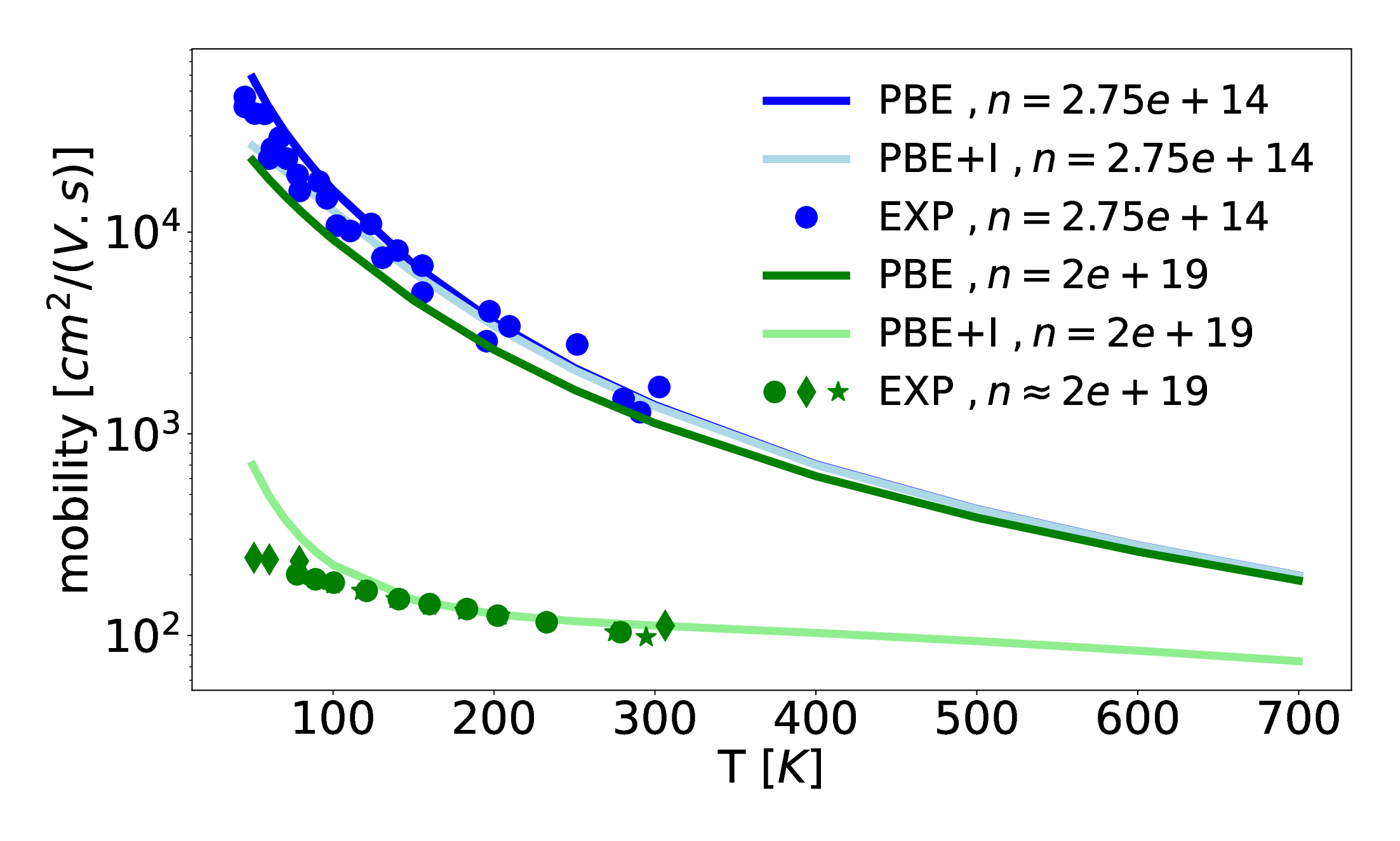}
    \caption{Mobility of silicon as a function of temperature for different carrier concentration. Dark blue and green lines show the results obtained for $n=2.75 \times 10^{14}$cm$^{-3}$ and  $n=2 \times 10^{19}$cm$^{-3}$  respectively. The light blue and green lines correspond to the same calculations but accounting for impurity scattering according to the Debye and Conwell\cite{debye} formula \label{mob_T}}
\end{figure}

\section{Spin Orbit Coupling \label{SOC}}

In this appendix we discuss the effect of spin-orbit coupling on electrical conductivity for Si and PbSe. 

In \cite{ponce}  it is shown that SOC  is important for $p$-type silicon. Using the experimental lattice parameter, we obtain, in the SERTA approximation, $586$ (PBE) and $638$ (PBE+SOC) in good agreement with the results obtained in \cite{2024_Roisin} $574$ and $611 \text{cm}^2/(\text{V.s})$ from an iterative solution of the Boltzmann equation. The experimental value is $502 \text{cm}^2/(\text{V.s})$, which can only be obtained when additional corrections are applied (see \cite{ponce}). This example reinforces our previous comment about SOC, it is not always beneficial at the level of semi-local functionals.



The results we obtain for PbSe are shown in Fig. \ref{PbSe_SO_cond}. It can be seen that including spin–orbit coupling leads to better agreement with experiment for PBEsol, but not for PBE. We can understand this result as a consequence of changes in the effective mass. As shown in Fig. \ref{PbSe_SO_PBE}, including SOC has little effect on the valance band for PBE. In constrast, Fig. \ref{PbSe_SO_PBEsol} shows that SOC increases the effective mass when PBEsol is used, which consequently leads to a reduction in electrical conductivity, as observed in Fig. \ref{PbSe_SO_cond}.

This improved agreement with experiment should, however, be interpreted with caution. The calculated effective mass is approximately $0.3$ for PBEsol and $0.6$ for PBEsol+SOC. The PBEsol+SOC result seems too large compared to the experimental value (0.28 at 300K in \cite{wang_2011}). Therefore the better agreement we obtain with PBEsol+SOC on the conductivity may only be accidental. When impurity scattering is considered in Fig. \ref{PbSe_SO_cond}, it is no longer obvious that the agreement with experiment is better for PBEsol+SOC+I, PBE+I may already be sufficient.

\begin{figure}
    \centering
    \includegraphics[scale=0.5]{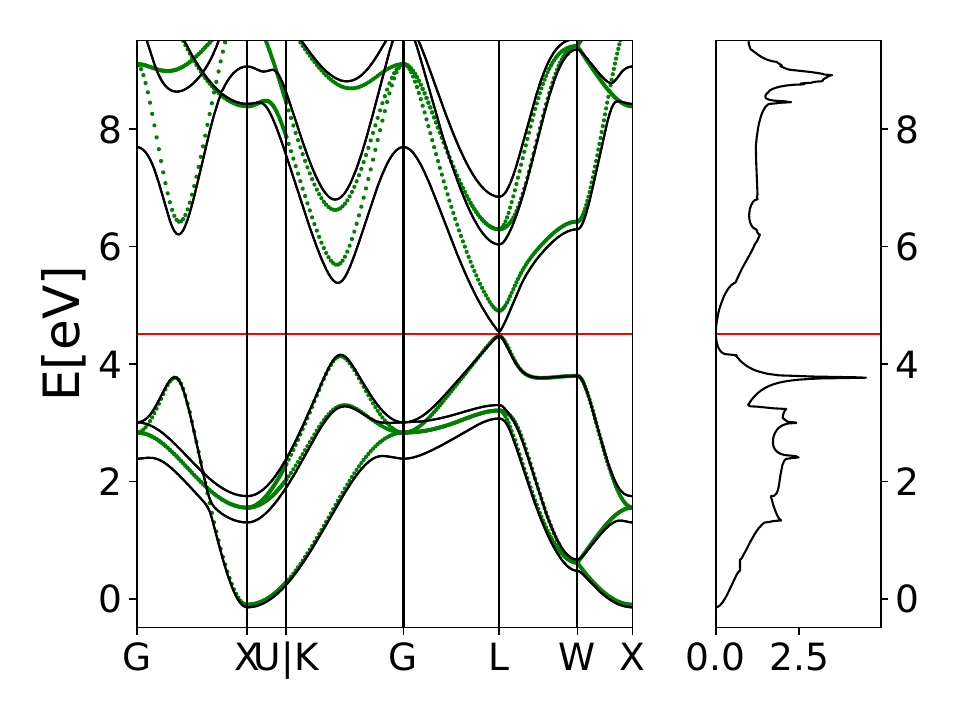}
    \caption{Electronic structure of PbSe using the PBE exchange correlation functional, with (black) and without (green) spin orbit coupling. \label{PbSe_SO_PBE}}
\end{figure}

\begin{figure}[H]
    \centering
    \includegraphics[scale=0.5]{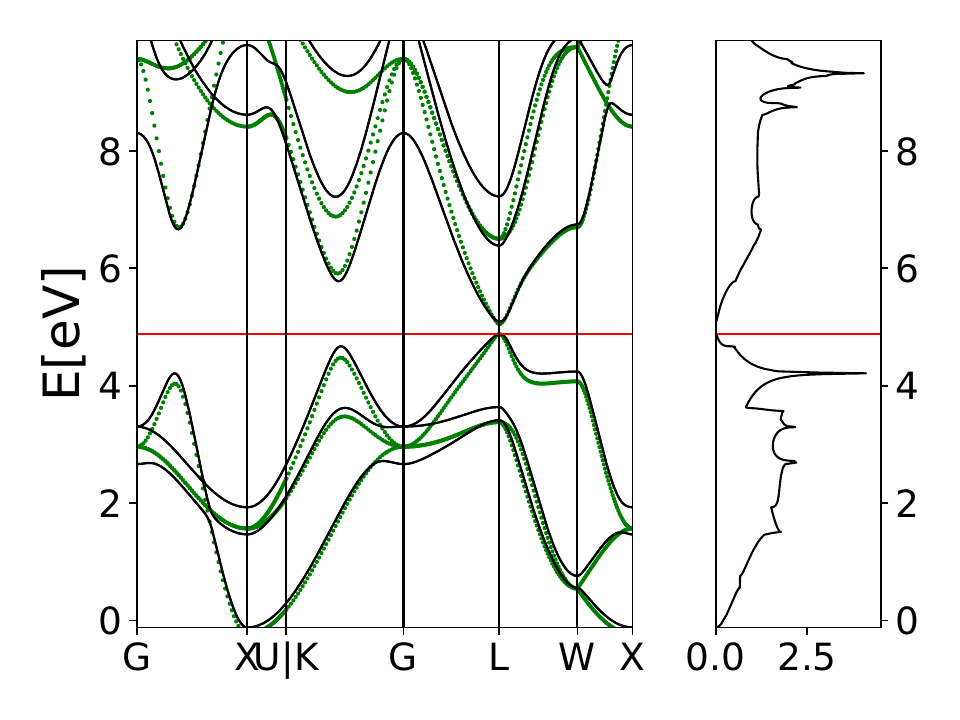}
    \caption{Electronic structure of PbSe using the PBEsol exchange correlation functional, with (black) and without (green) spin orbit coupling. \label{PbSe_SO_PBEsol}}
\end{figure}

\onecolumngrid

\begin{figure}[H]
    \centering
    \includegraphics[scale=0.25]{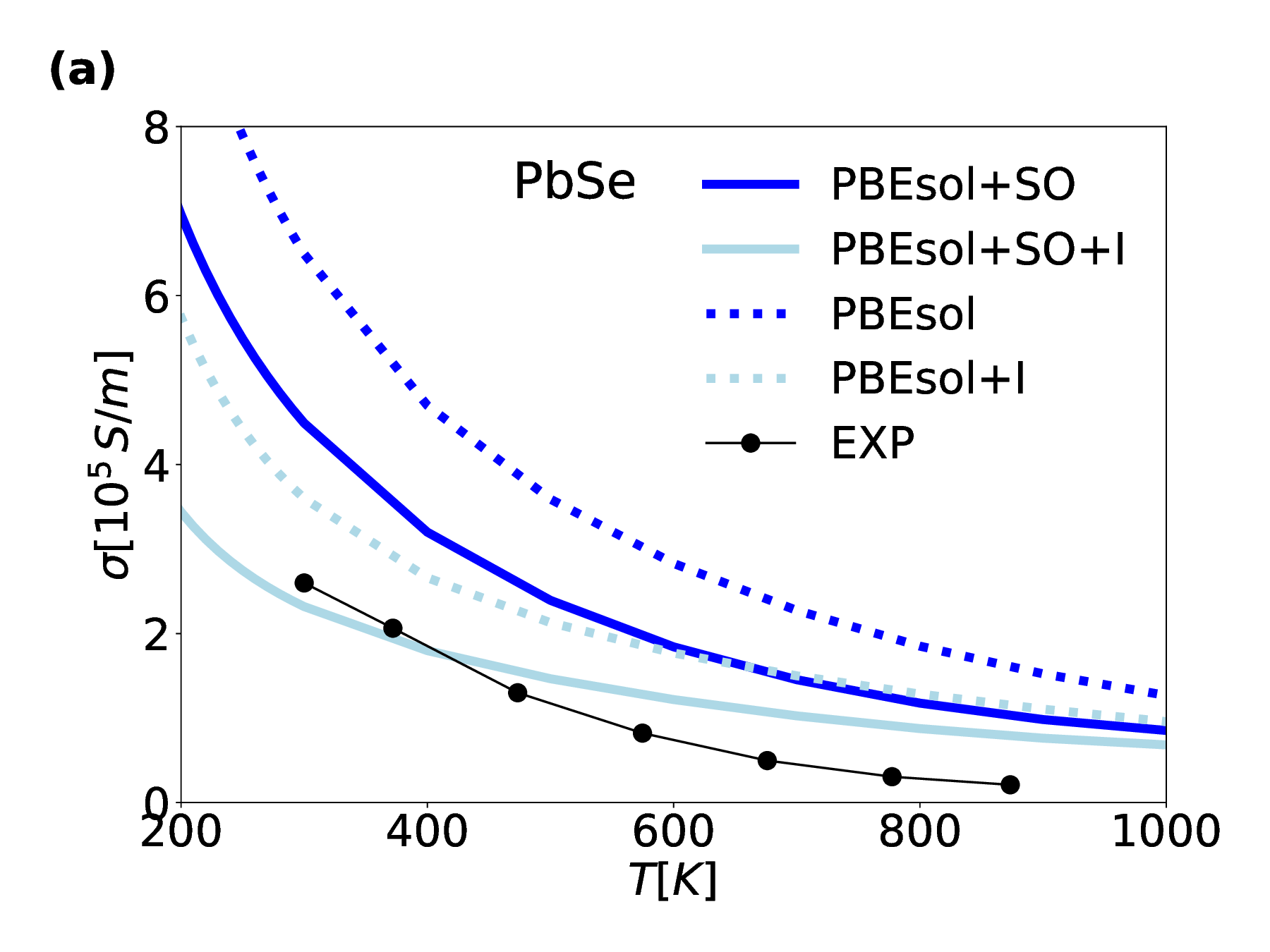}
       \includegraphics[scale=0.25]{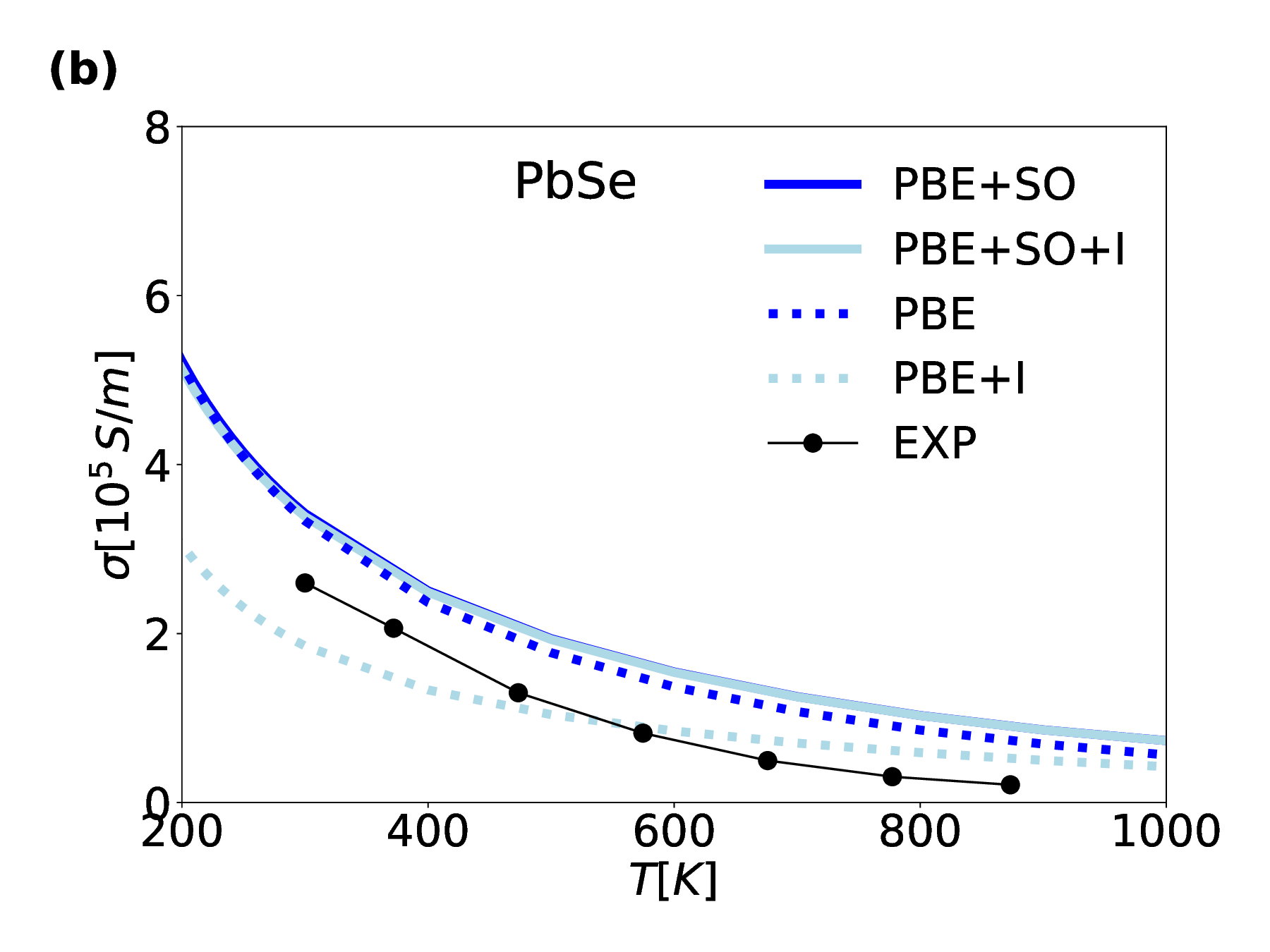}
    \caption{Electrical conductivity of PbSe computed with the PBEsol (a) and PBE (b) exchange correlation functionals. $+SO$ in the legend means that spin-orbit coupling is included. $+I$ in the legend means that impurity scattering has been considered using the Debye and Conwell\cite{debye} formula with the Brooks-Herring model, as explained in the main text. Details about the experimental values are given in Tab. \ref{roomT} \label{PbSe_SO_cond}}
\end{figure}

\twocolumngrid

\bibliographystyle{unsrt}
\bibliography{references}
\end{document}